# Olivine or Impact Melt: Nature of the "Orange" Material on Vesta from Dawn

## Lucille Le Corre

Planetary Science Institute, 1700 East Fort Lowell, Suite 106, Tucson, AZ 85719, USA.
Max-Planck-Institute for Solar System Research, Katlenburg-Lindau, Germany.
Email: lecorre@psi.edu

## Vishnu Reddy

Planetary Science Institute, 1700 East Fort Lowell, Suite 106, Tucson, AZ 85719, USA.
Max-Planck-Institute for Solar System Research, Katlenburg-Lindau, Germany.

## Nico Schmedemann

Freie Universitaet Berlin, Berlin, Germany.

## Kris J. Becker

Astrogeology Science Center, USGS, Flagstaff, Arizona, USA.

## David P. O'Brien

Planetary Science Institute, 1700 East Fort Lowell, Suite 106, Tucson, AZ 85719, USA.

## Naoyuki Yamashita

Planetary Science Institute, 1700 East Fort Lowell, Suite 106, Tucson, AZ 85719, USA.

## Patrick N. Peplowski

Planetary Exploration Group, Space Department, Johns Hopkins University Applied Physics Laboratory, Laurel, MD.

## Thomas H. Prettyman

Planetary Science Institute, 1700 East Fort Lowell, Suite 106, Tucson, AZ 85719, USA.

## Jian-Yang Li

Planetary Science Institute, 1700 East Fort Lowell, Suite 106, Tucson, AZ 85719, USA.

## Edward A. Cloutis

Department of Geography, University of Winnipeg, Manitoba, Canada

## Brett W. Denevi

Johns Hopkins University Applied Physics Laboratory, Laurel, MD 20723, USA

## Thomas Kneissl

Freie Universitaet Berlin, Berlin, Germany

## Eric Palmer

Planetary Science Institute, 1700 East Fort Lowell, Suite 106, Tucson, AZ 85719, USA.

## Robert W. Gaskell

Planetary Science Institute, 1700 East Fort Lowell, Suite 106, Tucson, AZ 85719, USA.

## Andreas Nathues

Max-Planck-Institute for Solar System Research, Katlenburg-Lindau, Germany.

## Michael J. Gaffey

Department of Space Studies, Room 518, Box 9008, University of North Dakota, Grand Forks, ND58202, USA.

## David W. Mittlefehldt

Astromaterials Research Office, NASA Johnson Space Center, Houston, Texas, USA.

## William B. Garry

NASA Goddard Spaceflight Center, Greenbelt, Maryland, USA.

## Holger Sierks

Max-Planck Institute for Solar System Research, 37191 Katlenburg-Lindau, Germany.

## Christopher T. Russell

Institute of Geophysics and Planetary Physics, University of California Los Angeles, Los Angeles California, USA.

## Carol A. Raymond

Jet Propulsion Laboratory, California Institute of Technology, Pasadena, California, USA.

**Proposed Running Head: "**Orange Material" on Vesta from Dawn

**Editorial correspondence to:**
Lucille Le Corre
Planetary Science Institute
1700 E Fort Lowell Rd #106
Tucson, Arizona 85719, USA
lecorre@psi.edu

**Abstract**

NASA's Dawn mission observed a great variety of colored terrains on asteroid (4) Vesta during its survey with the Framing Camera (FC). Here we present a detailed study of the orange material on Vesta, which was first observed in color ratio images obtained by the FC and presents a red spectral slope. The orange material deposits can be classified into three types, a) diffuse ejecta deposited by recent medium-size impact craters (such as Oppia), b) lobate patches with well-defined edges (nicknamed "pumpkin patches"), and c) ejecta rays from fresh-looking impact craters. The location of the orange diffuse ejecta from Oppia corresponds to the olivine spot nicknamed "Leslie feature" first identified by Gaffey (1997) from ground-based spectral observations. The distribution of the orange material in the FC mosaic is concentrated on the equatorial region and almost exclusively outside the Rheasilvia basin. Our in-depth analysis of the composition of this material uses complementary observations from FC, the visible and infrared spectrometer (VIR), and the Gamma Ray and Neutron Detector (GRaND). Several possible options for the composition of the orange material are investigated including, cumulate eucrite layer exposed during impact, metal delivered by impactor, olivine-orthopyroxene mixture and impact melt. Based on our analysis, the orange material on Vesta is unlikely to be metal or olivine (originally proposed by Gaffey, 1997). Analysis of the elemental composition of Oppia ejecta blanket with GRaND suggests that its orange material has ~25% cumulate eucrite component in a howarditic mixture, whereas two other craters with orange material in their ejecta, Octavia and Arruntia, show no sign of cumulate eucrites. Morphology and topography of the orange material in Oppia and Octavia ejecta and orange patches suggests an impact melt origin. A majority of the orange patches appear to be related to the formation of the Rheasilvia basin. Combining the interpretations from the topography, geomorphology, color and spectral parameters, and elemental abundances, the most probable analog for the orange material on Vesta is impact melt.

# 1. Introduction

Vesta is the largest differentiated asteroid with a basaltic surface that is mostly intact today. Since 1929, Vesta's surface has been extensively studied using ground-based telescopes in the ultraviolet, visible and infrared wavelength ranges (Bobrovnikov 1929; McFadden et al. 1977; Blanco et al. 1979; Taylor et al. 1985; Festou et al. 1991; Gaffey, 1997; Hendrix et al. 2003; Zellner et al. 2005; Rivkin et al. 2006; Reddy et al., 2010) and using Hubble Space Telescope (HST) data (Thomas et al. 1997; Li et al. 2010; Li et al. 2011). All these studies led to a better understanding of the mineralogy, space weathering effects and the global shape of Vesta. Especially, visible and near-infrared spectroscopy suggested that Vesta is one of the most diverse objects in the asteroid belt in terms of its surface composition and HST imagery showed its large-scale variations in albedo. In the meantime, laboratory analyses of reflectance spectra of meteorite samples and observations of Vesta have strengthened the idea of a genetic link between the Howardite-Eucrite-Diogenite suite of meteorites and Vesta itself (McCord et al. 1970; Binzel and Xu 1993; Burbine et al. 2001; McSween et al. 2011; Reddy et al. 2011a). However, some HEDs might originate from catastrophic impact on Vestoids as well (Moskovitz et al. 2010). These Vestoids correspond to fragments of Vesta excavated by previous impacts and ejected into space that are forming the majority of objects in the Vesta family of asteroids.

After a cruise of less than four years, NASA's Dawn spacecraft placed itself in orbit around Vesta on July 16, 2011 for a yearlong global characterization. The Framing Camera (FC) (Sierks et al. 2010) is one of three instruments onboard the Dawn spacecraft and mapped the asteroid through a clear filter and seven narrow-band filters from 0.4 to 1.0 μm. Dawn FC color data have confirmed several ground-based and HST observations of Vesta including its hemispherical dichotomy, rotational color variations (Reddy et al. 2012b, Reddy et al., in review; Li et al., in review) and the presence of dark carbonaceous chondrite xenoliths (Reddy et al. 2012d). The high-resolution color images obtained by the FC allowed the discrimination of a variety of color units and provided data to study their distribution on the vestan surface. We also use the visible and infrared spectrometer (VIR) an imaging spectrometer combining two data channels: the visible-infrared (VIS) from 0.25 to 1.07 μm, and the infrared (IR) from 0.95 to 5.1 μm (DeSanctis et al. 2010).

Finally, we use data from Dawn's Gamma Ray and Neutron Detector (GRaND), which measured the elemental composition of Vesta's regolith (Prettyman et al. 2011). Global Fe/O and Fe/Si ratios measured by GRaND show that Vesta's regolith is howarditic, strengthening the link between Vesta and the HEDs; and, measurements of the abundance and distribution of H by GRaND reveal the presence of H-rich materials likely delivered by the infall of carbonaceous chondrites (Prettyman et al., 2012). Recently, the GRaND team has completed global mapping of the abundance of total Fe ($w_{Fe}$), the thermal neutron absorption cross section ($\Sigma_A$), high energy gamma rays (HEGR), and the effective atomic mass of the regolith (Lawrence et al., in review; Peplowski et al., in review; Prettyman et al., in review; Yamashita et al., in review). A subset of these elemental parameters (Fe, neutron absorption, and HEGR) will be used to characterize selected, broad regions containing orange materials identified in FC data. Since GRaND is omnidirectional, the spatial resolution of elemental maps depends on altitude. Data acquired in Dawn's Low Altitude Mapping Orbit (LAMO) with a mean altitude of about 210 km are used. Maps determined from LAMO data have a spatial resolution of about 300 km (full-width-at-half-maximum of arc length on the surface). At this scale, distinct compositional regions can be distinguished. GRaND is further sensitive to bulk-regolith composition to depths of about 1 m.

Several color ratio and color parameters have been developed to map the distribution of HED terrains on Vesta (Le Corre et al. 2011). Reddy et al. (2012b) applied color ratios from the Clementine mission (Pieters et al. 1994) to distinguish color units using Dawn FC images. Clementine color ratio images are RGB color composites in which $C_R= R(0.75)/R(0.45)$, $C_G= R(0.75)/R(0.92)$, and $C_B= R(0.45)/R(0.75)$; where $R(\lambda)$ is the reflectance in a filter centered at $\lambda (\mu m)$ and $C_R$, $C_G$, $C_B$ are the red, green and blue channels respectively. Greener areas in this color ratio have deeper pyroxene absorption bands (typical of diogenites) and redder areas have steeper visible slopes relative to bluer areas. Reddy et al. (2012b) observed two large impact craters (Oppia and Octavia) that showed prominent red/orange ejecta in Clementine color ratio. They also noted several brighter orange patches around Oppia. The orange/red color of these craters and surrounding patches appears to come from a steeper visible slope $(R(0.75)/R(0.45))$ compared to surrounding areas.

Gaffey (1997) noticed a drop in pyroxene band area ratio (BAR) between 60-120°E longitude in rotationally resolved spectra of Vesta. He interpreted this drop in BAR as an indicator for the presence of olivine. Gaffey (1997) informally named this feature the "Leslie formation" and suggested it to be exposed olivine mantle material due to a large impact. A low albedo feature (#15) corresponds to this 'olivine unit' in HST maps of Vesta (Binzel et al. 1997; Li et al. 2010). HST color spectra showed that this unit had the reddest visible spectral slope compared to any other unit on the surface of Vesta (Li et al. 2010). This feature also had weaker 1-micron band depth compared to the average surface. Li et al. (2010) interpreted this red slope and weaker band depth as an indicator of lunar-style space weathering. Comparing HST albedo maps, ground-based compositional maps and Dawn FC data, Reddy et al. (in review) suggested that 'Leslie formation,' and feature #15 correspond to Oppia crater and its orange ejecta. This link is remarkable in the sense that orange material on Vesta was first detected by ground-based telescopes prior to the arrival of Dawn at Vesta, however its spatial context remained nebulous.

Here we present a detailed study of the nature of this orange material on Vesta. Combining multiple data sets from the suite of instruments on the Dawn spacecraft we provide a detailed geological description of the orange material, constrain its mineralogy, mineral chemistry, and meteorite affinities within the HED suite. Using this information, we test multiple hypotheses for the origin/formation of orange material on Vesta.

## 2. Data Description and Processing

### 2.1 Orbits and resolution of FC and VIR data

The FC dataset examined in this study spans a range of resolution from 487 to 16 m/pixel. It consists of FC color images, corresponding to filters F2 to F8 (Table 1) acquired from the approach phase at a distance of 5200 km from Vesta, to the High Altitude Mapping Orbit (HAMO) phase (Table 2). We also used clear filter images from the LAMO phase at the closest from the surface. The approach and orbital phases are summarized in Table 2. During the approach phase, there were three rotational characterizations (RC) consisting of the first observations of Vesta using the color filters (F2 to F8). In between each set of color images the FC also acquired clear filter (F1) frames, which have a much broader wavelength band pass. Next, the survey phase encompasses seven cycles of observations (one cycle per orbit) with the first, fourth and the seventh comprising color mapping (the other ones begin clear filter imaging). The first science phase at HAMO (HAMO 1) has six cycles with this time each cycle made of ten orbits. Among them, only cycle 1 and cycle 6 have color nadir imaging. The LAMO phase acquired 21 cycles with four cycles (cycle 9 to 12) that have color data for only three color filters (F2, F3 and F4). Clear filter images from LAMO were also used to investigate the morphology of the features studied in this paper. Another phase of HAMO observation (HAMO 2) was carried out after the LAMO, with a similar spatial resolution as that of the first HAMO observations, but will not be presented in this study. A sequence of seven color frames (or three color frames in the case of LAMO) acquired consecutively with the FC color filters is called a station.

VIR data used here consists of VIS and IR cubes from the survey phase of the mission. The best resolution obtained in survey is 676 m/pixel. Spectral cubes were selected from among the 513 VIR acquisitions from the survey phase in order to produce mosaics of different regions of interest.

*2.2 FC Color Images processing*

We present here a summary of the processing steps applied to the radiometrically calibrated FC color images (level 1c), but a detail explanation of each step from the level 0 is available in the supplementary material of Reddy et al. 2012b. The following processing steps were applied for creating color mosaic of each orbital phase. First, it is important to note that radiometrically calibrated FC imagery (level 1b) is affected by multiple reflections between the color filters and the CCD. Thus, all frames have been corrected using a complex stray light removal algorithm to produce the level 1c. Then we performed a selection of stations in order to discard the set of frames that don't have the seven corresponding filters for a given acquired color station, as well as the frames that are not pointing at Vesta. FC color frames were processed with the Integrated Software for Imagers and Spectrometers (ISIS) from United States Geological Survey (USGS) (Anderson et al., 2004). The ISIS system is designed to support various NASA spacecraft missions such as Voyager, Viking, Galileo, Mars Global Surveyor, Mars Reconnaissance Orbiter, MESSENGER, Lunar Reconnaissance Orbiter and many others. ISIS provides three major functions in the case of FC and VIR datasets: (1) ingestion and conversion of raw instrument data into ISIS format, (2) a camera/sensor model to geometrically render raw data into commonly used map projection products, and (3) photometric correction. The existence of a camera model in ISIS for both FC and VIR instruments also provides photometric orientation parameters for each image/cube observation. The combination of these features provides support for the FC and VIR data processing and analysis (Becker et al., 2012).

The input data in ISIS is what we call level 1c data. The calibrated images are first converted from Planetary Data System format (PDS) to ISIS images and relevant kernels are attached to each frame. Then they are photometrically corrected with a Hapke photometric model, which uses photometric parameters defined with both ground-based data of Vesta and vestoids, and Dawn approach and survey (Li et al., in review). Images are normalized to an incidence angle of 30° as is the case for all spacecraft data. Depending on which orbital phase the data is processed we adjusted incidence and emission angles used for trimming of the images. For HAMO data, images were trimmed to angles <80° as measured from a digital elevation model (DEM). The shapemodel used for the photometric corrections has been derived from Dawn FC clear filter images (Gaskell et al., 2012).

The color frames, for each set of acquisitions, are taken one after the other on the filter wheel of the FC, thus we need to align the color frames for each station and coregister them in order to create a multispectral cube. This is achieved by projecting the seven color frames to a common projection and performing a coregistration with the filter F3 (the least affected by stray light) as a reference frame. During the projection, all the frames are orthorectified using the Gaskell shape model derived from FC clear filter images. For each station, the 7 color frames are stacked together to create FC image cubes. Finally, FC color cubes can be assembled together to create color mosaics and other derived products (*e.g.* color ratio maps). All the maps are produced in the Claudia coordinate system (Reddy et al., in review) that is used by the Dawn Science Team.

*2.3 Visible and Infrared Mapping Spectrometer Data Processing*

VIR VIS and IR data was also processed using the ISIS software package. Processing scripts were developed in PERL and used to produce photometrically corrected reflectance maps used in this work. The FC and VIR data are processed in five main steps. Some of the processes are contained within PERL scripts that run a series of ISIS applications. Details of the first four processing steps are explained in appendix A1. This processing set the stage for successfully merging these data into one photometrically corrected, spatially aligned, spectrally combined map-projected product. The two input cubes (IR and VIS) are then stacked together. The end result is a single, combined VIR VIS and IR map-projected image cube ordered monotonically increasing by wavelength for each simultaneously acquired VIR VIS and IR observations. Additional VIR merged products can now be combined into one map product. This complete processing pipeline was repeated for all regions of interest analyzed in our study. At the spatial resolution of the survey phase, VIR data could be retrieved for 17 of the 23 sites for which FC color data was extracted from the HAMO phase.

## 3. Data Analysis

### 3.1 FC Color Ratios

FC color frames can be used to calculate various spectral parameters (such as spectral slopes or spectral ratios) that are useful for the analysis of the surface composition and lithology (Le Corre et al. 2011; Reddy et al 2012b) in a similar way to what has been applied to Clementine data of the Moon (Pieters et al. 1994). The vestan surface exhibits a 1-μm absorption band due to the presence of pyroxene. Despite the fact that FC filters do not cover the whole wavelength range that encompasses this band, the ratio of 0.75/0.92-μm can be used as a proxy for the band depth parameter (BD) as used similarly in Tompkins and Pieters (1999). Band depth can be sensitive to grain size, space weathering, illumination conditions and abundance of each component. In our case, it is used to map the diversity of geologic units as shown in Reddy et al. (2012b). In addition, a map of the slope between 0.44 and 0.75 μm provides information on the steepness of the visible slope. In order to capture most of the spectral diversity on the surface in one color composite, we combine the previous parameters to produce Clementine color ratio images as described in section 1. The identification and description of the orange material is based on this color mapping.

Other interesting color parameters are the 0.75/0.65-μm visible tilt (VT), the 0.98/0.92-μm ratio eucrite/diogenite (ED), the 0.92-μm "kink" (K920) and the 0.83-μm "kink" (K830). We will use those in combination with BD and ED parameters to analyze the composition of the terrain by comparing with laboratory spectra of HEDs. VT was previously used by Reddy et al. (2012a) to look for the absorption band of chromium in the FC color data (absorption band of $Cr^{3+}$ at 0.6 μm). ED was proposed in Reddy et al. (2012b) and is diagnostic of the eucritic- or diogenitic-like composition of the regolith. In the laboratory, spectra of eucrites exhibit band I center shift towards longer wavelengths relative to diogenites. After convolving the spectra to the FC band passes, the instrument responsivity results in a decrease of the slope between the two last filters. As described and analyzed in Reddy et al. (2012b), the higher the value the closer it gets to a diogenitic composition, and the lower the value the more eucritic is the surface. What we call "kink" parameters here were first described in Bell et al. (2000) and we adapt them to the FC filter wavelengths. The following equations correspond to the kink at 0.830 μm (K830) and 0.920 μm (K920), with $R\lambda$ being the reflectance at the wavelength λ:

$$K830 = R_{0.830}/(R_{0.750}+R_{0.920})$$

$$K920 = R_{0.920}/(R_{0.750}+R_{0.980})$$

The kink (or curvature) around the 1-μm band was first proposed by Isaacson and Pieters (2009) to quantify the overall shape of the 1-μm absorption band of pyroxene detected on the lunar surface. In the same way as ED, this parameter should be sensitive to the eucritic vs. diogenitic component of the surface of Vesta because the 1-μm pyroxene band curvature differs between eucrites and diogenites. K830 is also related to the composition of the surface. Laboratory spectra of eucrites and diogenites resampled to the FC band passes show that the kink at the 0.83 µm channel is linked to ED. Therefore we can use the value of the 0.83-µm kink calculated with the filters F3 (0.75 µm), F4 (0.92 µm) and F6 (0.83 µm) to assess the surface compositional trends.

*3.2 VIR Band Parameters*

Diagnostic spectral band parameters (band centers) are critical for the identification of specific minerals in a surface assemblage. Band center is the wavelength position of the minimum reflectance in a continuum removed absorption feature. Typically, the continuum is a straight line across the shorter and longer wavelength shoulders of the absorption band. Band centers are more robust than band minima as the influence of spectral slope is removed and it facilitates comparison with laboratory spectral data. In our analysis we will also use BAR, the Band Area Ratio, corresponding to the ratio of the area of 2-µm ferrous iron absorption band in pyroxene to the area of 1-µm absorption band.

HED meteorites differ from each other primarily based on their pyroxene composition and their pyroxene to plagioclase feldspar ratio. These ratios were estimated to be 3:1 for howardites, 1:1 for eucrites, and 30:1 for diogenites (Gaffey, 1976; Mittlefehldt et al., 1998). The Fe content in HED pyroxenes increases from diogenites ($Fs_{20-30}Wo_{1-3}$), to cumulate eucrites ($Fs_{30-44}Wo_{6-10}$), to basaltic eucrites ($Fs_{43-55}Wo_{9-15}$) (Mittlefehldt et al., 1998). Pyroxenes typically have band I centers between 0.90-0.94 µm for low-Ca pyroxenes, 0.91-1.07 µm for type B high-Ca pyroxenes and band II centers between 1.78-1.97 µm for low-Ca pyroxenes and ~1.97-2.36 µm for type B high-Ca pyroxenes (Cloutis and Gaffey, 1991). Diogenites have band I centers between 0.918-0.924 µm and band II centers between 1.88-1.92 µm, in contrast eucrites have band centers at much longer wavelengths (band I centers 0.931-0.950 μm; band II centers 1.97-2.05 μm) due to higher iron abundance in their pyroxene (Burbine et al. 2010). Changing plagioclase feldspar abundance does not affect pyroxene band centers, as plagioclase is nearly spectrally neutral (Adams and Goullaud, 1978).

Using band centers, mean pyroxene chemistry can be estimated for HED meteorites to distinguish eucrites from diogenites. Within eucrites, basaltic eucrites have higher iron and calcium abundance than cumulate eucrites. No clear trend is observed in the BAR values of HEDs as band area/band depth is not diagnostic of ferrous iron abundance and can be influenced by particle size, phase angle, temperature, and opaque abundance (for example Reddy et al. 2011a, 2012b, c, d). Similarly, spectral slope is not a diagnostic parameter and can be influenced by phase angle, particle size, temperature, presence of metal, and space weathering.

Surface temperature also has an influence on band centers. Temperature causes broadening or narrowing of absorption features (affecting BAR), and shifting of band centers (Singer and Roush, 1985; Hinrichs and Lucey, 2002; Hinrichs et al., 1999; Moroz et al., 2000; Schade and Wäsch, 1999). Reddy et al. (2012c) quantified the influence of temperature on spectra of HEDs and developed equations to correct remotely sensed spectra. Typically, temperature effects on HED band I centers are negligible and slightly larger for Band II centers. Temperature corrections will enable spectral data of Vesta obtained between 200-270K (Capria et al. 2012) to be interpreted using laboratory spectra taken at room temperature (300K). Compared to room temperature spectra (300 K), band I and II centers of spectra of howardite/eucrite terrains on Vesta obtained at 200 K have negative offset of about 0.005 μm and 0.016 μm, respectively. Similarly, band I and II centers of diogenite-rich terrains on Vesta obtained at 200 K will show a change of about 0.016 μm and 0.013 μm compared to laboratory spectra obtained at room temperature (300 K). Temperature effects on band I and II centers translate to about 10 mol. % and 3 mol. % overestimation of ferrosilite (Fs) and wollastonite (Wo) content in diogenites, and 4 mol. % and 1 mol. % overestimation in ferrosilite (Fs) and wollastonite (Wo) content in eucrites and howardites using Burbine et al. (2010) equations. The overestimation of Fs (4 mol. %) and Wo (1 mol. %) content for howardites and eucrites due to temperature stays within the uncertainties from spectral calibration (3.3 mol. % for Fs and 1.1 mol. % for Wo) by Burbine et al. (2010) and can essentially be ignored. Additionally, all our studies sites fall outside the diogenite-rich areas making temperature correction a lesser concern. It is also interesting to note that Mayne et al. (2011) argues that temperature corrections are not needed. Therefore, given that we observed only small shifts in band centers due to temperature, we will not apply any temperature correction to the spectral data.

## 4. Description of Orange Material

*4.1 Types of Orange Material Deposits: Morphology and Color Properties*

### 4.1.1 Diffuse orange material

The first diffuse deposit of orange material described on Vesta using FC color data from the approach and survey phases was the ejecta of Oppia crater (Reddy et al. 2012b). This ejecta stands out even in low-resolution color ratio maps produced with FC color data. As seen in Fig. 1, the non-circular crater shape, local topography and asymmetric ejecta indicates that this 34 km crater is the result of an impact on a slope. The local topography has been influenced both by the set of troughs associated with Divalia Fossa at the equator and ~50 km to the north of Oppia, and by the set of troughs associated with Saturnalia Fossa (Buczkowski et al. 2012). For location of all the geologic features on a global map, the reader is referred to the USGS webpage dedicated to Vesta's nomenclature (http://planetarynames.wr.usgs.gov/Page/VESTA/target). A significant topographic depression is present between the two sets of troughs, just north of Oppia, called Feralia Planitia. To the south of Oppia is the Rheasilvia ridged terrain bordered by the eroded basin rim. This configuration suggests the pre-impact surface to be a slope as displayed by the gradient of color (Fig. 1 B and D). The ejecta made of orange/red material in the Clementine color ratios is mostly distributed on the southern part of the crater, deposited on the rim of Rheasilvia and its ridged terrain, but it also covers the floor of the Oppia crater and most of the walls (except the upper part of the crater walls which have higher reflectance). It also has a somewhat ellipsoidal shape with a major axis in the northwest-southeast direction of about 200 km and a minor axis of about 130 km.

Octavia's ejecta appears to have the same kind of asymmetrical shape and direction of deposition (northwest-southeast) but the majority of the ejecta material is deposited on the northern part of the crater (Fig. 1E). The local topography is relatively flat around the crater rim, suggesting an oblique impact may explain the asymmetry of the ejecta. The crater is located in the equatorial region and seems to have impacted on top of the set of giant troughs (Fig. 1F) that is associated with Divalia Fossa (Buczkowski et al. 2012). The orange/red ejecta spans ~190 km in its longest axis around the 30 km diameter crater. The orange material is also present on the crater wall and on the large slump of material that has collapsed toward the center of the crater. Dark material (dark bluish in the Clementine color ratio images) forms debris flows along the crater rim and some dark material is present in the ejecta field in the form of small craters (either secondary craters or subsequent craters) or stripes (preferentially oriented east-west). Some of the small craters located in the orange ejecta excavated brighter orange material than the surroundings, likely corresponding to fresher material from within the ejecta deposit or to sampling of the surface below. Both Octavia and Oppia lack prominent bright or dark crater rays extending from the crater rim such as the ejecta rays that can be seen for other smaller but clearly fresh impact craters on Vesta (e.g., Cornelia). The lack of rays could be indicative of differences in the conditions of emplacement of the craters or in the underlying stratigraphy of the terrain, or that such features have already been erased by impact gardening due to the relatively older age of Oppia and Octavia compared to craters with distinct ejecta rays.

### 4.1.2 "Pumpkin" patches of orange material

In addition to the diffuse deposits described previously, other orange deposits stand out in the FC color ratio images with a great diversity of shapes (Fig. 2 and 3). Most possess clearly visible edges, which indicate sharp margins for these formations, along with lobate morphology resembling flow-like features. The range of dimensions of these orange patches is from about 150 m to 62 km for the largest deposit. When compared with topographic information, no clear relationship is found with a specific relief feature or common elevation that could point to a specific vestan lithology. This class of features can be divided further: those that do not exhibit a specific link with any nearby impact crater (Fig. 2) and the patches that seem to have formed in association with a nearby impact crater (Fig. 3). The morphology and the geologic settings of these patches give us indirect clues about the nature and origin of this orange material: it does not seem to be linked with specific geologic features or elevation but it is often associated with impact craters; and it does not seem to be confined in a stratigraphic layer in the subsurface.

Lobate orange patches are also found in association with the diffuse orange ejecta of the Oppia crater (Fig. 1 A and C). They are mostly located to the west and the north of the crater rim where the orange diffuse deposit is scarce. It is difficult to distinguish if the patches are pre-existing or were deposited during the impact event. One of the patches overlaps a part of the crater rim of Oppia. North of Oppia, a smaller and likely more recent impact crater excavated material with higher reflectance on top of some orange patches. Most of the patches found here have roundish shapes except one with an arcuate shape that possibly follows an ancient/buried crater rim. The largest of these patches are often interconnected with filaments of the same orange material. All of them possess characteristic distinctive margins with the surroundings, even when they are found in the orange/red diffuse ejecta deposit. This suggests that the orange material in the patches could have been emplaced subsequent to the orange/red material in the diffuse ejecta but before the small fresh crater that impacted north of Oppia (Fig. 1 A and C). This also indicates that the patches are not really thick if a small impact crater (~5 km in diameter) could excavate the underlying lithology. In addition, we also observed orange patches in the diffuse ejecta material of Octavia, but they are much more scarce and smaller (only few kilometers) (Fig. 1E).

We used high spatial resolution LAMO clear filter images (~16 m/pixel) to investigate whether the orange patches are associated with any specific texture or small-scale morphological features. Fig. 4 shows that no differences in the geology of these units are found relative to the surrounding terrains (usually associated with bluish color in the Clementine color ratio images). They are overlapping terrains formed by ejecta and craters <1km in diameter are able to excavate the underlying layer of fresher (bright) material.

### 4.1.3 Orange material in crater ejecta rays

Cornelia (Fig. 5A) and Rubria (Fig. 5B) craters expose orange material but in a different way than the previously discussed diffuse material and patchy materials. 15 km-crater Cornelia has some orange material on the crater walls (overlaying bright or dark material), at the center of the crater where the pitted terrain has been observed (Denevi et al. 2012b), and distributed radially from the crater rim. The ejecta was formed with both diffuse orange material and patches with distinct boundaries with the surroundings (on crater wall and outside the crater rim). The diffuse orange material could correspond to an impact melt sheet and the patches to larger amounts of impact melt deposited, producing flow-like morphologies. Some of the ejected orange material outside the crater exhibits lobate fronts that could suggest melted material emplaced after the impact. The impact melt interpretation would be consistent with the observation of a pond of orange material on the pitted terrains, which represent terrains that lost volatiles after the impact and have been melted (Denevi et al. 2012b). Orange material is also found where morphologic evidence for impact melt has been observed outside the crater rim (Denevi et al. 2012b). Other orange patches can be found further away from the crater ejecta and could be related to the Cornelia impact or pre-existing units.

Based on Fig. 5A, the orange material in Cornelia has likely been emplaced last during the impact event as it is largely overlying bright and dark ejecta units with only dark and faint ejecta rays deposited on top of it. For 10 km-crater Rubria (Fig. 5B), the dark material in the ejecta is clearly emplaced on top of the orange material (forming ejecta rays) in the upper half of this asymmetric crater. Therefore, the configuration of the units might depend on the conditions of the impact (oblique, impact on slope) and pre-existing stratigraphy. Another recent looking impact crater, 11.2 km-diameter Arruntia (39.4°N and 71.6°E), also has orange/red ejecta rays and diffuse orange ejecta (like Oppia and Octavia) but is not discussed in detail. Located at high northern latitudes, it has been observed with high sun incidence angles only during HAMO1, resulting in a FC color data quality that is not well-suited for analysis of this crater. Observations from the HAMO2 phase, which are not included here, would be required for precise analysis.

### 4.1.4 Similarities with specific landforms on the Moon

The generally sharp boundaries and the roundish shape of the orange deposits are quite similar to the morphology of mare basalt units on the nearside of the Moon seen with the ultraviolet and visible camera (UVVIS) onboard the Clementine spacecraft (Fig. 6 A). In particular, the filaments of material connecting orange patches and the roundish edge of the deposits, sometimes filling depressions/old craters are identical to the lunar volcanic landforms. Some orange material forming an arcuate shape (like one North of Oppia crater in Fig. 1 A and C) appear similar to arcuate shapes seen in the lunar mare basalts in Fig. 6 A. In the same way as for the Vestan orange deposits, the volcanic deposits on the Moon have been impacted by small craters excavating brighter and possibly fresher underlying material. However, the lunar mare basalt presents some key differences with the orange deposits, for example the lunar mare basalts have a topography signature whereas the vestan orange deposits have no apparent thickness (Fig. 4). Only based on their appearance in color data, we considered that some of the patches could be of volcanic origin. We searched for possible source vents or fractures in the FC clear and color images from HAMO and LAMO in the vicinity of the orange material but none could be identified. There is also no evidence for discrete lava flows units overlapping each other like on the lunar mare basalts. Most of the orange patches seems rather well preserved and are probably too young to be related to any volcanic activity. Therefore, from a geomorphology point of view, this explanation is unlikely.

The thin and lobate patches of orange material are reminiscent of the appearance of thin impact melt flows forming sheets on the Moon (Öhman et al. 2012, Denevi et al. 2012a). The striking resemblance with impact melt can be also highlighted by comparing with Copernicus crater on the Moon (Fig. 6B). This 93 km crater has already been analyzed with Clementine UVVIS color ratio images and color spectra by Pieters et al. (1994). They noted that the impact melt glass identified in previous spectroscopic studies appears as red/orange deposits in color ratio composite. It is not distributed uniformly, concentrated mostly in the northwestern part of the crater floor and wall. Comparative analysis of the impact melts' spectra in Copernicus crater (from telescopic data) with mixtures of melt + crystalline lunar samples demonstrated the presence of at least 50% of glass (Tompkins and Pieters, 2010). The impact melt material seems to be also deposited radially from the crater and further away from the crater rim to the northwest (see arrows Fig. 6B). Despite the difference in diameter, Copernicus appears to be a good lunar analog for Cornelia crater in the Clementine color ratios, with orange deposits on the floor, walls and ejecta of the crater as well as brighter material on the wall of the crater and dark ejecta rays outside the crater rim. By using Clementine color ratios, we can infer impact melt as a possible hypothesis. Nonetheless, as shown in Fig. 4, the orange patches do not display a smoother texture relative to the surroundings or levees like those observed on the Moon (Öhman et al. 2012). This could be due to the spatial resolution of the FC data that is not sufficient to observe these features, or that gravity driven flows are harder to form in low gravity environment, or maybe that the volume of melt is not sufficient to really flow.

Like Tycho crater on the Moon (Pieters et al. 1994), Oppia and Octavia craters might represent a case in which impact melt material is prevalent in the vicinity of the crater rim. Neither of these craters shows the bright or dark material found on the crater wall excavated close to the crater rim, as we would expect to see with an inverted stratigraphy emplaced after an impact. If these orange deposits are actual impact melt products, we could infer the same conclusions as Pieters et al. (1994) concerning Tycho and Copernicus craters: these large and rather uniform deposits of melted material imply high energy impact events. In order to assess the composition, we will investigate the color properties in the next section.

### 4.1.5 Color properties of orange material

Simple color properties of the orange material can be derived from the FC color spectra average on specific areas (Fig. 7). All the distinct orange/red color units identified in the mosaic of Clementine color ratios (Fig. 7A) has been mapped by creating shapefiles in ENVI with the ROI tool. This allows the extraction of statistics for each unit and the average spectra for each of them. Fig. 7A shows average spectra normalized at 0.75 µm for the Oppia ejecta (spectrum 1), Octavia ejecta (spectrum 14), for two distinct cluster of patches (group 4 located in the eastern hemisphere and group 17 located in the darker western hemisphere), as well as the global average spectrum for Vesta for comparison. Color properties of Oppia ejecta consist of a steep visible spectral slope (the cause of the orange color in Clementine ratios) and a shallower 0.90-micron pyroxene band compared to the global average spectrum of Vesta. Octavia is distinct from Oppia in that its ejecta has a shallower band depth for band I and a visible slope not as steep as for Oppia. Also, its crater walls have lower reflectance at all wavelengths than for Oppia. Octavia is located in the dark hemisphere of Vesta (Reddy et al., 2012b and 2012d), whereas Oppia is in the brighter hemisphere. This could explain the weaker 0.90-micron pyroxene band depth and the overall lower reflectance of Octavia if the material from the ejecta has mixed with the local regolith during impact.

The orange material in the pumpkin patches has similar color properties to the diffuse orange material found in the Oppia ejecta. They exhibit lower 0.75/0.92 µm band ratio than the average spectrum of Vesta and a steeper visible slope. We analyzed all of the clusters of patches mapped in Fig. 8 C and D and found no significant differences among them. Only the depth of band I changes from cluster to cluster. Group 4 and 17 from Fig. 7 are two examples showing the maximum variability we can see, with group 4 having a band I depth almost identical to average Vesta whereas group 17 has a lower band depth. Again, this could be explained by the mixing with the local regolith (in either the dark hemisphere or bright hemisphere). In 0.75 µm-albedo maps such as the ones presented in Reddy et al. (2012b,d) the patches are difficult to distinguish from the surrounding regolith given that they do not have a specific texture and that they have usually only slightly higher albedo. Meanwhile, the orange diffuse ejecta from Oppia and Octavia have a lower albedo relative to the nearby terrains. Overall the orange patches have a 0.75 µm-albedo of 0.17, which is higher than the ejecta deposits of Oppia (0.16) and Octavia (0.14). In some cases we observed that the sharpness of the edges and color contrast of the patches in Clementine color ratios are less pronounced and might indicate a weathering process in which the orange patches are fading away in the background terrain with more diffuse edges the older they become (like some patches in Fig. 2C), similar to how bright, newly-exposed material on Vesta "fades" into the background due to regolith mixing processes (Pieters et al., 2012). Some of the orange patches could correspond to thinner deposit, thus eroding more quickly with the effect of impact gardening, or alternatively they could represent older deposits. Orange patches in the vicinity of recent craters, such as Cornelia for example, appear less evident because they are overlapped by ejecta rays and likely mixed with local regolith. The crater density does not appear to be different between the orange patches and the nearby terrain so it is not possible to date the orange patches.

The average spectra for area 7a, the ejecta of a recent crater that impacted in the orange ejecta of Oppia, still shows the steep visible slope. But in contrast with the other examples discussed, it has a deeper band I depth than the average of Vesta. This crater has excavated fresher material (with higher band depth and higher 0.75 µm-albedo) that has mixed with the orange material. Therefore we can conclude that in general, orange material has a steeper visible slope while the 1 micron-pyroxene absorption can be deeper or shallower than the average value for Vesta. This likely depends on the characteristics of the pre-existing geologic unit, in particular the strong East-West dichotomy of Vesta.

*4.2 Mapping the Distribution of the Orange Material*

The global mosaic in Clementine color ratios (Fig. 8 A and B) has been used to map the distribution of all the different types of orange material units in a Geographic Information System (GIS). Preliminary mapping was done with RC3b color mosaic and the units were refined and completed with the global color mosaic from HAMO1. The corresponding polygons overlaid on a color-coded topographic map are shown Fig. 8 C and D. Noteworthy, the majority of the orange patches as well as the Oppia and Octavia ejecta are confined to 30°S and 30°N. The Octavia and Arruntia ejecta and some cluster of light orange patches are found in the low albedo hemisphere of Vesta (in dark blue-violet in Fig. 8 A and B). This dark terrain has a lower albedo than the rest of Vesta, especially at 0.75 µm (Reddy et al. 2012b,d). On the other hand, Oppia ejecta and numerous clusters of orange patches are present in other types of terrain with higher reflectance. Therefore, the orange material is distributed across all longitudes on Vesta with the exception of the longitude range 30°E-90°E where material from Rheasilvia ejecta seems to have covered the terrain (Reddy et al. 2012b), and is not associated with any specific preexisting terrains.

Some groups of orange patches appear to form spatial patterns. For example, around 0°E orange patches have formed following the topography of some of the troughs of Divalia Fossa and old craters present in the same area (see in Fig. 2A, C and G). Another example is on Vestalia Terra, where two major groups of patches (located at 0-10°S; 85-130°W) are oriented parallel to the Saturnalia Fossa set of giant troughs. So their emplacement must have been affected by the topography of both sets of troughs (Saturnalia and Divalia), giving a hint on their relative ages. Numisia crater (partially visible in Fig. 3A and B) impacted in the middle of one of the groups of troughs and it is not evident to distinguish if the orange patches were there before or after (possibly formed by the Numisia impact). Some orange patches between 130°E and 140°E cross the rim of the Veneneia basin and others are located in its basin floor. Along longitudes 0°, ~115°E and 40°W, clusters of orange patches have been deposited at the same longitude, like a ray of material ejected from the South and draped over the topography were they landed (e.g. crest, floor or walls of troughs). In some cases, orange material appears as if it has flowed down the crater rim and wall or local topography downslope (example in Fig. 3E), so it would have stayed fluid for some time after deposition. Moreover in some areas, patches seem to have been deposited preferentially on the south-facing flanks of old craters. But it is not the case for all the orange material as some rays and patches are clearly linked to an impact (like for the Cornelia and Rubria crater). Except for the southern part of Oppia's diffuse orange ejecta, none of the orange units cross the rim of the Rheasilvia basin. These observations could suggest that most of the orange material found in patches would have been deposited by the Rheasilvia impact and could represent impact melt and/or excavated material. For Oppia, Octavia and Arruntia, the diffuse ejecta could be either material excavated from the Rheasilvia impact that was covered up by regolith over time, or produced during the individual impact events as either impact melt due to a violent impact or as remnants of the impactor materials.

For the first hypothesis, that would imply that the orange layer deposited by Rheasilvia is not uniform or at constant depth given that not all fresh craters possess this kind of ejecta. In the south polar projection (Fig. 8D), the distribution of orange material forms a broad pattern around Rheasilvia basin with material deposited radially from the basin rim. Hence it could correspond to the deposition of ejecta/melted or excavated material from the south pole. Mass wasting of material in the ridge and groove terrain of the basin floor might have covered the orange material in this region. Alternatively, the orange material could have been mixed with the regolith breccia during impact making it undetectable in the basin floor.

*4.3 Formation Age of Oppia and Octavia craters*

### 4.3.1 Oppia crater

Oppia is only ~20 km away from the rim of the Rheasilvia basin and formed on top of the Rheasilvia ejecta blanket. We measured crater frequencies in two areas at about opposite directions of the Oppia crater, on the southern and northeastern ejecta blanket (Fig. 9A), in order to determine the formation age of Oppia. The methodology of the crater counting technique is explained in appendix A2 and the selected sites are described in appendix A3.

The northeast area shows an intense resurfacing event, which destroyed craters below 2 km in diameter. The derived surface age from craters larger than 2 km is 3.62 +0.05/-0.09 Ga. This would be the age of the surface in which Oppia formed. The small craters give an age for the resurfacing event of 276 +/- 23 Ma (Fig. 9C). This would be the age of the formation of Oppia. The southern counting area reveals the same result within the error bars: 3.63 +0.06/-0.1 Ga for the base age, and 272 +/-9.3 Ma for Oppia resurfacing (Fig. 9B). This area also shows an additional resurfacing before the formation of Oppia. This resurfacing could have been caused by another impact, which blanketed the surface before Oppia formed. This resurfacing event can be dated to 1.83 +/- 0.3 Ga.

Since Oppia very likely formed on the Rheasilvia ejecta blanket, the derived base ages also imply an age of at least 3.6 Ga for the Rheasilvia ejecta. Using a different methodology, Marchi et al. (2012) found a much younger age for Rheasilvia, closer to 1 Ga.

### 4.3.2 Octavia crater

Octavia is located ~20 km away from the rim of the Veneneia basin and formed on top of the Veneneia ejecta blanket. Its pristine morphology also indicates that it is very likely not blanketed by Rheasilvia ejecta either. Therefore, Octavia, like Oppia, was formed on top of the Rheasilvia stratigraphic layer, which is underlain by the Veneneia strata. At the position of Octavia relative to the Rheasilvia and Veneneia basin we would expect thinner Rheasilvia ejecta and thicker Veneneia ejecta in contrast to the position of Oppia. In order to determine the formation age of Octavia, we measured crater frequencies in two areas on the southern and southeastern ejecta blanket (Fig. 10A). The methodology of the crater counting technique is explained in appendix A2 and the selected sites are described in appendix A3.

The larger area south east of Octavia shows an intense resurfacing event, which destroyed craters below 2.5 km diameter. The derived surface age from craters larger than 2.5 km is 3.78 +0.04/-0.06 Ga. This would be the age of the surface in which Octavia formed. The small craters give an age for the resurfacing event of 372 +/- 30 Ma (Fig. 10C). This would be the age of the formation of Octavia. The smaller counting area reveals the same result within the error bars for the Octavia formation of 356 +39/-38 Ma (Fig. 10B). The base age derived from larger craters could not be derived without a huge error, because the area is too small to capture a significant number of craters, large enough to be visible through the thick ejecta blanket of Octavia at this area. Both measurements give reliable resurfacing ages for the formation of Octavia around 360 to 370 Ma.

The high base age measured in the larger area is significantly older than what has been found for the base age of Oppia, which probably dated the Rheasilvia ejecta blanket. Given the close proximity to the Veneneia basin and larger distance to the Rheasilvia basin, the derived base age of around 3.78 Ga dates the ejecta blanket of Veneneia, rather than that of Rheasilvia. In case the area around Octavia was covered by Rheasilvia ejecta, it was not thick enough to blanket craters larger than 2 km. Given a depth to diameter ratio of 0.2, this converts into a layer thickness of about 400 m (Jaumann et al., 2012). The later Octavia ejecta has to be thick enough to cover all the craters on the putative Rheasilvia ejecta blanket to erase the Rheasilvia resurfacing from the cratering record. This can be accomplished for instance, if the Octavia ejecta thickness is larger than the Rheasilvia ejecta in the measured area.

## 5. Composition of the Orange Material

### 5.1 Spectral Band Parameters

Table 3 lists all the sites (identified from FC Clementine color ratios map) from which average VIR spectra from the survey phase were extracted for mineralogical analysis along with their spectral band parameters. For each of the pyroxene absorption bands, band center, band depth, band continuum slope, band area were calculated in addition to BAR and visible slope (0.55-0.65 μm). Band I center for orange material ranges from 0.930-0.937 μm, band II center from 1.955-1.996 μm, and BAR from 1.45-1.98. The narrow range of these three key spectral parameters suggests compositional homogeneity among the 17 sites despite being geographically dispersed across the entire surface of Vesta.

Fig. 11 shows band I center vs. BAR from Gaffey et al. (1993) along with zones for various S-type asteroid subtypes. The sinuous line is the mixing line for olivine and orthopyroxene (OPX) with BAR increasing to the right for higher OPX abundance. As noted previously, the location of Oppia coincides with that of the proposed olivine-rich unit observed by Gaffey (1997) from ground-based telescope and with feature #15 observed in HST albedo maps (Li et al. 2010). To quantitatively verify this link, we plotted band I center and BAR of the orange material from all our sites (including Oppia ejecta) and the "Leslie feature" from Gaffey (1997) in Fig. 11. Band parameters of these two features are consistent with each other confirming the link between the 'Leslie feature,' feature #15, and Oppia. The band I center and BAR of the orange material also plots on the left corner of the basaltic achondrite zone (rectangular box) with a few sites spreading towards S(V) subtype region. The band parameters for orange material are clearly offset from the olivine + orthopyroxene mixing line suggesting the presence of clinopyroxene, which typically increases the band I center without an increase in BAR (Gaffey et al. 1993).

*5.2 Pyroxene Chemistry*

Using band parameters listed in Table 3, we calculated several pyroxene chemistry parameters based on calibrations from Gaffey et al. (2002), Burbine et al. (2010) and also developed new equations for estimating total iron (Fe wt.%). Ferrosilite (Fs), enstatite (En) and wollastonite (Wo) abundances were calculated independently using two sets of calibrations from Gaffey et al. (2002) and Burbine et al. (2010). We performed the analysis from these average Fs, En and Wo values that were calculated. Mg # was calculated using the calibration from Burbine et al. (2010). Then total Fe (wt.%) was calculated based on a spectral calibration we developed using spectral parameters from 45 HED meteorites and microprobe measured values of Fe (wt.%). Two calibrations were developed for band I and II centers and finally, an average Fe wt.% was calculated from these values (see Section 5.3). The two equations along with their coefficient of determination ($R^2$) values are given below. The total Fe (wt.%) values have a 1-σ error of 0.21 wt.%.

Fe (wt.%) = 111.3.α − 90.073 with α = band I center (Eq. 1) ($R^2$ = 0.74)

Fe (wt.%) = 19.814.β − 25.248 with β = band II center (Eq. 2) ($R^2$ = 0.72)

The range for pyroxene chemistry that we obtained for the 17 sites is between $Fs_{38-46}Wo_{8-11}$ with an average of $Fs_{42}Wo_9$. Fs and Wo values have calibration uncertainties of 3.3 mol. % and 1.1 mol. %. Consistent with band parameters, the narrow pyroxene chemistry range observed among the 17 orange material sites suggests similar composition and formation mechanism.

*5.3 Relationship with HEDs*

Establishing the genetic relationship between the orange material and HED suite of meteorites is critical for understanding their origin. Using near-IR spectra, comparison of band parameters, mineralogy and mineral chemistry derived from these parameters is an established protocol for identification of meteorite analogs. Fig. 11 plots band I center vs. BAR of orange material along with spectral parameters for basaltic and cumulate eucrites, and diogenites. Since diogenites contain near-pure orthopyroxene, their band I centers and BAR plot close to the olivine + orthopyroxene mixing line. Basaltic eucrites span the entire length of the basaltic achondrite zone into the S(V) region. Fig. 11 confirms that the band I center and BAR values of the orange material sites are most consistent with those of basaltic eucrites and not with diogenites.

The pyroxene chemistry (Fs, Wo) is calculated from band I and II centers for all the orange material sites using different methodology described in section 5.2. Fig. 12A shows the pyroxene quadrilateral with the four end-members enstatite (Mg-rich), ferrosilite (Fe-rich), diopside (Mg-Ca-rich) and hedenbergite (Fe-Ca-rich). The pyroxene chemistry ranges for diogenites, howardites, cumulate eucrites and basaltic eucrites are also indicated. The pyroxene chemistries for the orange material plot in the transition region between cumulate eucrites and basaltic eucrites and at the upper end of howardites (basaltic eucrite-rich). Based on the pyroxene chemistry, two out of 17 sites have pyroxene chemistry similar to basaltic eucrites and the rest (15 sites) could be a) cumulate eucrites, b) cumulate eucrite + diogenites, c) cumulate + basaltic eucrite, d) diogenite + basaltic eucrite, and e) cumulate + basaltic eucrite + diogenite. Given that two sites of orange material fall in the basaltic eucrite region of the quadrilateral, it is possible that the other sites are a mixture of basaltic eucrite and diogenite material or cumulate eucrite excavated from the Rheasilvia impact basin to the south (i.e. howardite dominated by a basaltic eucrite component).

In order to investigate if the orange material has a composition similar to cumulate eucrites, we used a selection of spectral parameters from FC (described in section 3.1) in scatterplots to compare the color data from the all the orange material sites along with eucrites, cumulate eucrites and diogenites from the RELAB database. Fig. 12B and C show clearly that the orange material fall in the areas dominated by eucrites but that does not exclude a possible diogenitic component mixed in this material. Orange material is plotting close to two of the cumulate eucrites in Fig. 12B and one in Fig. 12C. But most of the cumulate eucrite are otherwise scattered in the eucrite zone of the plots and towards the diogenite zone, demonstrating possibly a quite diverse composition. Therefore, cumulate eucrites could be a component of the orange material but Fig. 12B and C does not give a definitive answer.

Using preliminary total iron (wt.%) calibration developed with HED spectral data we estimated the iron abundance for the 17 sites for which we extracted VIR spectra. The range of iron abundance for these sites is very narrow (13.5-14.25 wt.%) with a mean of 13.8 wt.%. In our sample of 45 HED meteorites used for creating this spectral calibration, cumulate eucrites had the lowest total Fe with a range between 9.7-12.7 wt.%, followed by diogenites with 11.5-13.3 wt.% and basaltic eucrites with 13.1-15.4 wt.%. Howardites, which are regolith breccias, had total Fe between 12.3-14.8 wt.%. The mean value of 13.8 wt.% for the orange material is consistent with basaltic eucrites and/or howardites.

This is unlikely that a heterogeneous crust formed through fractional crystallization of isolated magma chambers would produce regions on the surface with varying spectral/color signatures like the orange material. Distinct compositional units formed by this process would not survive eons of impact induced mixing unless they are as large as the diogenite-rich unit deposited in the Rheasilvia basin (Reddy et al. 2012b). Moreover, the distribution of orange material as discussed in the section 4.2, would not fit the idea of localized magma chambers. Therefore, we will not investigate further this option in our analysis of the orange material.

*5.4 Elemental composition of the orange regions*

Nuclear spectroscopy measurements indicate that regions containing orange deposits are consistent with howardite, with a broad mix of basaltic eucrite and diogenite. To demonstrate this, we compare mapped data for the abundance of Fe (total Fe) and the macroscopic thermal neutron absorption cross section measured by GRaND (Yamashita et al., in review; Prettyman et al., in review). For HED-compositions, neutron absorption is sensitive to the abundance of Fe, which is found in pyroxene and other mafic mineral phases, as well as the abundance of elements such as Ca and Al, which are prominent constituents of plagioclase found in eucritic materials (Prettyman et al. 2011; Prettyman et al. 2012; Prettyman et al., in review). Thus, Fe and absorption provide complementary measurements of the petrology of Vesta's regolith. We further compare these data with maps of high-energy gamma rays, which are sensitive to the atomic number and atomic mass of regolith materials (Peplowski et al., in review).

### 5.4.1 Fe versus neutron absorption

To estimate the elemental compositions for the three distinctive orange regions on Vesta (namely the Arruntia, Octavia, and Oppia craters and their ejecta blankets) by nuclear spectroscopy, the Fe counting rate, neutron absorption, and high-energy gamma ray derived $C_p$ values observed by GRaND (Prettyman et al., in review; Yamashita et al., in review; Peplowski et al., in review) were spatially smoothed by 30˚-radius boxcar filtering and then were rebinned to 15˚ equal area pixels. The resultant distribution of the Fe counting rate and neutron absorption on Vesta are shown in Fig. 13A and 13B, respectively, with the corresponding pixels containing the ejecta blankets annotated with white lines.

The total elemental abundance of Fe for the Oppia region was estimated to be 7.16 +/- 0.08 [10^-2 count/s] with the neutron absorption of 65.1 +/- 0.4 [10^-4 cm^2/g]. The orange regions plots in accordance with the major trend in the scatterplot of Fe and neutron absorption orbital data, as shown in Fig. 13C. Under conditions in which the mean Fe counting rate can be normalized to the mean Fe abundance in HED meteorites (Yamashita et al., in review), the Fe counting rate near Oppia is translated to 13.5 +/- 0.1 wt.%, which is roughly consistent with the VIR estimation mentioned above.

These orbital observation results would compare most likely to howardites based on the elemental compositions of HED meteorites (Usui and McSween 2007; Usui et al., 2010), as shown in Fig. 13D. However, possibility of the Oppia ejecta region being rich in cumulate eucrite cannot be completely ruled out solely from Fig. 13D. The other two impact ejecta near Arruntia and Octavia were also identified in the scatterplots with distinct Fe abundances and neutron absorption from one another. This reveals that the orange materials in the impact ejecta have a fairly wide variety in regard to elemental abundances among Vesta's surface compositions, and that the orange diffuse regions do not match a unique HED meteorite type.

### 5.4.2 Constraints from GRaND Fe versus $C_p$ measurements

GRaND measurements of Fe abundances as a function of the composition parameter ($C_p$) can be used to test the hypothesis that the orange material includes a contribution from cumulate-eucrite-like material (section 5.3). As discussed in Peplowski et al. (in review), $C_p$ is directly proportional to the heavy element (e.g. Ca, Ti, Fe) content of the regolith and therefore increases with the eucritic content (see Fig. 14) of the sampled material. Fe abundances are derived from Yamashita et al. (in review) as discussed in section 5.4.1. Using the same regions of interest for the Arruntia, Octavia, and Oppia craters and ejecta blankets shown in Fig. 13A and B, we characterized the Fe abundances and $C_p$ values of these regions for comparison to the HED compositions (Fig. 14). These values group in a region that is largely consistent with howardite-like compositions, a finding that is in agreement with those drawn from the Fe versus neutron absorption data.

The Fe versus $C_p$ measurements can be extended to estimate the maximum amount of cumulate-eucrite like material that is allowed by the GRaND measurements. This procedure begins with the 57 HED whole-rock compositions listed in Usui and McSween (2007). The $C_p$ values for these compositions are calculated following the procedure detailed in Peplowski et al. (in review) and are plotted versus Fe abundance in Fig. 14. In order to characterize the cumulate eucrite content of the regions of interest, we define the cumulate-eucrite endmember as being the average of the Moama and Serra de Magé meteorites. These two samples were selected because they have lowest Fe content of the cumulate eucrites, and we note that Usui and McSween (2007) also used Serra de Magé as the cumulate eucrite endmember. We linearly add the elemental composition of the cumulate eucrite endmember to each non-cumulate-eucrite HED composition to define the cumulate eucrite (C.E.) admixture trendlines shown in Fig. 14. For example, a fit of Fe versus $C_p$ for the unaltered non-C.E. HEDs produced the 0% C.E. trendline. Next, a mixture of 75% non-C.E. and 25% C.E. was created for each sample, and the fitted trendline for those values are shown as the 25% C. E. trendline. The procedure was repeated for 50, 75, and 100% C. E. admixtures.

As shown in Fig. 14, the Arruntia and Octavia regions of interest have Fe versus $C_p$ values that are consistent with the 0% C. E. trendline. This supports the conclusions drawn from the Fe versus neutron absorption values that these regions have indistinct, howardite-like elemental compositions. The Oppia region falls on the 25% C. E. admixture line, indicating the possibility that the orange material at this site may contain a ~25% C. E. admixture. When considering the large spatial response of GRaND (full-width half-maximum of 300 km) as compared to the extent of the Oppia ejecta blanket (~150 km), higher C.E. admixtures may be allowed. However, this conjecture is tempered by the Fe versus neutron absorption measurements, which do not indicate a notable cumulate eucrite signature in the Oppia region. We therefore conclude that the Oppia ejecta has a howardite-like elemental composition with a cumulate eucrite admixture of ≤ 25%.

## 6. Origin of the Orange Material

### 6.1 Olivine Option

#### 6.1.1 Background

McFadden et al. (1977) first postulated the presence of olivine on the surface of Vesta using ground-based visible wavelength spectra. Using the symmetry of the 1-μm pyroxene absorption feature they concluded that little or no olivine was present on the surface of Vesta. The olivine spectrum is a composite of three overlapping individual bands and an increase in the abundance of olivine in a mixture of olivine + pyroxene causes asymmetric widening of the 1-μm pyroxene absorption feature towards longer wavelength (1.3 μm). Larson and Fink (1975) studied the near-IR spectrum (1.1-2.5 μm) of Vesta and noted that no spectral features due to olivine were observed. Similarly

Feierberg et al. (1980) studied the Fourier interferometer spectrum of Vesta and concluded that no olivine was detected. Gaffey (1997) suggested that the short wavelength of the pyroxene absorption feature and the high BAR puts the upper limit for average olivine abundance on Vesta at 5%. However, a decrease in BAR by 10-15% compared to the average surface at a certain rotation aspect suggested the possible presence of an olivine unit. Gaffey (1997) suggested that this feature ("Leslie formation") would occupy 10% of the visible surface at this rotation phase (~150 km in diameter) if it were pure olivine. The asymmetric orange ejecta from the Oppia crater that corresponds to the location of the "Leslie formation", spans between 130-200 km and is consistent with Gaffey (1997) estimates.

Olivine is relatively rare in HED meteorites, observed only in a handful of diogenites. Diogenites are orthopyroxenites composed entirely of orthopyroxene and are thought to have formed in the lower crust/upper mantle of Vesta. Diogenites with olivine are harzburgites containing 10-30 vol. % olivine, the rest being orthopyroxene (Beck and McSween, 2010). Petrologic models suggest that the olivine formed in the mantle of Vesta below the orthopyroxenite layer during magma ocean fractionation. Among the HEDs, only one dunite (MIL03443) composed of pure olivine is known (Mittlefehldt et al. 2008) suggesting limited sampling of Vesta's mantle during excavation of the two large impact basins (Rheasilvia and Veneneia) in the south pole. A search for large olivine deposits within the Rheasilvia basin has provided negative results so far (McSween et al. 2013).

### 6.1.2 Olivine Calibration: Detecting Harzburgites on Vesta

The olivine abundance in vestan harzburgites provides important constraints for their detection in Dawn data. In an effort to test the olivine hypothesis we investigated the effect of 10-30% olivine on the spectral band parameters of orthopyroxenes. Mixtures of olivine and orthopyroxene were created in five particle size ranges (0-38, 38-53, 45-63, 63-90, 90-125 microns) at 10% intervals. Olivine used in our study is a magnesium-rich ($Fo_{90.4}$) peridot from San Carlos County, Arizona, and the orthopyroxene is an enstatite ($Fs_{13}$) from Bamle, Norway. Reflectance spectra in the 0.35-2.50 μm wavelength range were measured at the University of Winnipeg's HOSERLab with an ASD FieldSpec Pro HR spectrometer, equipped with a fiber optic cable to direct reflected light to the spectrometer's three internal detectors (Cloutis et al., 2010).

Spectral band parameters were extracted for these data using a Matlab-based code and by averaging band parameters from the five particle size ranges we derived a mean spectral parameter. Variations in band parameters due to particle size effects are quantified by taking the standard deviation of that parameter from all particle size bins. Cloutis et al. (1986) noted changes in band I center and BAR with increasing olivine abundance in an olivine + OPX mixture. Orthopyroxenes have a band I center close to 0.90 μm and olivine close to 1.05 μm (see for example Reddy et al. 2011b) and with increasing olivine abundance the band I center of a mixture is shifted to longer wavelengths. Olivine has a composite absorption band at 1-μm and no 2-μm absorption band. Therefore, with increasing olivine abundance the band II feature in an olivine + OPX mixture gets weaker, lowering the BAR. This change in band I center and BAR is illustrated in Fig. 11 by the sinuous trend line with pure olivine having BAR = 0 and pure orthopyroxene having a BAR of 3.

We developed two calibrations based on band I center and BAR for olivine + OPX mixtures and used these calibrations to extract olivine abundance from the spectral parameters of the orange material. The first calibration is shown in Fig. 15A and plots band I center vs. olivine abundance with error bars depicting one-sigma errors due to particle size effects. As evident in the plot, particle size has a significant effect on the Band I center position in olivine + OPX mixtures. No clear linear relationship exists between olivine abundance and band I center. Using a third order polynomial fit we developed an equation (Eq. 3) in order to estimate the olivine abundance of the orange material. Using Eq. 3, we have estimated the olivine abundance of orange material assuming that their band I center is shifted to longer wavelength due to the presence of olivine. The olivine abundance estimated with Eq. 3 for orange material ranges from 61-65 vol. %; this is significantly higher than what has been observed in HED harzburgites (10-30 vol. %). As noted earlier the large shift in band I center due to particle size also makes Eq. 3 a weaker algorithm for estimating olivine abundance. The discordant results from these two approaches reinforce the fact that multiple spectral parameters provide more robust results than single parameters, as well as the fact that olivine is not a viable option to explain the spectral properties of the orange material.

Olivine abundance = $37506.\alpha^3 - 111474.\alpha^2 + 110588.\alpha - 36540$ with $\alpha$ = band I center (Eq. 3)

Olivine abundance = $-46.52\ \gamma - 94.01$ with $\gamma$ = BAR (Eq. 4)

A second calibration developed using BAR (Fig. 15B) show a linear relationship between the spectral parameter and olivine abundance. Using Eq. 4 we estimated the olivine abundance of the sites of orange material to be between 2 and 26 vol. % with a majority falling between 10-26 vol.%. These olivine abundance values are consistent with laboratory-measured range (10-30 vol. %) for HED harzburgites suggesting the possibility that some of the orange material could be olivine-rich.

As a further test, we compared the band I centers and BAR for the orange material from several sites with howarditic background terrain to test its affinity to olivine. As noted earlier, if orange material is dominated by olivine then their Band I centers would moved to longer wavelength and BAR would be reduced compared to the background terrain. Our tests indicate that the Band I centers and BAR values of orange material do not show the expected shift in these parameters if olivine was a significant component. This makes it highly unlikely for the orange material to be associated with olivine.

### 6.1.3 Comparison between FC color data and laboratory spectra

We used a selection of the spectral parameters from FC (described in section 3.1) in scatterplots to compare the color data from the orange material sites along with eucrites and diogenite from the RELAB database for context, and especially for comparison with the olivine + OPX mixtures that we measured in the laboratory. Orange material seems to be distinct from an olivine-bearing composition when visible tilt (VT) and eucrite-diogenite (ED) ratios are plotted. In the same comparison, the olivine + OPX mixtures cluster below the zone for eucrites and then extend into the diogenite zone as the abundance of OPX increases (consistent with pure OPX composition of diogenites). Orange material data points are grouped in one area in the eucrite zone and do not overlap with the olivine + OPX data points. Thus, according to our analysis using FC color data it is highly unlikely that the presence of olivine could be an analog for orange material. In a similar comparison of K830 (kink) and ED there is no overlap between the olivine + OPX mixtures and the orange material.

In addition, several lines of morphological and meteoritical evidence counter the olivine-orange material association. As noted earlier in section 4, the morphology of the orange material suggests that it is primarily associated with diffuse impact ejecta around moderately large craters (Oppia and Octavia), with ejecta rays in smaller craters (Cornelia and Rubria) and with lobate orange patches. The widespread distribution of orange material on the surface of Vesta is in contradiction with the relative rarity of harzburgites and dunites in the HED collection. Vestan harzburgites and dunites would likely be much more common in the HED collection given the widespread distribution of the orange material. The diffuse red/orange ejecta of Oppia and Octavia would be difficult to explain if it was olivine, unless these two craters are excavating sub-surface deposits of orange material or correspond to secondary impact craters following the formation of the Rheasilvia basin.

*6.2 Metal Option*

The effect of metal on spectral properties of silicates has been studied extensively over the years (e.g., Cloutis et al. 1990). The presence of metal in a metal + silicate mixture reduces absorption band depths, moves the band centers to shorter wavelengths and adds an overall red slope to the spectra. Metal also lowers the albedo of the silicate material. Meteoritic metal powder has red-sloped featureless spectrum with a slight change in slope at 0.6 and 1.3 µm (Cloutis et al., 2010). Spectra of orange material also show similar properties as metal–silicate mixtures, with a red slope, weaker absorption bands and lower albedo compared to pure silicate phases.

Metal is relatively rare in HED meteorites due to metal-silicate fractionation on Vesta. The original presence of metal is inferred by the abundance of platinum group elements (PGEs), which partition into nickel-iron metal during differentiation (Wee et al. 2010). But some howardite breccias show relatively high abundance of PGEs and this has been attributed to exogenic impactor material (Chou et al. 1976; Tagle and Hecht 2006; Wee et al. 2010). Laboratory studies of howardite and polymict eucrites have led to the identification of at least seven different exogenic meteorite types within the vestan regolith. These include carbonaceous chondrites (CM, CI and CK), ordinary chondrites (H, L, and LL), and enstatite chondrites (EL) (Wee et al. 2010). The Kapoeta howardite shows the highest abundance of PGEs with at least 10% chondritic material (CK, CM, H and LL) in its matrix (Wee et al. 2010). The presence of exogenic PGEs in howardites, and therefore possibly in Vesta's regolith, and the similarities in spectral properties of metal + silicate mixtures with orange material motivated us to explore this option further.

We produced laboratory spectra of mixtures of metal and silicates and extracted their band parameters to compare with those of orange material. The samples used to produce metal + silicate mixtures include metal from the Odessa IAB coarse octahedrite (7.3% Ni), and a terrestrial low calcium pyroxene (PYX042; $Fs_{12.8} Wo_{0.4}$). The mixtures were produced as weight percent mixtures at 10% intervals with each mixture consisting of at least 1 gram and the spectra were measured at $i=30°$ and $e=0°$ and spectral band parameters were extracted (Reddy et al. 2010).

Fig. 16A shows a plot of Band I center and Band Area Ratio of HED meteorites, orange material from Dawn, and metal + orthopyroxene (OPX) intimate mixtures. The metal + OPX mixtures plot on the olivine + OPX mixing line and away from the orange material band parameters. We observed a similar trend for areal mixtures of metal + orthopyroxene. The spectral slope of orange material is redder than the average vestan surface as observed by Dawn. Fig. 16B shows a plot of Band I continuum slope as a function metal abundance along with similar parameters for metal + OPX mixtures from our laboratory spectra. Using the relationship between Band I continuum slope and metal abundance in the laboratory mixtures we estimated the corresponding metal abundance for orange material. As seen in Fig. 16B, the Band I continuum slope gives unrealistic metal abundance values, indicating that metal might not be responsible for the spectral characteristics of the orange material. Reproducing the same experiment with Dawn FC data where visible slope is the ratio of 0.75/0.44 μm filters (Fig. 16C) also leads to unrealistic metal abundances and diametrically opposite trendlines.

As discussed in section 5.4, the variability of the elemental composition of the orange regions is within the range of howardites. Elevated abundances of Fe and neutron absorption on a 300-km regional scale, consistent for example with mesosiderite compositions tabulated by Prettyman et al. (2010), are not observed (Prettyman et al., in review; Yamashita et al., in review; Peplowski et al., in review). Thus, synthesizing the data from all three instruments on Dawn, we can conclude that exogenous PGEs (metal) are unlikely to be a large component of the orange material on Vesta.

*6.3 Impact Melt/Shock Option*

**6.3.1 Spectral evidence**

The formation of impact craters like Oppia can significantly alter the target material spectral properties. The effect of impact shock on spectral properties of target material is well documented. Gaffey (1976) and Adams et al. (1979) note that impact shock lowers the albedo and reduces the band depth of the original target material. Several non-HED meteorites show the effects of impact shock including shock-blackened ordinary chondrites (Britt and Pieters 1994). However, meteorites similar to shock-blackened ordinary chondrites are not seen in the HED collection. This could be due to lower impact velocities or differences in composition of ordinary chondrites and HED meteorites. However, shocked eucrites have significantly lower albedo (0.27) than unshocked eucrites (0.55). The impact process also creates glass that is ubiquitous in shocked meteorites. Lunar remote sensing data allowed McEwen et al. (1993) to observe that impact melt haloes around craters generally have a redder slope in the visible near-IR than the surroundings; and they usually have a lower albedo as seen with the Clementine data (Pieters et al. 1994). Tompkins and Pieters (2010) noticed that the presence of glass in Apollo impact melt samples produced a reddening of the continuum (i.e. a red slope) in the visible wavelength range. Increasing amount of glass fraction in mixtures of glass and crystalline lunar samples modifies the 1-$\mu$m pyroxene band so that it appears broader and the band center is shifted to longer wavelength (Tompkins and Pieters 2010). The same authors also noted that the slow cooling of impact melt from the Moon could produce mineralogy identical to igneous lunar rock of the same composition. Therefore, by analogy with the case of the Moon, it seems the presence of glass in impact melt could account for the redder slope of the orange material on Vesta. In the case of the diffuse orange ejecta (e.g. Oppia), a lower albedo (at 0.75 $\mu$m) than the surroundings is also observed as for lunar impact melt deposits. This interpretation would be in agreement with the pyroxene chemistry we found for two of the orange material sites with a composition consistent with basaltic achondrite. However, in HEDs impact glass is relatively rare with most howardites having <1 vol.% and a few with as high as 15-20 vol% (Desnoyers and Jerome, 1977).

The brecciated eucrite Macibini contains an impact melt clast with 50% devitrified glass and 50% silicates (Buchanan et al. 2000). Monomict eucrite Padvarninkai is the most heavily shocked eucrite known with most of the plagioclase converted to maskelynite, and with significant impact melt glass (Hiroi et al. 1995). LEW85303 is a polymict eucrite with an impact melt matrix. Jiddat al Harasis (JaH) 626 is a polymict eucrite that contains significant impact shock and melt quenching (Irving, 2011). The compositional diversity of impact melts from Vesta is evident from our study of HED meteorites.

To test the similarities between orange material and shocked eucrites and impact melt we obtained spectra of a Macibini clast and the Padvarninkai eucrite from the RELAB database. We also measured the visible-near-IR spectra of JaH 626 in five particle size ranges. All spectral data were resampled to Dawn FC filter band passes to get their color spectra. We used the visible spectral slope (0.75/0.44 μm) and BD (0.75/0.92 μm) as parameters to compare the laboratory data with orange material (Fig. 17A). Most of the meteorite data plots around the orange material except for one sample (JaH 626 <250 μm). This JaH 626 sample plots in the middle of the orange material data suggesting a possible similarity between the two. However, band depth and visible slope are not uniquely diagnostic parameters and this match is tenuous at best.

To further test the affinity of HED impact melts with orange material we compared the spectra of the Macibini impact melt clast with those of orange material from Oppia and orange patches. Since the Macibini clast was extracted from a polymict eucrite matrix, some contamination with other HED material is possible (Burbine et al. 2001). Fig. 17B shows the continuum-removed spectra of orange material and potential meteorite analogs. Band depths of the Macibini clast matches with that of orange material, however, the Band I center and width do not match as well. The cause of this mismatch could be a difference in composition or temperature between the laboratory and spacecraft data. An increase in temperature moves the pyroxene band centers to longer wavelengths, which would be the case for laboratory data obtained at room temperature (300 K) and also increases the band width. However, both Macibini and the orange material show a weak absorption feature in the 1.2-μm region. This is a spin-allowed feature due to $Fe^{2+}$ in M1 site of pyroxene that intensifies and moves to longer wavelength with increasing iron abundance (Klima et al. 2008). The band depth of this absorption band has been used as a proxy for estimating the cooling rates (burial depth) in HED meteorites (Klima et al. 2008). The spectrum of orange material shows a much weaker 1.2-μm than a typical basaltic eucrite suggesting contamination by either cumulate eucrites or diogenites. A spectrum of a 50:50 areal mixture of howardite Y-790727 and Macibini impact melt clast provides a better match with the orange material (Fig. 17B).

Fig. 17C shows the spectral band parameters of orange material and shock/impact melts on the Band I center vs. Band Area Ratio plot with 1-σ error bars. JaH 626 samples plot away from the orange material due to their lower BAR and Band I centers. Padvarninkai samples have overlapping BAR but their Band I centers are lower than those of orange material. The two samples that plot closest to the orange material in both parameter spaces are LEW85303 (impact melt matrix) and Macibini (impact melt clast).

The diversity of band I center positions of the samples could reflect the different nature of the original material that has been impacted and melted. As Tompkins and Pieters (2010) proposed in the case of the Moon, the spectral properties of impact melt is likely dependent on the nature of the starting material and conditions of the impact. Combining the interpretations from topography, geomorphology, color parameters from the FC, curve matching and band parameters from the VIR spectral data, the most probable analog for the orange material on Vesta is impact melt dominated howardite material.

### 6.3.2 Impact melt production on Vesta

Because the mean impact velocity in the asteroid belt (~5 km/s) is low compared to that for the Moon and terrestrial planets, it has generally been assumed that impact melt production will be negligible on asteroids (Keil et al., 1997). However, there are several factors that can allow for melt production in some asteroidal impacts. Vesta has a mean impact velocity with other asteroids of ~4.75 km/s (slightly lower than the main-belt average value of 5.14 km/s), but the velocity distribution has a high-velocity "tail" such that about 6% occur at more than 8 km/s and about 2% occur at more than 10 km/s (O'Brien & Sykes, 2011). These relatively uncommon high-velocity impacts may lead to melt production, especially if the impact angle is nearly vertical (the vertical velocity component is the most important for melt production, and is decreased as the impact angle becomes more oblique). In addition, porosity of the surface material can significantly increase melt production (e.g., Wünnemann et al., 2008), since energy is dissipated by compaction of pore space and is deposited in a more localized region around the impact point, compared to a non-porous target.

Williams et al. (in review) surveyed Vesta for evidence of flow-like features. While they found many examples, the vast majority seemed to be due to solid impact ejecta or erosional processes (e.g. landslides). However, they do find evidence for impact melt flow deposits in and around Marcia and Calpurnia craters (the largest "Snowman" craters). The total volume of those putative impact melt deposits is estimated to be ~0.4-11 km$^3$. Marcia also has fairly unique pitted terrain (Denevi et al., 2012b) that has been interpreted to be the result of volatile release, possibly aided by significant impact heating.

Williams et al. (in review) use the results of recent hydrocode simulations of impact heating and scaling laws for crater size and volume to calculate the amount of melt that could be produced for a range of crater sizes, impact velocities, and target porosities (0 and 20%). For a crater the size of Marcia, they estimate that ~0.025 km$^3$ to 32 km$^3$ of impact melt could be produced, with the lower end for 0% porosity and a 5 km/s (vertical) impact velocity, and the upper end for 20% porosity and 10 km/s (vertical) impact velocity. This compares favorably to the estimate of ~0.4-11 km$^3$ they got from the deposits mapped using FC clear filter images. Using the results of Williams et al. (in review, their Table 1), a 35 km diameter crater, roughly the size of Oppia and Octavia, could have impact melt deposits ranging from 0.008 to 10 km$^3$ in volume, depending on the impact velocity and the porosity of Vesta's surface where the impacts occurred. A 500 km diameter crater, roughly the size of Rheasilvia, could have impact melt deposits ranging from 47 to 59,000 km$^3$ in volume.

Osinski et al. (2011) discuss how melt may be emplaced in and around a crater. The majority of impact melt deposits remains within the crater, and for simple craters this will be in deposits on the crater wall or mixed into breccia in the floor of the crater. For larger complex craters, a thicker melt sheet may form on the crater floor, although for Vesta the total amount of melt produced is unlikely to result in coherent sheet like on the Moon. A small fraction of the melt may be ejected outside of the crater, lying on top of the more solid ballistically emplaced ejecta. Oblique impacts lead to melt being preferentially ejected in the downrange direction. However, oblique impacts also have a smaller vertical velocity component, and thus may produce less melt that an impact close to vertical.

Oppia crater orange diffuse deposit would be consistent with an impact melt sheet deposited on an impact on a slope and could have been a nearly vertical impact. Given Octavia's orange ejecta asymmetry and relatively flat local topography, this crater could have been formed by an oblique impact therefore it would be difficult to explain the formation of such a large impact melt deposit. Typically, only as small amount of impact melt is predicted to lie outside the crater (Osinski et al., 2011) and is observed for lunar craters (e.g. Öhman et al. 2012; Denevi et al. 2012a). However, as stated earlier, pre-existing ejecta material can have higher porosity, facilitating the formation of impact melt. Oppia crater, being nearby the edge of the Rheasilvia basin, can form a significant amount of impact melt to account for the diffuse orange ejecta if it impacted through a layer of Rheasilvia ejecta material. The same could be true for Octavia impact if it excavated material from the older layer of the Veneneia ejecta. This would be consistent with our conclusions drawn from the crater counting results indicating a young age for Oppia and Octavia. Alternatively, Oppia and Octavia ejecta impact craters could have excavated impact melt material from Rheasilvia and Veneneia. In this case, melt formation is not required during these impact events to account for the extended diffuse orange ejecta, but only the excavation of ancient impact melt material and deposition as breccia. This would be consistent with the specific diffuse ejecta that is observed only of Oppia and Octavia. Another fresh-looking crater, Arruntia, displays the same kind of orange material ejecta and could have similar origin but is located above 30°N. Thus, in order to account for a similar origin the orange material identified as impact melt from Rheasilvia or Veneneia ejecta should have been deposited above 30°N.

Smaller craters like Cornelia and Rubria exhibit obvious orange ejecta rays and ponds. They might represent cases where impact melt is more likely formed during the impact event. According to the results of Williams et al. (in review, their Table 1), a 10-km crater like Rubria would produce impact melt deposits ranging from 0.00007 to 0.088 $km^3$ in volume, again, depending on the impact velocity and the porosity of Vesta's surface where the impacts occurred. The orange deposits around these craters appear more reasonable relative to the size of the crater and their morphology and distribution is more consistent with impact melt formation during impact.

## 7. Conclusion

Our comprehensive analysis of the morphology, topography, elemental composition, color and spectral properties of the orange material on Vesta observed by the Dawn spacecraft allowed us to infer the following results:

• The orange material has three different type of settings with each different morphological characteristics: the diffuse ejecta deposits of Oppia and Octavia craters, the orange material found in ejecta rays such as for Rubria and Cornelia craters, and the distinctive lobate orange patches. Observation of orange material in impact crater rays of ejecta lobes indicates a possible impact melt (and/or exogenic) origin. This material is not observed for older craters with eroded rims, suggesting that the melt has been degraded and is fading away in the background material (regolith) over time.

• The orange patches have roundish shapes and do not exhibit any specific texture. They drape the local topography, and in some cases, they seem to have flown downslope of crater walls or depressions. This would suggest material was emplaced as a fluid. Some of these patches seem to be linked to the formation of young craters close-by whereas others cannot be traced back to any crater in the immediate surroudings. Maybe the low-gravity of Vesta means that impact melt can be deposited much farther from the crater.

• Redder slope in the visible is a critical parameter for identifying the orange material especially with the FC color data. This material has the reddest slope of any terrain on Vesta and its band depth (BD) depends on the geologic context and types of deposits: BD will be weaker if the orange unit is deposited in the darker hemisphere of Vesta (e.g. for Octavia ejecta which likely mixed with the preexisting terrain during impact), also albedo at 0.750 μm is generally brighter for patches and darker for diffuse ejecta.

• We found an interesting analogy with the orange/red terrains that has been observed on the Moon with similar color ratio composite using Clementine data. This similarity suggests an impact melt origin when looking at the Copernicus crater example.

• Orange material is concentrated within 30°S-30°N at all longitudes except between 30°E and 90°E, which lack orange patches. All these units, as well as the Octavia diffuse ejecta and other craters with orange ejecta rays are distributed outside the Rheasilvia basin. Only Oppia orange ejecta is partially overlapping the Rheasilvia basin rim. Characteristics of the orange patches (morphology resulting from deposition of melted material) along with their distribution indicate that some of them could originate from melted material excavated from the Rheasilvia impact basin.

- Crater size–frequency distributions for Oppia and Octavia ejecta using two counting areas for each crater yield model ages that are quite young (around 270 Ma for Oppia and 360-370 Ma for Octavia) and suggest that these craters formed on top of the Rheasilvia ejecta material. These craters could be excavating orange material that has been previously deposited during the Rheasilvia impact basin and subsequently buried by impact gardening/regolith mixing and ejecta deposition.

- We confirmed the link between Oppia crater ejecta, the "Leslie feature" from Gaffey (1997), and the feature #15 from Li et al. (2010). Using band parameters we showed that this geologic feature is inconsistent with the presence of olivine.

- When using VIR data, all orange material sites appear compositionally homogeneous. We obtained an average pyroxene chemistry of $Fs_{42}Wo_9$ with a narrow range of values suggesting a common formation mechanism for all sites. A narrow range of values is found for the iron content as well with a mean abundance of 14 wt.%.

- The spectrally derived iron content, which is in agreement with the GRaND measurements, would be consistent with a composition close to basaltic eucrites and/or howardites. The pyroxene chemistry suggests that 2 of the 17 sites are similar to basaltic eucrites while the others could correspond to a mixture of basaltic eucrite and diogenite material (or cumulate eucrite) excavated from the Rheasilvia impact basin (i.e. howardite dominated by basaltic eucrite component). The elemental abundance derived from GRaND observations suggests Oppia ejecta has ~25% cumulate eucrite component in a howarditic mixture, whereas Octavia and Arruntia ejecta do not show a cumulate eucrite component. This might represent the influence of the material inherited from different type of impactors and/or the different terrains impacted.

- The possibility of the orange material having a cumulate eucrite component is proposed after analysis of the VIR data. To further test this option, the Fe versus $C_p$ measurements from GRaND were used to determine the possible content on the diffuse orange ejecta deposits. Arruntia and Octavia do not contain any cumulate eucrite component whereas Oppia is possibly consistent with ~25% cumulate eucrite admixture. Oppia being closer to the rim of the Rheasilvia ejecta basin, could have excavated material from Rheasilvia's ejecta that originally came from a deeper lithology in the vestan crust.

- Combined analysis of VIR spectra and FC color data for the orange material along with the observations from GRaND revealed that a large abundance of metal is not a valid explanation for the composition of this material.

- Laboratory measurements of olivine + OPX mixtures spectra allowed us to develop two spectral calibrations in order to retrieve the olivine abundance. Scatterplots from FC color data argue against the olivine option for the composition of the orange material. Along with considerations from morphologic and meteoritical evidence, we inferred that the olivine + OPX mixture hypothesis is not valid for explaining the spectral properties of the orange material.

- Comparison of impact melt and shocked eucrites spectral properties with the vestan orange material showed that a mixture of Macibini-type material (impact melt clast) and howarditic material is the most probable composition that can account for the orange material color and spectral properties. The morphology and distribution of this material across Vesta is also consistent with this hypothesis.

• Production of impact melt on Vesta is thought to be rare, nonetheless, recent modeling work shows that there is a possibility of impact melt formation in the case of high velocity and nearly vertical impact. If the large orange diffuse ejecta of Oppia and Octavia include a high proportion of impact melt, its production could have been facilitated by the a pre-existing low porosity sub-surface layer made of ejecta material from Rheasilvia in the case of Oppia and from Veneneia in the case of Octavia or higher than average impact velocities. Typically, in-situ impact melt emplacements are limited to small deposits, ponds and sheets close to the crater rim. Therefore, another plausible explanation for the diffuse orange deposit is the excavation of pre-existing impact melt material formed by Rheasilvia and/or Veneneia basins. Smaller young craters such as Rubria and Cornelia seems to exhibit melt directly formed during the impact in the form of orange material in rays, wall deposits and ponds.

Further analysis and modeling will be required to fully differentiate between the possible options for the formation of the impact melt. Another interesting question that will need further study is that if Rheasilvia (and maybe Veneneia) produced some of the orange material found in the subsurface (and that was excavated by Oppia and Octavia impacts) and most of the orange patches, why is none of the orange material detected in the basin itself?

**Acknowledgements**

The authors would like to thank the Dawn Flight Operations team for a successful Dawn at Vesta mission. L.L. work is supported by Dawn UCLA subcontract# 2090-S-MB170. V.R. work is supported by NASA Dawn Participating Scientist Program grant NNH09ZDA001N-DAVPS and NASA Planetary Geology and Geophysics grant NNX07AP73G. D.O. work is supported by Dawn at Vesta Participating Scientist Program grant NNX10AR21G. The research utilizes spectra acquired with the NASA RELAB facility at Brown University. The authors would like to thank Guneshwar S. Thangjam for his help gathering laboratory spectral data from RELAB. E.A.C. thanks CSA, CFI, MRIF and the University of Winnipeg for supporting HOSERLab.

**Table 1.** List of FC filters with their band pass center, peak and full width at half maximum (FWHM). F1 is a broad band filter and is also called the clear filter.

| Filter name | Wavelength center (μm) | FWHM (μm) |
|---|---|---|
| F1 | 0.740 | 0.371 |
| F8 | 0.430 | 0.040 |
| F2 | 0.550 | 0.043 |
| F7 | 0.650 | 0.042 |
| F3 | 0.750 | 0.044 |
| F6 | 0.830 | 0.036 |
| F4 | 0.920 | 0.045 |
| F5 | 0.980 | 0.086 |

**Table 2.** List of FC observational phases and their respective characteristics. The orbital phases are listed in chronological order. The orbital phases actually used in this study comprise RC3b, HAMO1 and LAMO (clear filter images only). RC stands for rotational characterization, which was the main goal of the observations during the approach phase. One station represents a sequence of seven color frames acquired with the FC color filters, except for the LAMO in which one station consists of three-color acquisitions only.

| Orbital Phase | Best Resolution (m/pixel) | Sub-Spacecraft latitude | Distance to Vesta (km) | Color stations/cycles included |
|---|---|---|---|---|
| RC1 (approach) | 9 067 | 32°S | 100 000 | 12 stations |
| RC2 (approach) | 3 382 | 54°S | 37 000 | 12 stations |
| RC3 (approach) | 487 | 25°S | 5 200 | 15 stations |
| RC3b (approach) | 487 | 25°S | 5 200 | 16 stations |
| Survey | 252 | 50°N to 90°S | 2 700 | In cycles C1 to C4, C7 |
| HAMO (1) | 61 | 66°N to 87°S | 660 to 730 | Cycles C1, C6 |
| LAMO | 16 | 85°N to 90°S | 190 to 240 | Cycles C9 to C12 |
| HAMO (2) | 60 | 85°N to 85°S | 640 to 730 | Cycle C6 |

**Table 3.** List of the 17 sites showing orange material in the FC Clementine maps for which VIR data has been extracted and analyzed (see online supplementary material for their location).

| Site # | BI area | BI center (µm) | BI slope | V. slope | BI depth (%) | BII area | BII center (µm) | BII slope | BII depth (%) | BAR |
|--------|---------|----------------|----------|----------|--------------|----------|-----------------|-----------|---------------|------|
| 1 | 0.092 | 0.933 | 0.184 | 0.175 | 45.0 | 0.145 | 1.981 | 0.067 | 26.1 | 1.57 |
| 2 | 0.095 | 0.931 | 0.172 | 0.192 | 45.3 | 0.162 | 1.969 | 0.059 | 27.6 | 1.71 |
| 3 | 0.091 | 0.934 | 0.191 | 0.178 | 44.0 | 0.139 | 1.990 | 0.084 | 24.9 | 1.52 |
| 4 | 0.084 | 0.935 | 0.190 | 0.179 | 44.2 | 0.136 | 1.987 | 0.086 | 25.9 | 1.62 |
| 5 | 0.097 | 0.932 | 0.193 | 0.199 | 46.9 | 0.153 | 1.972 | 0.079 | 27.5 | 1.58 |
| 6 | 0.102 | 0.934 | 0.197 | 0.192 | 46.2 | 0.158 | 1.981 | 0.069 | 27.4 | 1.54 |
| 7a | 0.107 | 0.935 | 0.208 | 0.220 | 48.9 | 0.185 | 1.979 | 0.072 | 30.9 | 1.74 |
| 7b | 0.090 | 0.932 | 0.180 | 0.196 | 44.8 | 0.160 | 1.982 | 0.063 | 27.0 | 1.77 |
| 8 | 0.096 | 0.933 | 0.185 | 0.197 | 45.1 | 0.149 | 1.978 | 0.062 | 25.9 | 1.55 |
| 9 | 0.095 | 0.936 | 0.203 | 0.213 | 45.4 | 0.171 | 1.996 | 0.076 | 28.8 | 1.80 |
| 10 | 0.094 | 0.934 | 0.205 | 0.173 | 45.9 | 0.164 | 1.969 | 0.073 | 27.5 | 1.74 |
| 11 | 0.096 | 0.932 | 0.184 | 0.186 | 45.2 | 0.153 | 1.965 | 0.065 | 26.5 | 1.60 |
| 13 | 0.087 | 0.937 | 0.125 | 0.149 | 40.9 | 0.126 | 1.994 | 0.062 | 23.0 | 1.45 |
| 14 | 0.074 | 0.935 | 0.131 | 0.147 | 40.0 | 0.120 | 1.966 | 0.064 | 22.0 | 1.62 |
| 18 | 0.077 | 0.933 | 0.226 | 0.183 | 44.3 | 0.152 | 1.959 | 0.076 | 26.9 | 1.98 |
| 19 | 0.096 | 0.932 | 0.203 | 0.197 | 46.0 | 0.146 | 1.955 | 0.059 | 26.3 | 1.52 |
| 21a | 0.107 | 0.934 | 0.222 | 0.240 | 48.8 | 0.177 | 1.969 | 0.074 | 30.4 | 1.65 |

**APPENDIX**

**A1. Details of the VIR data processing**

The four processing steps described here have the following functions: 1) ingestion of calibrated level 1B data (*dawnimport*), 2) conversion to I/F and photometric correction (*dawnpho*), 3) map projection and mosaicking, 4) photometric calibration of the VIR VIS. The VIR data ingested in our pipeline is in Planetary Data Systems (PDS) format (McMahon et al., 1996) and is in the Claudia coordinate system (Reddy et al., in review). The PERL script, *dawnimport*, was used for ingestion of VIR data products. This script imports VIR IR/VIS 1B PDS EDR image cubes into ISIS using the application *dawnvir2isis*. Then, *spiceinit* was applied to each image cube for geometric computations. This initializes the camera models for both data sets that will support photometric processing and map projection of the data. We applied R. Gaskell's digital elevation model (DEM) for these images as parameters to the *dawnimport* script that establishes a common shape model for use in the Dawn camera models for geometric operations and calculations.

VIR data requires an additional housekeeping file for level 1A and 1B as input into the *dawnvir2isis* application. The 1A housekeeping files should also be used for 1B data. This is required for import into ISIS as it contains information about the scan mirror positions of the VIR during data acquisition. This data is crucial to proper functioning of the VIR camera model. FC level 1B files are converted to ISIS format using the *dawnfc2isis* application.

The script, *dawnpho*, applies photometric correction to VIR IR image cubes (VIR VIS bands have no derived photometric correction parameters). Critical to photometric correction calculations is an accurate shape model for the planetary body being observed. ISIS has the ability to calculate local photometric angles (emission and incidence) from the DEM for highly accurate photometric correction using the photometric parameters from Li et al. (2013). Using the Gaskell DEM, we photometrically corrected the VIR IR using the ISIS *photomet* application. This application applies photometric parameters to single image bands, which becomes particularly challenging for large multi-spectral data sets (VIR has 432 IR bands each). I/F solar flux normalization is applied to the VIS and IR image cubes using the *fx* application. The solar flux, normalized to 1 AU, for each of the VIS and IR data sets is stored in a 1 sample, 1 line, 432 band ISIS cube. *campt* is used to determine the distance from Vesta to the Sun. The conversion from radiance values in level 1B VIR data to reflectance (I/F) using solar flux values is calculated using the *fx* equation:

$$(I/F)^i = (\pi(d_V)^2 I^i)/(F_{Sun})^i$$

Where i = 1-432 for each VIS and IR band, $d_V$ is the distance of Vesta to the Sun in AU at the time of observation, $I^i$ is the radiance value in that VIR band and $(F_{Sun})^i$ is the solar flux at 1 AU for each band i.

For each band, the script determines the corresponding photometric parameter file and then applies it using *photomet*. Once all bands are processed to individual files, they are then stacked back into a multi-spectral cube. Now the photometric image cube can be easily projected to a common map product. Note that simultaneously acquired VIR VIS and IR image cubes cannot be merged at this stage since line scan geometry differs slightly for these products. This will be addressed in the map projection and VIR merging processes that follow. Pixels were trimmed at the edges of the VIS and IR cubes in order to exclude pixels with anomalous values.

To support FC and VIR analysis, the data we used was projected using the same mapping parameters (resolution and latitude/longitude extents). This provides a common map environment to combine data sets from independent sources – in our case the FC and VIR instruments. First, the ISIS application *mosrange* was used to determine the best spatial resolution, the coordinate of the center of the dataset, and latitude/longitude minimum and maximum ranges for all images in our dataset. Once this has been determined, we project each product into an equirectangular projection using the same mapping parameters with the *cam2map* ISIS application. Then, an FC color mosaic from the HAMO phase was reprojected using these mapping parameters with the *map2map* application (this data will be used for VIR VIS photometric calibration in the next step).

An important additional step is the photometric calibration of the visible part of VIR. We do not have photometric parameters directly derived from VIR VIS therefore corresponding FC photometrically corrected data is used to correct the VIR VIS data. Both scaling at 0.750-$\mu$m and slope correction are carried out using the *fx* application in order to photometrically correct the VIS data. The slope correction factor has been determined by comparing FC color spectrum and VIR spectrum on several sites. Finally, the IR part of VIR is scaled to the VIS to obtain a consistent spectral dataset. The VIS and IR containing several overlapping wavelengths (from 1.02 to 1.071 $\mu$m), we cropped the visible range at 1.02 $\mu$m.

**A2. Crater counting methodology**

We used Dawn projected FC clear filter data, and the mapping software ArcGIS in combination with the CraterTools plug-in for map projection independent crater mapping (Kneissl et al., 2010). The statistical analysis of the measured craters has been carried out utilizing the craterstats software (Michael and Neukum, 2010). The measured crater frequencies were converted to the normalized crater frequency at 1 km crater size (N(1)) and absolute crater retention ages as described in Schmedemann et al. (in prep). Where the measurement has been fitted in order to derive a surface age, the craters showed a random spatial distribution within the counting area, based on a randomness analysis (Michael et al., 2012).

### A3. Crater counting sites description

For the Oppia crater, we selected two sites on the ejecta to perform the crater counting. The areas were mapped in order to cover representative parts of the Oppia ejecta, which should be a good average of the whole ejecta blanket and can be characterized by a random crater distribution. A morphological inspection of the counting areas shows fresh and pristine small craters but also heavily degraded larger craters, which very likely pre-date the Oppia formation. The larger craters therefore give information about the age of the surface in which Oppia was formed. The small crater fraction of the pre-existing surface however, was completely destroyed or blanketed in the course of the Oppia formation. Thus, the small craters were formed on top of the Oppia ejecta blanket and can be used for dating the Oppia impact event, if a resurfacing correction is applied (Michael and Neukum, 2010).

For the Octavia crater, we also chose two sites for crater counting, both in the southern part of the ejecta blanket because the northern part is probably contaminated by Marcia secondaries as indicated by linear groove-like features in alignment with Marcia. The areas were mapped in order to cover representative parts of the Octavia ejecta, which should be a good average of the whole ejecta blanket and be characterized by a random crater distribution. A morphological inspection of the counting areas shows fresh and pristine small craters in both areas. Those craters were formed on the Octavia ejecta blanket and can be utilized to derive the age of Octavia. Heavily degraded larger craters were also measured, which very likely pre-date the Octavia formation. As discussed for Oppia, larger craters give information about the age of the surface Octavia was formed in. Regarding the underlying craters, we found very few in the smaller area, located right at the crater rim. In this area the Octavia ejecta blanket is thicker than in the larger counting area, further to the east.

In all counting areas, the crater size-frequency distributions are presented as standard cumulative plots (Crater Analysis Techniques Working Group et al., 1979).

FIGURES

Fig. 1: Perspective views of Oppia and Octavia craters. A to D images are views of the Oppia crater (27.70°S-6.07°N, 288.68°E-331.78°E) and E and F correspond to Octavia crater (18.02°S-17.92°N, 121.12°E-164.25°E). The images A, C and E are made by draping a mosaic of HAMO (~60 m/pixel) color ratio images on the topography of the area. The other images (B, D and F) show the color-coded topography relative to Vesta ellipsoid in the same configuration as the top images. N indicates the direction to the North. DF indicates terrain shaped by the troughs from the Divalia Fossa formation, F indicates the location of Feralia Planitia, and R corresponds to a portion of Rheasilvia basin and its eroded rim. AT points to a hill with dark material called Aricia Tholus. Shading is also added to render the images more realistic. Topographic exaggeration of 1.5 was added to enhance the relief of the geologic features. Dark areas are present when there is no data available in the HAMO 1 phase. A to D: E and F:

Fig. 2: Perspective views of orange material that forms "pumpkin" patches on the surface of Vesta with FC color data from HAMO (~60m/pixel). These 4 examples of deposits (A, C, E, G) and their corresponding color-coded topography (B, D, F, H) are not associated with impact craters. A and B: The orange material can be found on the crest and walls of the equatorial troughs (DF) and partly filling some craters. Most of the patches seem to follow the local topography formed by the two sets of crisscrossing troughs Divalia Fossa (DF) and Saturnalia Fossa (SF); others (like at the bottom of the image indicated by P) have a more complicated pattern, possibly following the rim of eroded craters. C and D: The orange patches are visible along the slope of ancient crater walls and troughs. E and

F: Orange material forms a cluster of patches located preferentially on south-facing slopes on the walls of several old craters and topographic highs. The latter example (E and F) would suggest that this material could have been brought by a large impact located further south (possibly Rheasilvia), which would have ejected orange material in the general direction of the north on the pre-existing surface features. One of the orange patches that has the most striking color contrast with its surrounding (G) is also one of the largest deposits. Unlike the other orange patches, this particular patch is following a similar elevation for most of the deposit with the orange/yellow color in the color-coded topography (H). However, the westernmost part of the patch then spreads to various elevations while forming smaller lobate-like shaped orange patches interconnected with narrower corridors of material. N indicates the direction to the North. Shading is added to render the images more realistic. Topographic exaggeration of 1.5 was added to enhance the relief of the geologic features. Dark areas are present when there is no data available in the HAMO1 phase. The latitudinal and longitudinal range of each image are for A and B: 24.13°S-5.52°S, 0.21°E-18.43°E; for C and D: 5.66°S-3.59°N, 348.20°E-359.85°E; for E and F: 10.67°N-25.93°N, 311.68°E-329.16°E; for G and H: 20.93°S-11.27°S, 348.59°E-359.98°E.

Fig. 3: Perspective views of orange material that forms "pumpkin" patches on the surface of Vesta with FC color data from HAMO (~60m/pixel). These 3 examples of deposits (A, C, E) and their corresponding color-coded topography (B, D, F) seem to be related to nearby impact craters. A to D: This is a rather large area that encompasses part of Vestalia Terra and contains several recent impact craters and associated orange units.

These may be linked to the nearby young impact crater Drusilla (Dr) and possibly Numisia (Nu) because they seem to be distributed radially around these craters. However, their distribution is not symmetrical around these craters possibly to due to the local topography and/or an oblique impact. It is not clear if the set of lighter orange patches were formed by Teia crater (T) or were preexisting and then covered by bright material ejected from the crater. Nonetheless, smaller orange patches of material can be found all around this crater and suggest emplacement during the impact event. E shows a large orange patch draping the rim and part of the walls of an unnamed impact crater found on a slope (F). Other smaller orange deposits are also found further away from the crater rim as well and all together they are elongated in the downslope direction. N indicates the direction to the North. Shading is also added to render the images more realistic. Topographic exaggeration of 1.5 was added to enhance the relief of the geologic features. Dark areas are present when there is no data available in the HAMO1 phase.The latitudinal and longitudinal range of each image are for A, B, C and D: 14.97°S-5.29°N, 249.05°E-276.04°E; for E and F: 17.19°S-3.40°S, 324.81°E-342.99°E.

Fig. 4: Close-up views of some orange patches at two different locations on Vesta. Clementine color ratios images from FC data (HAMO phase) are overlaid in transparency over mosaics of FC clear filter images (LAMO phase). No texture difference is visible between the orange deposits and the surrounding terrain.

Fig. 5: Perspective views of orange material found in crater rays of fresh craters with FC color data from HAMO (~60m/pixel). N indicates the direction to the North. A represents

the crater Cornelia (centered at 9.3°S, 225.2°E). Orange deposits are found in the center of the crater where pitted terrains have been identified (Denevi et al. 2012b), covering part of the crater walls overlapping either the bright or the dark material, and finally outside the crater rim covering some of the dark ejecta rays and distributed radially from the crater. B corresponds to the crater Rubria (centered at 7.4°S, 18.4°E), which shows similar deposits of orange material despite being smaller than Cornelia. It is also a fresh crater with mostly sharp rims and prominent bright and dark crater rays. Unlike Cornelia, Rubria does not exhibit any sign of pitted terrains. Shading is also added to render the images more realistic. Topographic exaggeration of 1.5 was added to enhance the relief of the geologic features.

Fig. 6: Color ratio images of selected sites on the Moon from the UVVIS camera onboard the Clementine spacecraft. Images are retrieved from the PDS Imaging node and are in simple cylindrical projection with a spatial resolution of 100 m/pixel. The false-color composites are using combination of color ratios such that red is 750/415 nm, green is 750/950 nm and blue is 415/750 nm. Image (A) corresponds to volcanic landforms on the nearside of the Moon in the mare basalt that have similar morphology to some Vestan orange material deposits. Image (A) is comprised between 17.02°S-12.25°S and 307.30°E-313.77°E. Image (B), framed between 4.56°N-16.10°N and 334.38°E-345.53°E, corresponds to the Copernicus crater (93 km in diameter) and its ejecta with a non uniform distribution of impact melt in orange/red material (Pieters et al. 1994). The white arrows indicate the orange material deposits that have similar morphology to the orange patches or orange ejecta material found on Vesta.

Fig. 7: A: Average FC color spectra from HAMO1 phase calculated on specific areas of interest along with the mean color spectra of the entire surface of Vesta (including data with phase angle < 90°). All color spectra have been normalized at 0.75 $\mu$m. B: Color spectra of the same areas of interest but divided by the average color spectra of Vesta. Numbers identifies the units and their location is available in the online supplementary material. Unit 1 (3 136 254 color spectra) corresponds to the Oppia ejecta, unit 4 (110 063 color spectra) corresponds to a cluster of orange patches located just north of Oppia crater; unit 7a (60 279 color spectra) corresponds to a recent crater in Oppia's ejecta, unit 14 (2 816 157 color spectra) is for Octavia's ejecta and unit 17 (25 405 color spectra) is for a group of orange patches west of Octavia but still in the dark hemisphere of Vesta.

Fig. 8: Global maps of Vesta in cylindrical projection centered at 180° and in polar projections centered on the south pole. (A) is a global mosaic in cylindrical projection in the Clementine color ratios using data from HAMO 1 observations. (B) is the equivalent but in south polar projection. (C) is a cylindrical map with color coded topography (relative to the Vestan ellipsoid) with the mapping of the orange material: the black polygons represent all the orange patches and orange crater rays (such as seen in Cornelia), black dotted polygons are used for Octavia and Arruntia orange ejecta, and white dotted polygon is used for Oppia orange ejecta. The basin rim of Veneneia is delineated in red and the Rheasilvia basin rim is depicted as a black line. (D) is the same data set and mapping as in (C) but in south polar projection. A high-resolution version of Fig. 8 is available in the online supplementary material. The numbering for the orange

material units (clusters of orange patches and diffuse orange ejecta) analyzed with VIR and FC color data is also available in the online supplementary material.

Fig. 9: A: Map of counting areas northeast and south of Oppia, outlined in white. We used HAMO clear filter image FC21A0009693 for the northeast area and FC21A0027572 for the south area. B: Cumulative plot of the area south of Oppia. Open squares represent the measured cumulative crater frequency per $km^2$. The measurement shows two kinks which are interpreted as resurfacing events, which destroyed craters < ~2 km (older event) and < ~600 m (younger event). Craters in the resurfaced diameter ranges formed after the end of the respective resurfacing event. Due to the cumulative characteristics, crater frequencies in the resurfaced diameter ranges need to be corrected for the contribution of the larger craters, which survived the resurfacing event (Michael and Neukum, 2010). The corrected crater frequencies are indicated by filled triangles (craters formed after older event) and filled circles (craters formed after younger event). The derived base age from the craters that survived both resurfacing events is 3.63 +0.06/-0.1 Ga. This is the age of the stratigraphic horizon below Oppia (likely Rheasilvia ejecta). The two resurfacings occurred at 1.83 +/- 0.3 Ga and 272 +/-9.3 Ma. The most recent resurfacing is the formation age of Oppia. The older resurfacing is probably caused by ejecta blanketing of a nearby crater, which formed around 1.8 Ga. C: Cumulative plot of the area north east of Oppia. Same symbology as panel B. Only one resurfacing kink is observed at ~2 km. The derived base age is 3.62 +0.05/-0.09 Ga and is again the age of the stratigraphic horizon below Oppia (likely Rheasilvia ejecta). Only one resurfacing

occurred at 276 +/-23 Ma, which indicates the formation of Oppia. The intermediate resurfacing is not present, indicating that it is a local effect at the southern counting area.

Fig. 10: A: Map of counting areas southwest and south east of Octavia, outlined in white. We used LAMO clear filter image FC21A0015915 for the southwestern area and FC21A0026315 for the southeastern area. B: Cumulative plot of the area south west of Octavia. Same symbology as Fig.9B. No base age could be derived because of a too small area for the given ejecta thickness. A resurfacing age could be derived, which likely marks the formation of Octavia at 356 + 39/-38 Ma. C: Cumulative plot of the area south east of Octavia. Same symbology as Fig.9B. One resurfacing kink is observed at 2 km. The derived base age from craters > 2 km is 3.78 +0.04/-0.06 Ga, which is the age of the stratigraphically resolvable horizon below Octavia (likely Veneneia ejecta). Only one resurfacing is detectable at 372 +/-30 Ma, which indicates the formation of Octavia. The intermediate resurfacing expected from Rheasilvia ejecta is not observed, indicating it is probably thinner than the Octavia ejecta at the southeastern counting area.

Fig. 11: Plot of the band I center vs. BAR including the orange material sites analyzed in this study and the "Leslie feature" observed by Gaffey 1997. Band parameters for cumulate eucrites, basaltic eucrites and diogenites are derived from spectral data from RELAB. Regions for different S asteroid sub-types are shown in the plot in grayish ellipses, and the region corresponding to the basaltic asteroids is displayed as a rectangle.

Fig. 12: Plots comparing the composition derived from VIR and the spectral parameters derived from FC for orange material sites with eucrites and diogenites. (A) Pyroxene quadrilateral including the VIR data points for the orange material sites and the range of values for cumulate eucrites (CE), basaltic eucrites (BE), diogenites (D) and howardites (H). 2D scatterplots using FC spectral parameters (described in section 3.1) with eucrites, cumulate eucrites, diogenites and orange material sites: (B) plot of K920 vs. ED, and (C) plot of K920 vs. K830. Spectral data of eucrites and diogenites is from RELAB and has been resampled to FC filter band passes for plots (B) and (C).

Fig. 13: Maps of (A) the fully corrected Fe counting rate and (B) neutron absorption on Vesta observed by GRaND on Dawn (Prettyman et al., in review; Yamashita et al., in press). Both maps were spatially smoothed over 30° in radius and rebinned to 15° equal-area pixels. The pixels corresponding to the diffuse orange deposits in the ejecta blankets of Arruntia, Octavia, and Oppia (see Fig. 8E) are delineated by white lines. The Rheasilvia basin is outlined by a white curve. (C) Scatterplot of the Fe counting rate and neutron absorption determined by GRaND which were used in Fig. 14A and 14B. The mean values and population standard deviations (1σ) for the three orange regions are overplotted. All the three orange deposits plot in the general trendline but are separated from each other. (D) Scatterplot of the Fe abundance and neutron absorption for various HED meteorites (Usui and McSween 2007; Usui et al. 2010) and for the three orange regions (Arruntia, Octavia and Oppia ejecta) observed by GRaND. The Fe counting rates were translated to abundances (see text; Yamashita et al., in press). The error bars for the orbital data are smaller than the markers. The three sites fall in the howardites/polymict

eucrite area of the scatterplot, with Octavia plotting close to some basaltic eucrites and Oppia close to cumulate eucrite data points.

Fig. 14: Fe abundances plotted against the high-energy gamma-ray derived Cp values for 57 HED whole-rock elemental compositions (Usui and McSween 2007; Usui et al., 2010). The Moama and Serra de Magé meteorites, which were used to define a cumulate eucrite (C.E.) elemental composition endmember are labeled. Arruntia, Octavia, and Oppia regions of interest (shown in Fig. 14) are represented in black diamonds. Black dash lines indicate trendlines for non-C.E. and C.E. admixtures (see section 5.4.2) and are used to constrain the possible contributions of C.E.-like material to the orange material in each region. The Arruntia and Octavia regions are found to have compositions that are consistent with howardites with no C.E. admixture. The Oppia region also has a howardite-like composition, however a C.E. admixture of ~25% is allowed by the GRaND measurements.

Fig. 15: Plots used to analyze the olivine option for the orange material. The plots correspond to the calibrations we developed to retrieve olivine abundance from the VIR spectra of the orange material sites based on band I center and BAR of olivine + OPX mixtures: (A) plot of olivine content vs. band I center, (B) plot of olivine content vs. BAR. For (A) and (B), the error bars for the olivine + OPX mixtures indicate the range in band I center for particles sizes ranging from 38 $\mu$m to 125 $\mu$m.

Fig. 16: Plots used to investigate the presence of PGEs in the orange material. (A) Plot of Band I center and BAR of orange material from VIR, and metal + orthopyroxene (OPX) intimate mixtures. Regions for different S asteroid sub-types are shown in the plot in grayish ellipses, and the region corresponding to the basaltic asteroids is displayed as a rectangle. Increasing metal content in the mixtures is indicated by a dashed arrow. (B) Plot of Band I continuum slope as a function metal abundance with metal + OPX mixtures from our laboratory spectra and orange material data from VIR. Increasing metal content in the mixtures is indicated by a dashed arrow. (C) Similar plot as for (B) but using the ratio of $0.75/0.44$ µm from FC color data for the visible slope of the orange material sites. All the error bars plotted are 1-σ.

Fig. 17: Plots used to test the affinity of shocked eucrites and impact melt with the orange material. (A) Plot of the visible spectral slope ($0.75/0.44$ µm) and band depth ($0.75/0.92$ µm) of laboratory spectra of impact melts and shocked eucrites resampled to FC filter bandpasses and FC color data of the orange material sites. The error bars plotted are 1-σ. (B) Comparison of the spectra of a Macibini impact melt clast with two of the orange material sites: site 10 with a "pumpkin patch", and site 7a with orange material mixed with fresh material from a recent impact (and therefore has deeper pyroxene bands) close the Oppia crater rim (see Fig. 8 for location). A slightly better match is found when mixing impact melt with howarditic composition (from howardite Y-790727). All the spectra are continuum-removed. (C) Plot of the Band I center vs. Band Area Ratio showing the spectral band parameters of all the orange material sites and shock/impact melts samples with 1-σ error bars. Regions for different S asteroid sub-types are shown in the plot in grayish ellipses and polygon, and the region corresponding to the basaltic asteroids is displayed as a rectangle.

Figure 1.

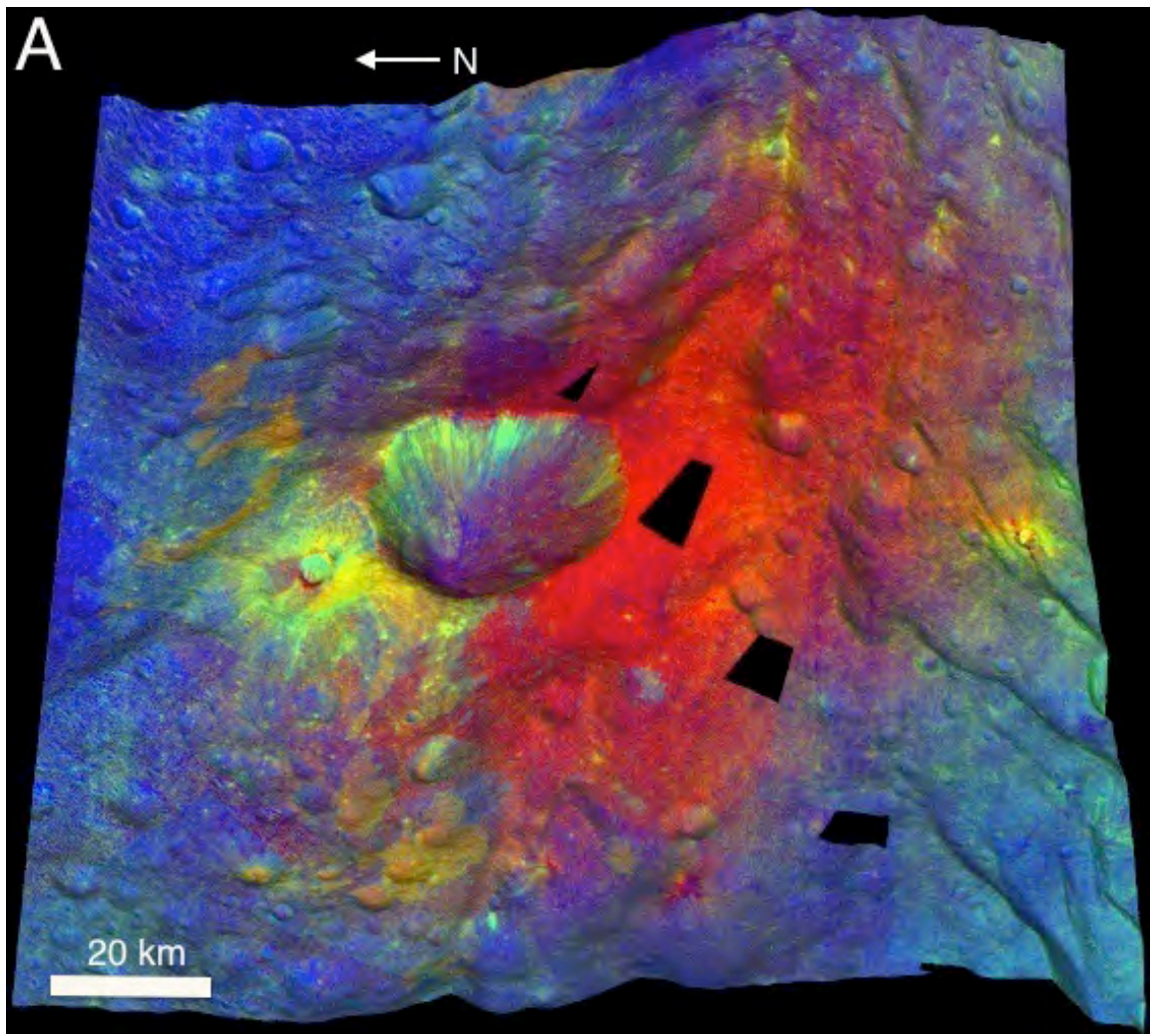

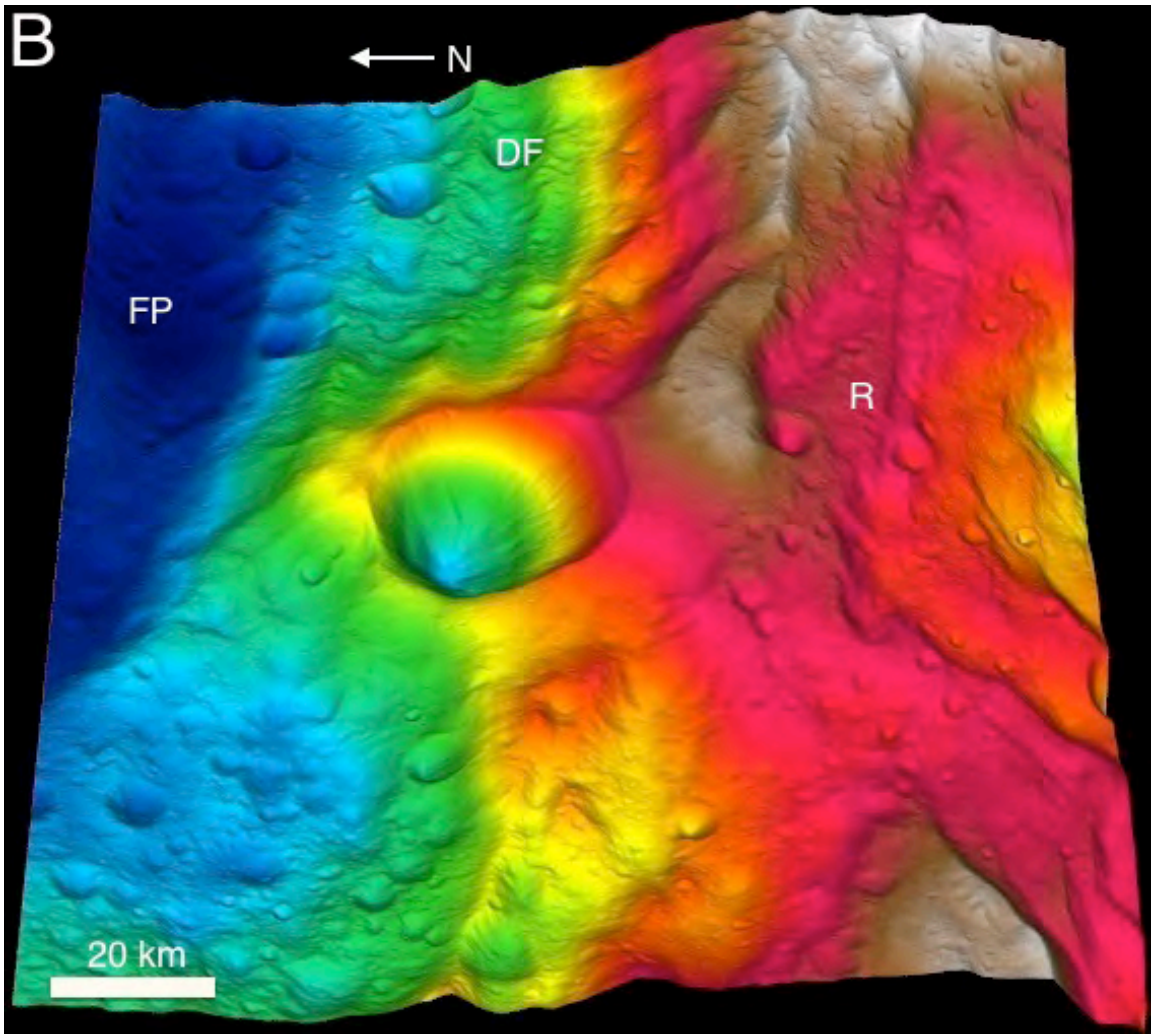

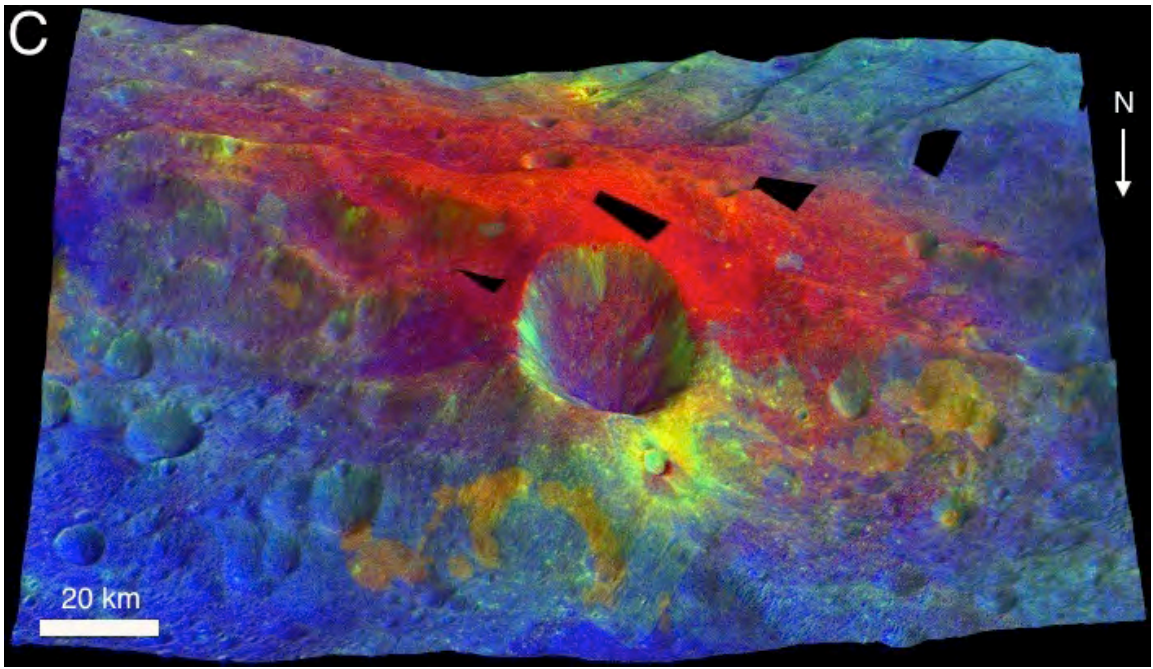

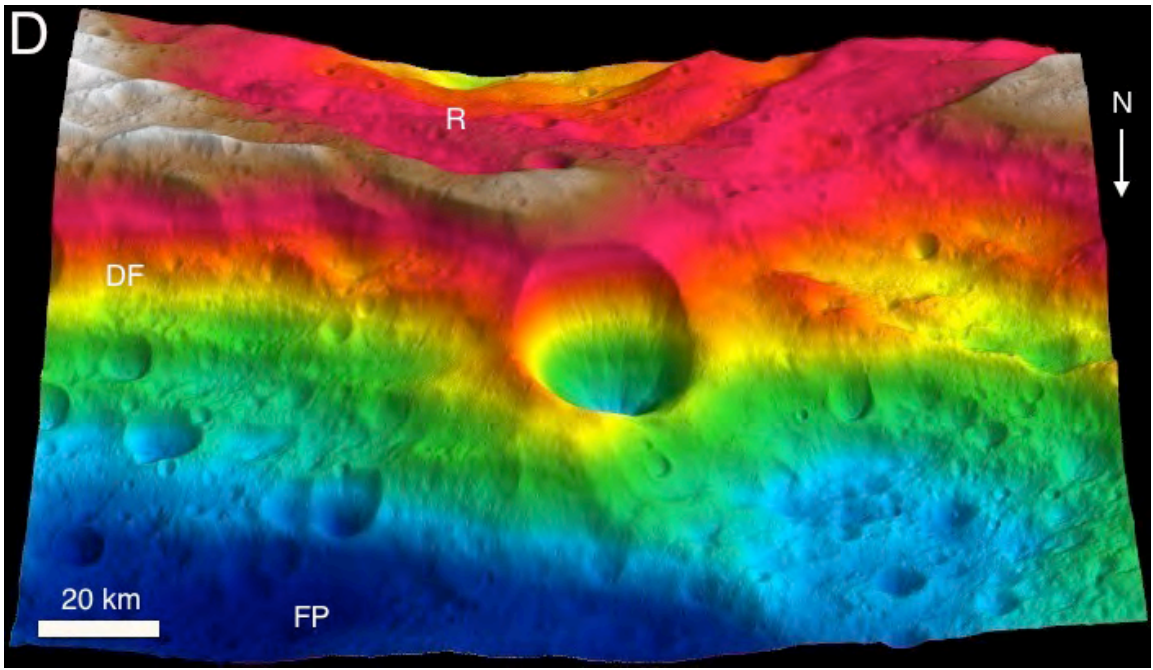

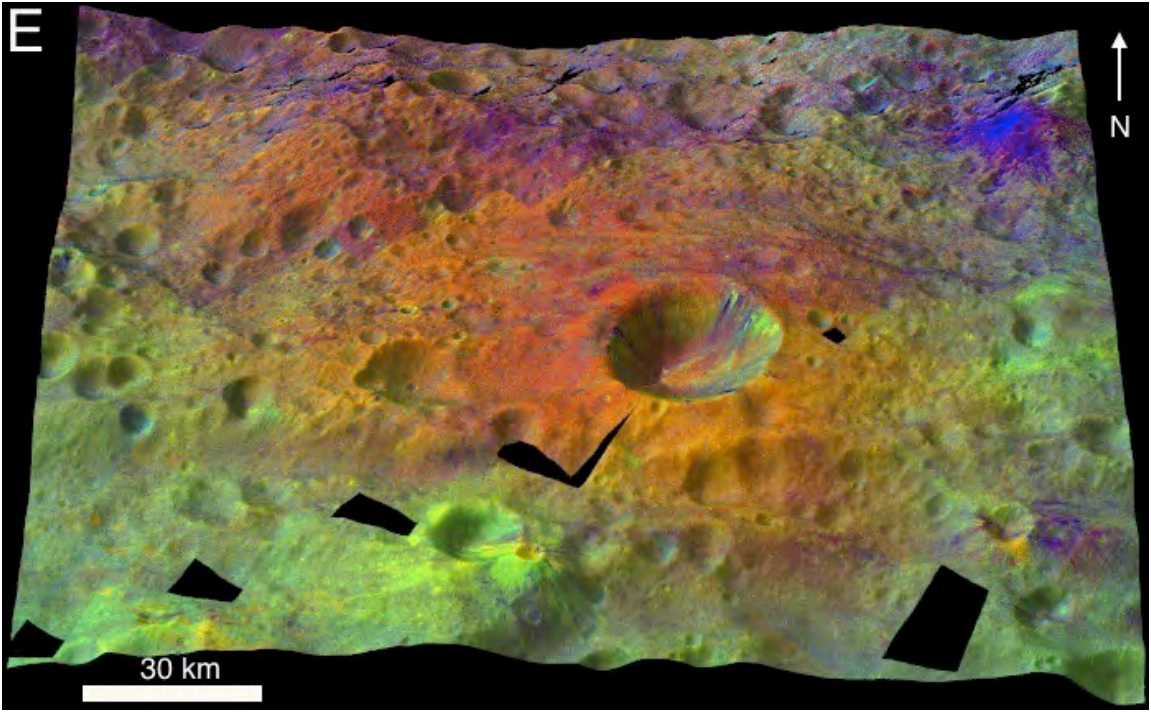

E

30 km

N

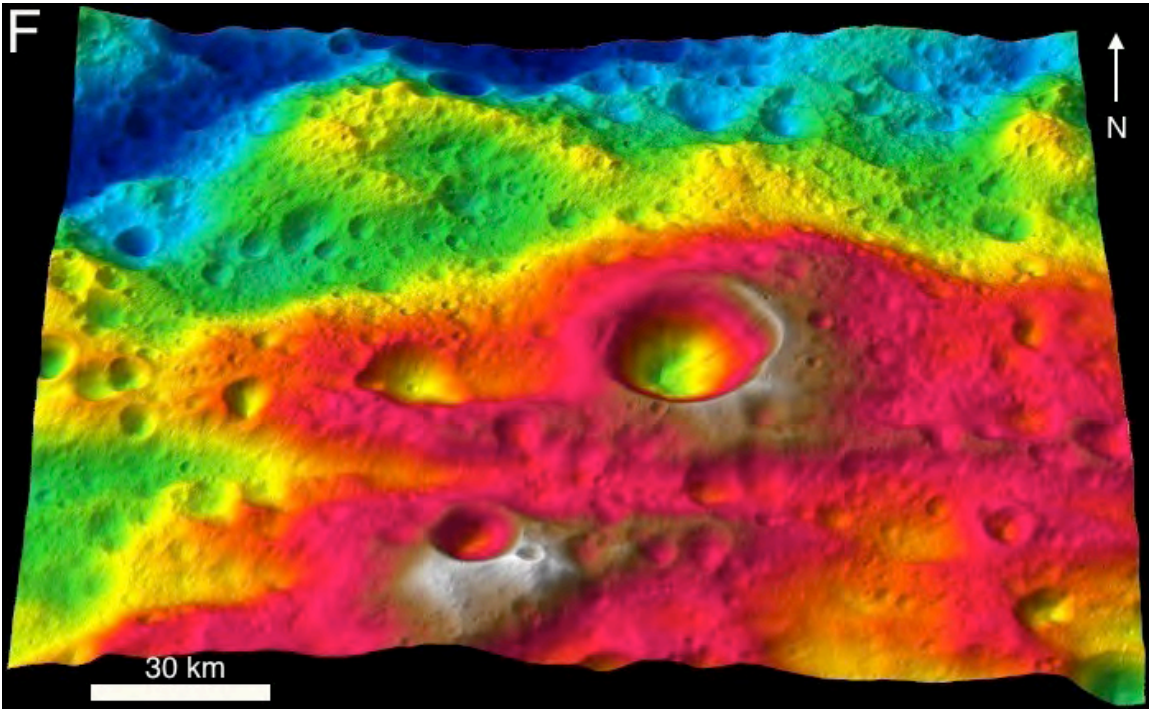

F

30 km

N

Figure 2.

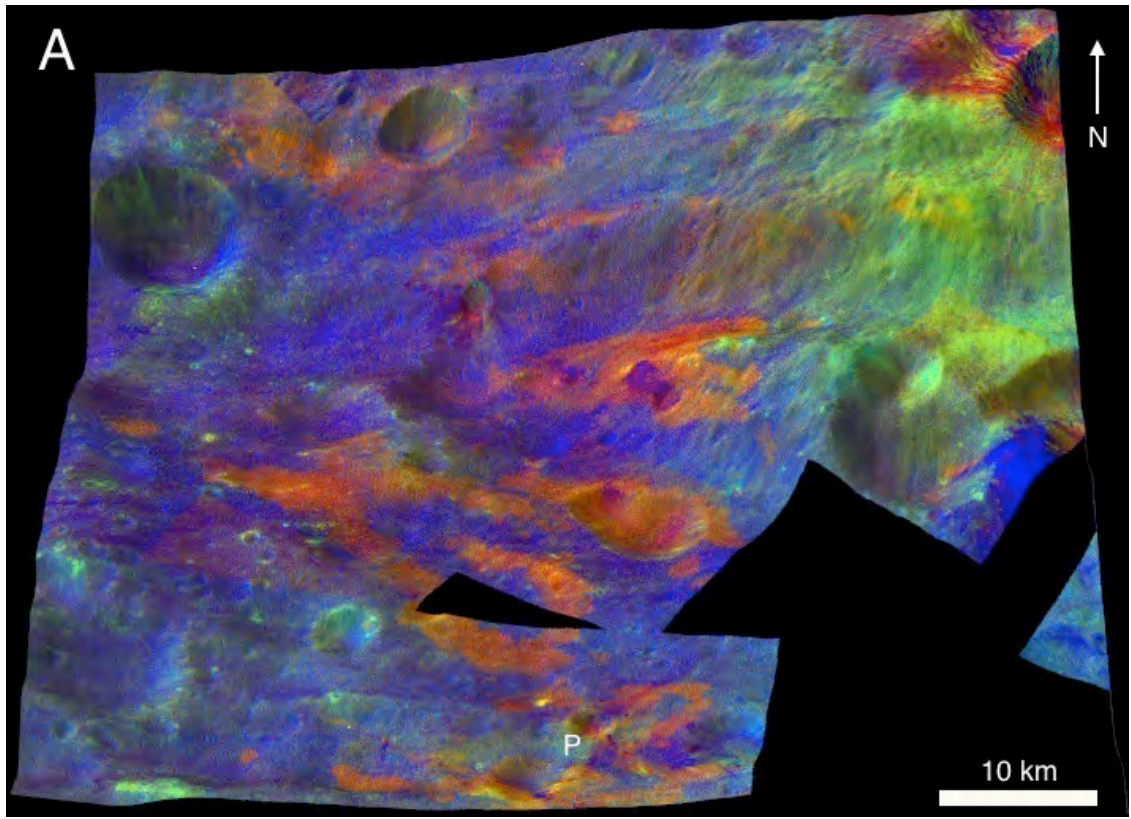

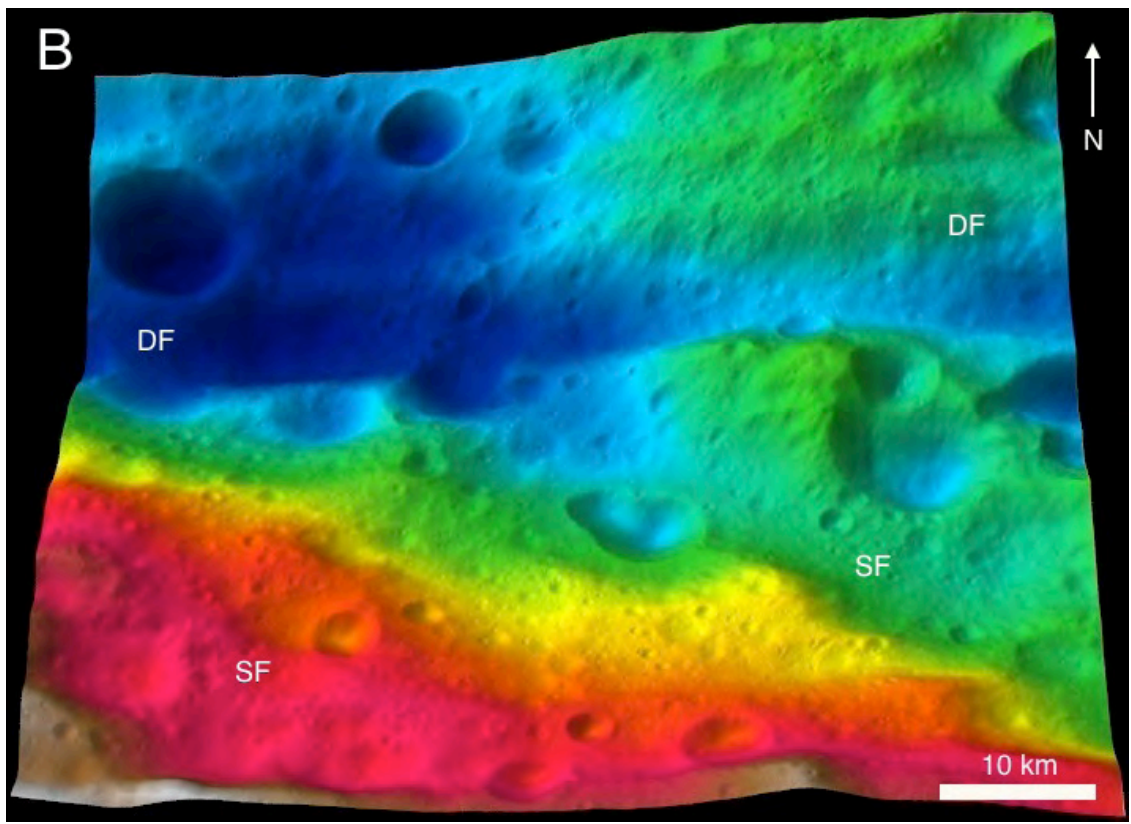

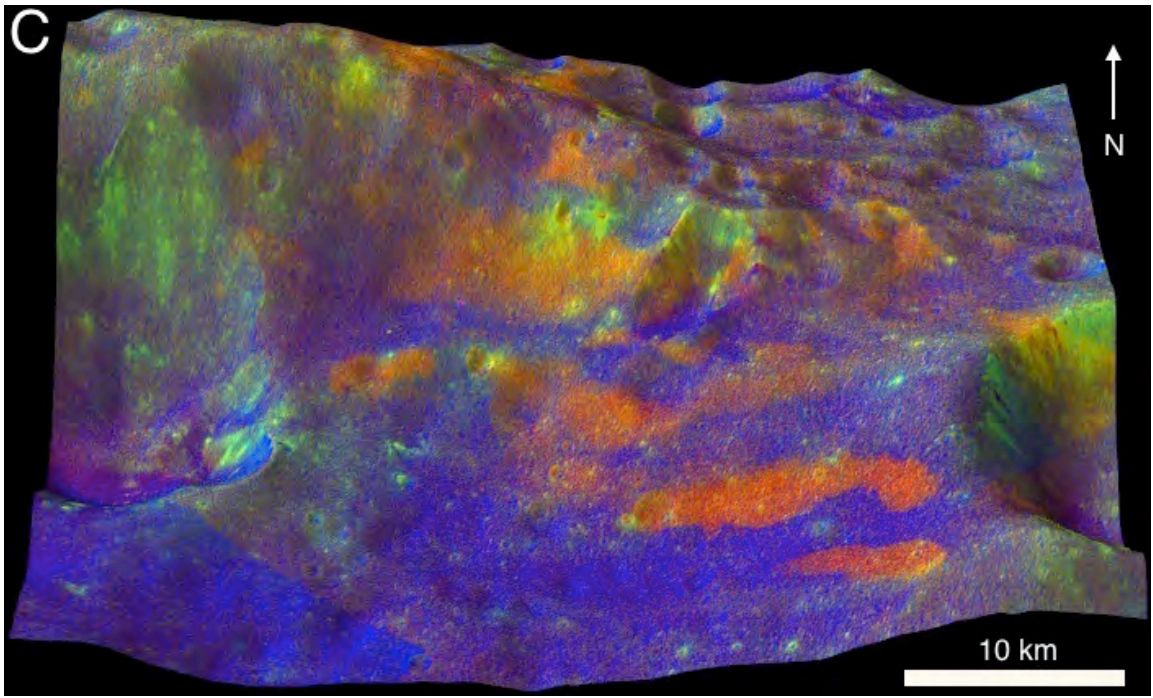

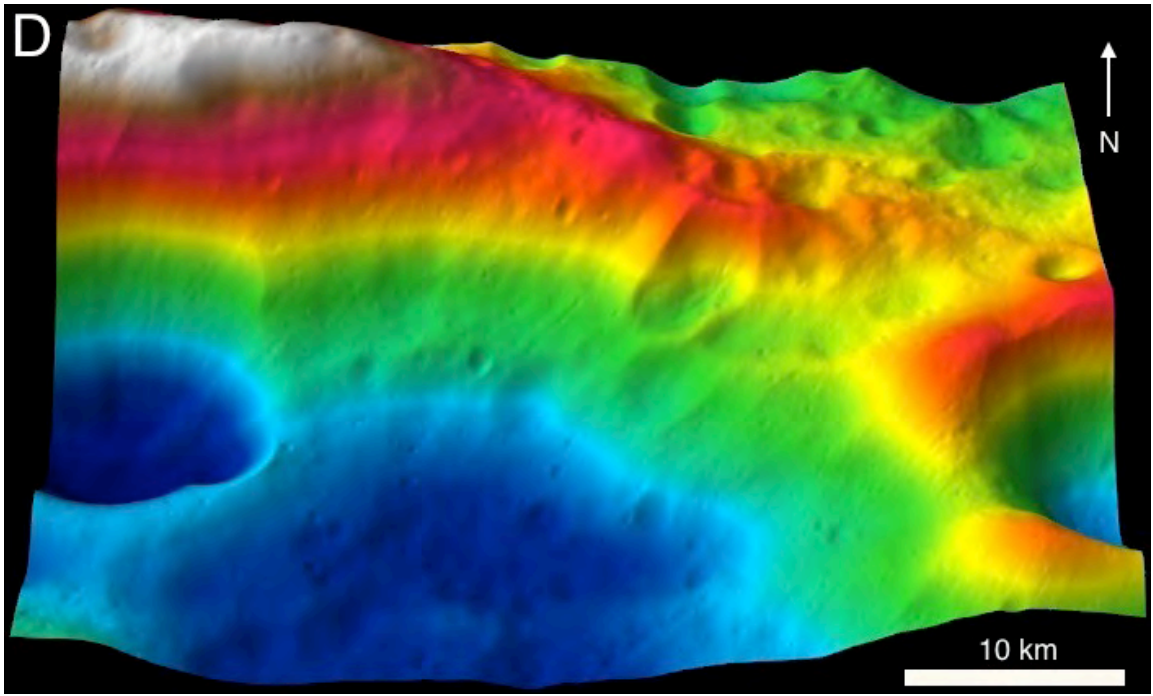

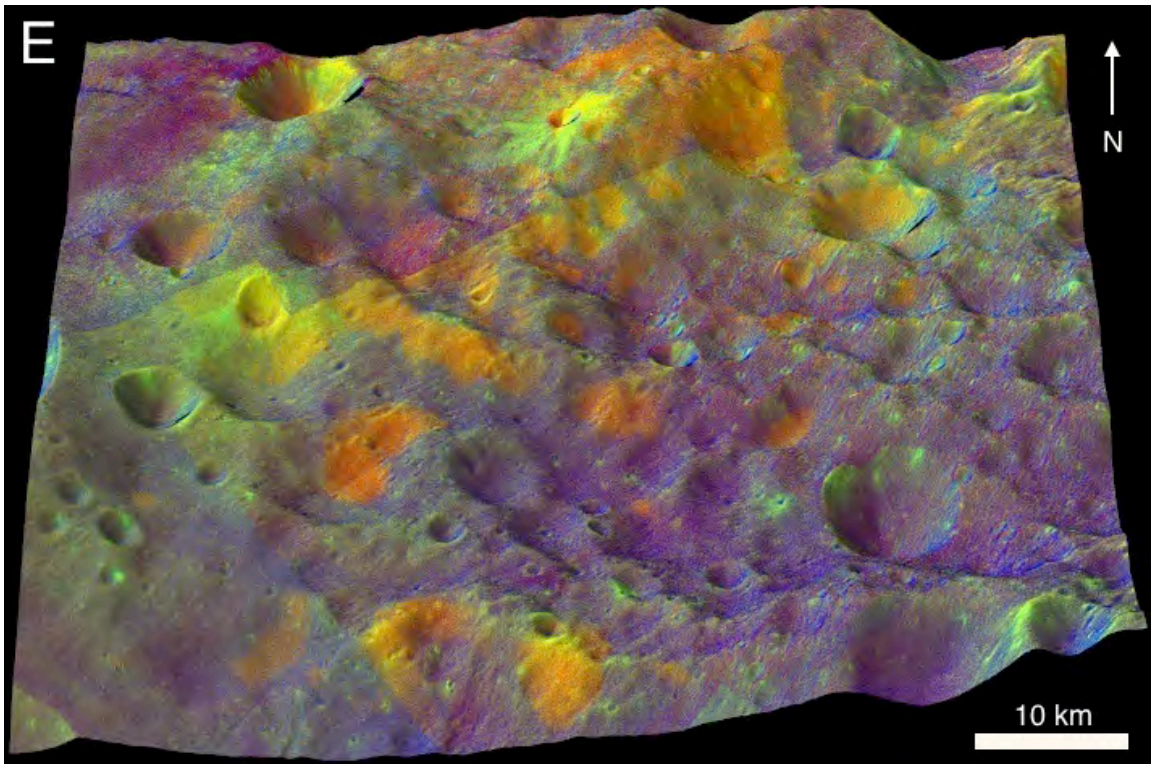

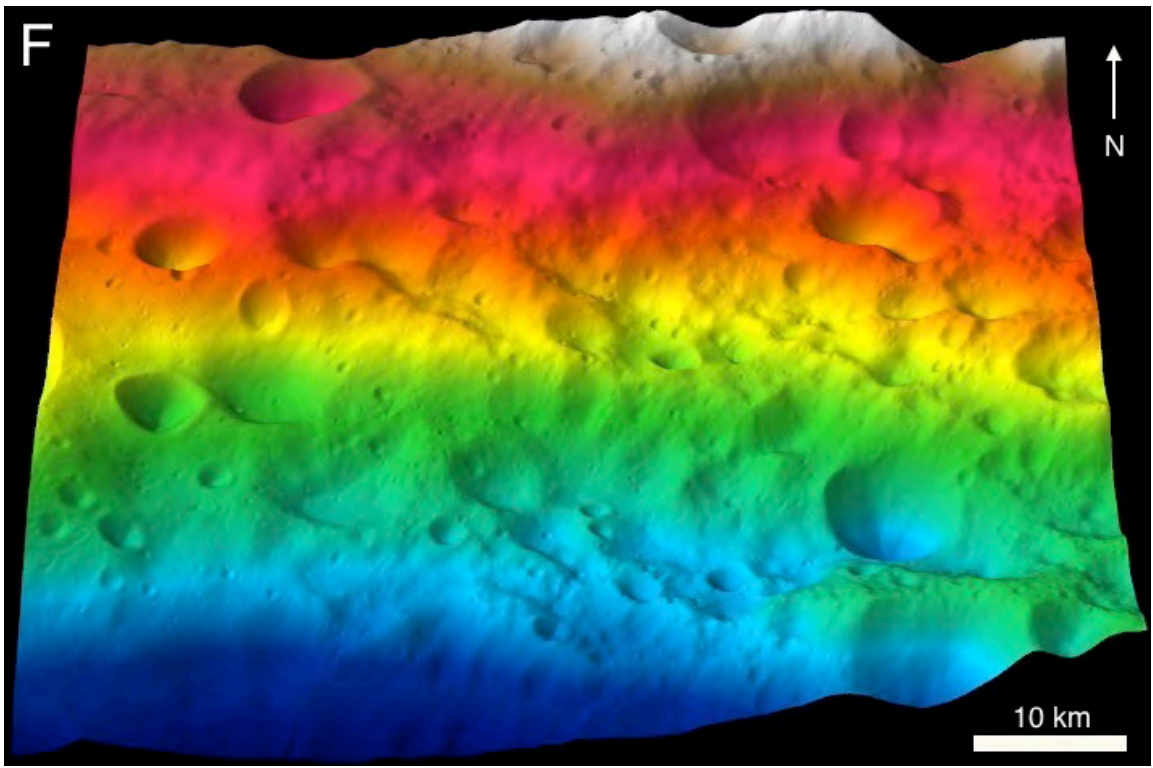

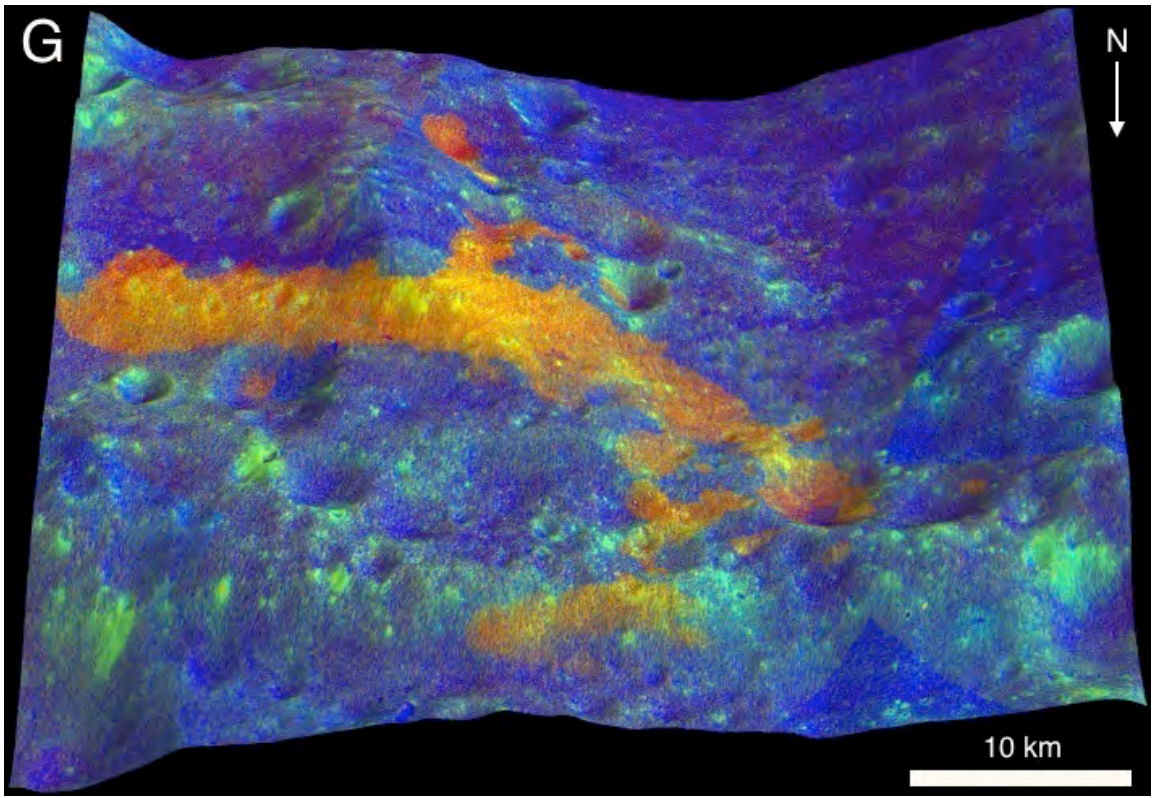

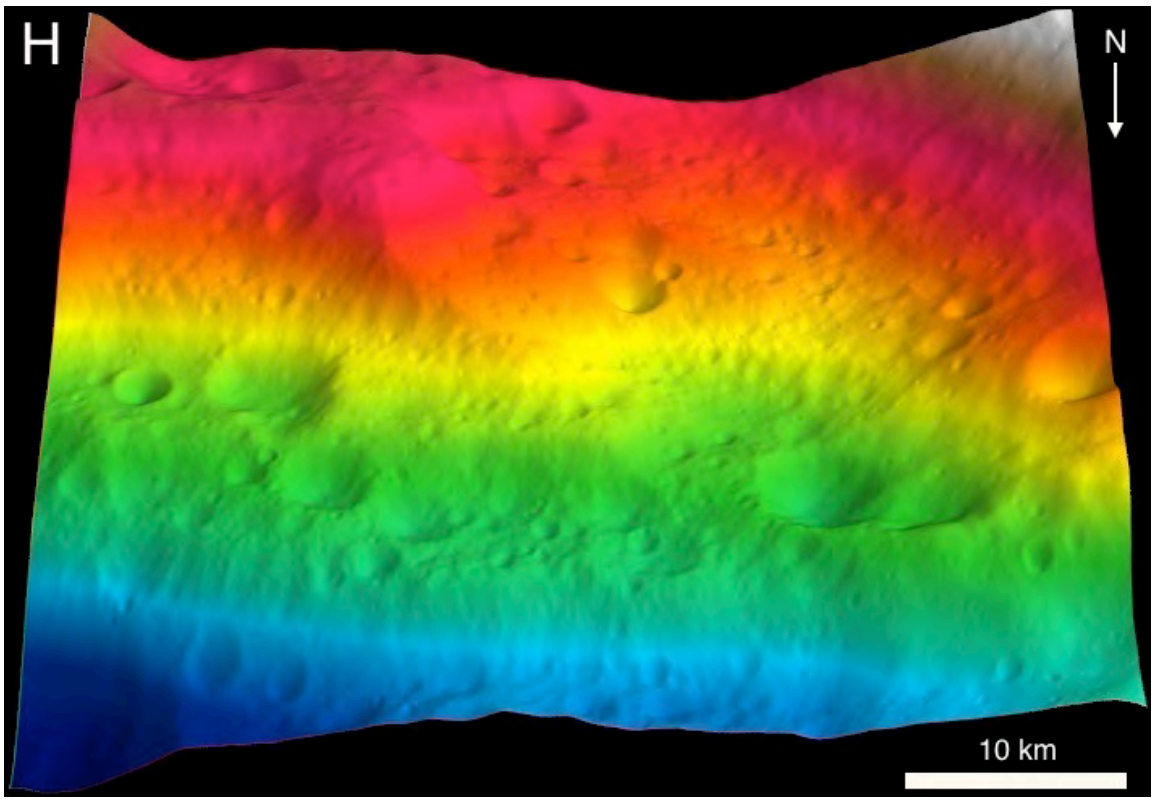

Figure 3.

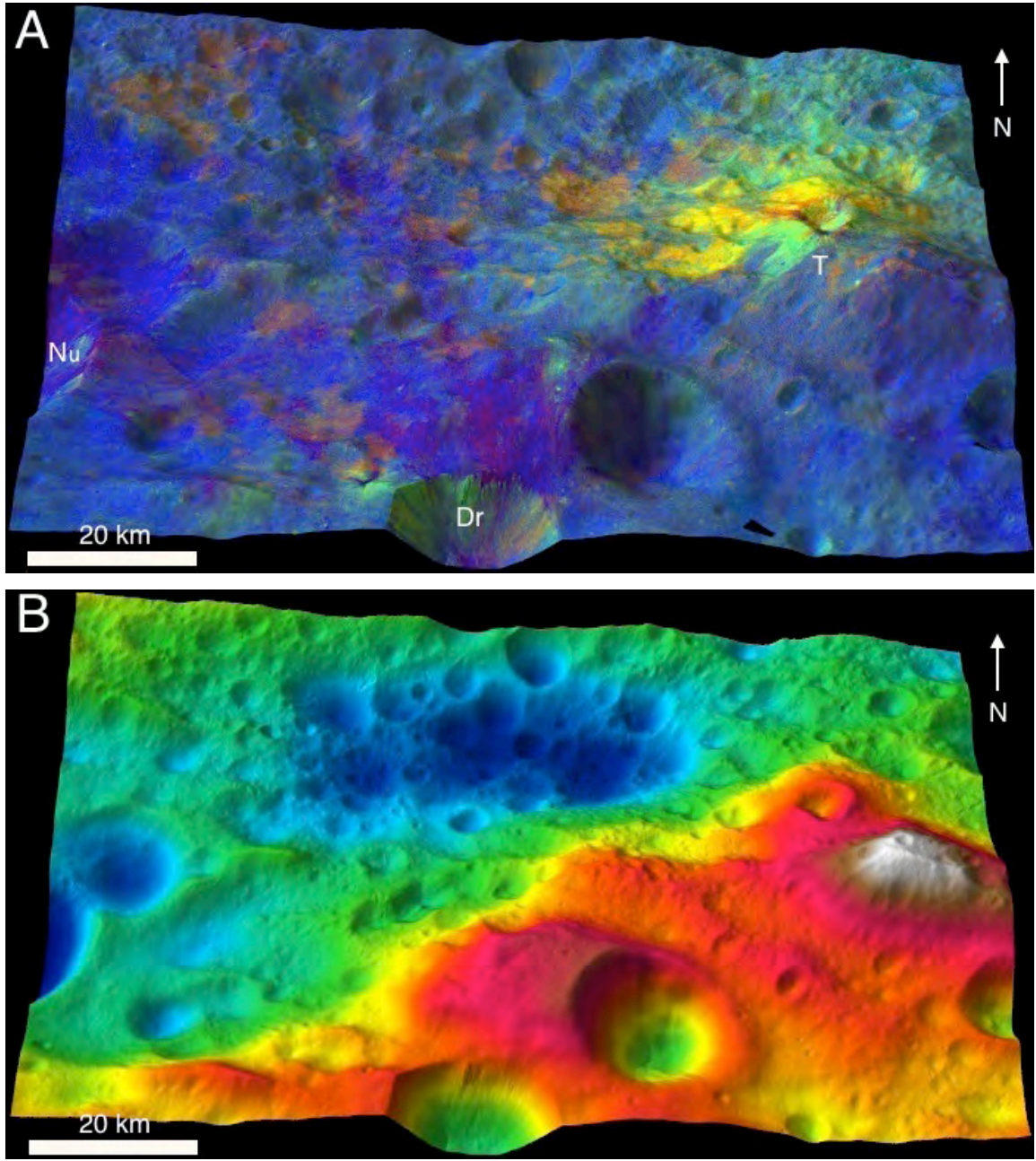

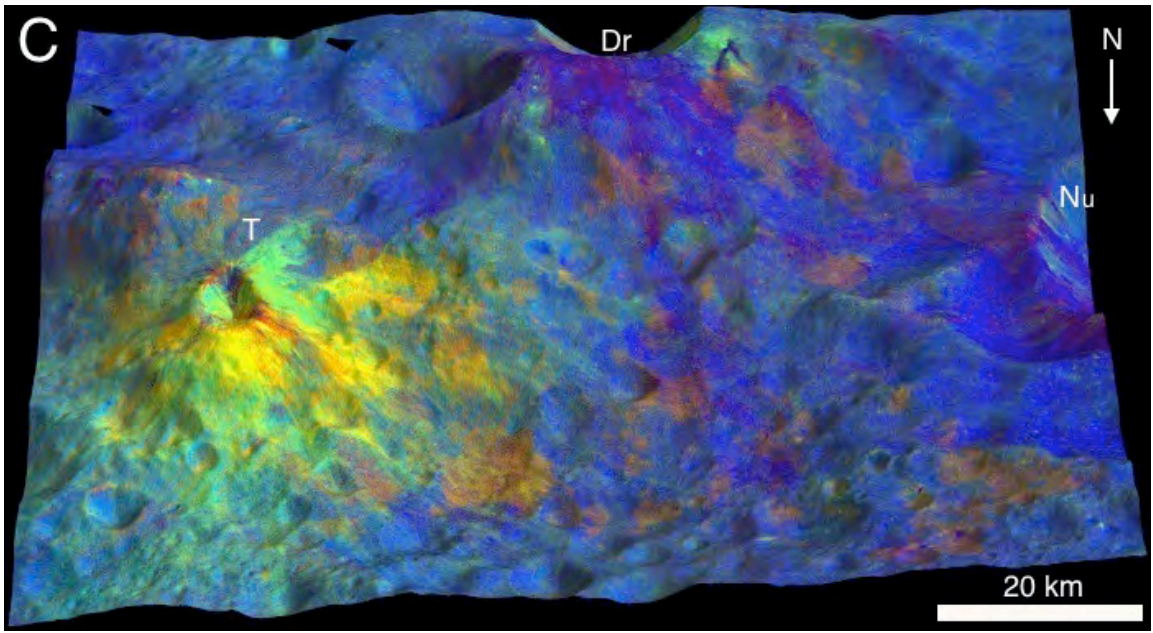

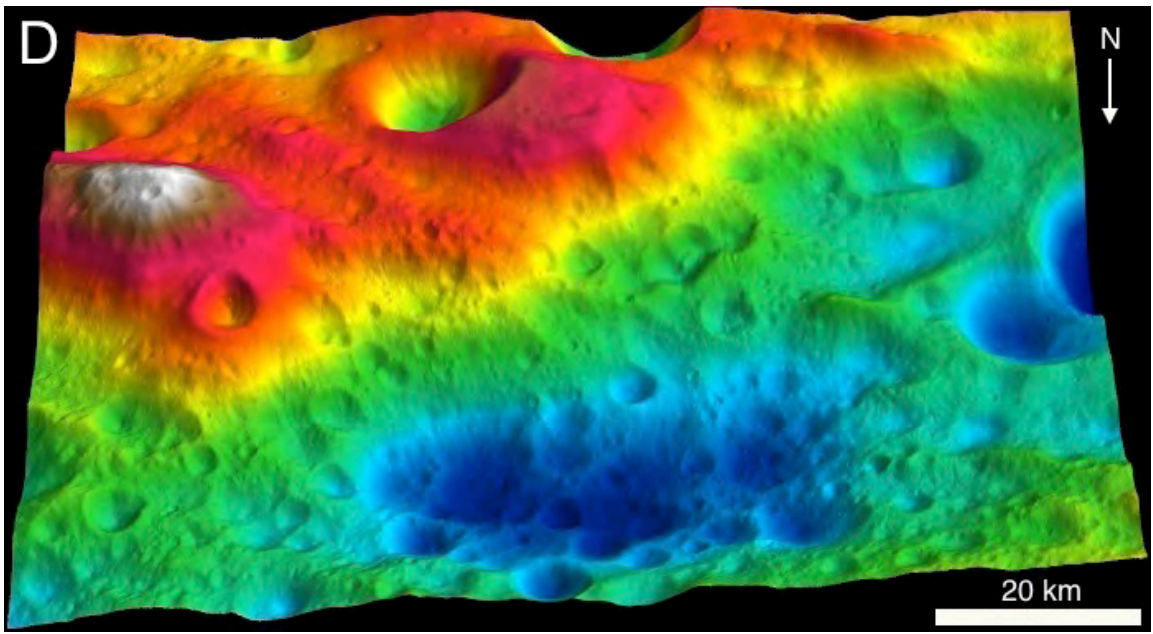

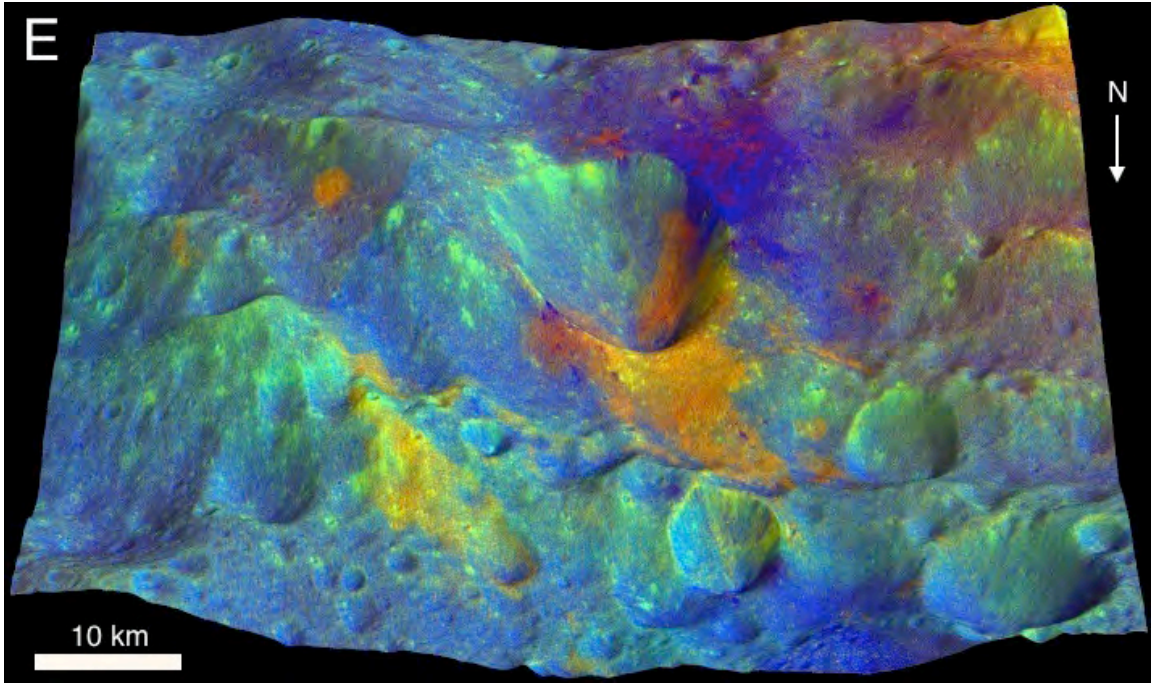

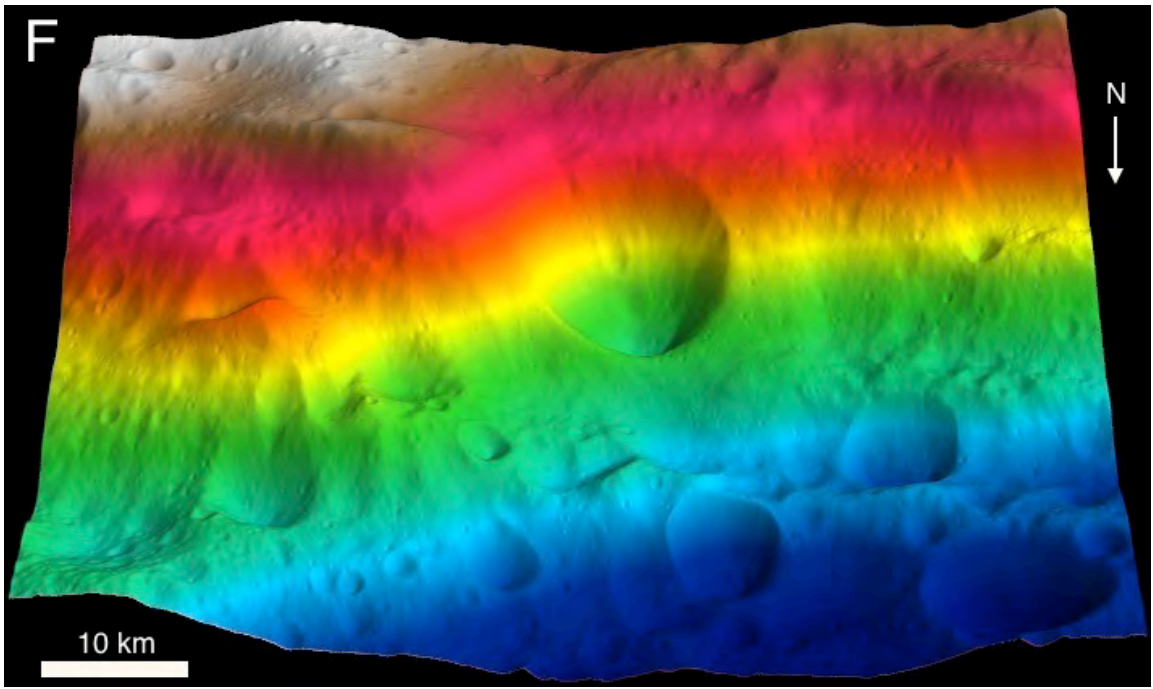

Figure 4.

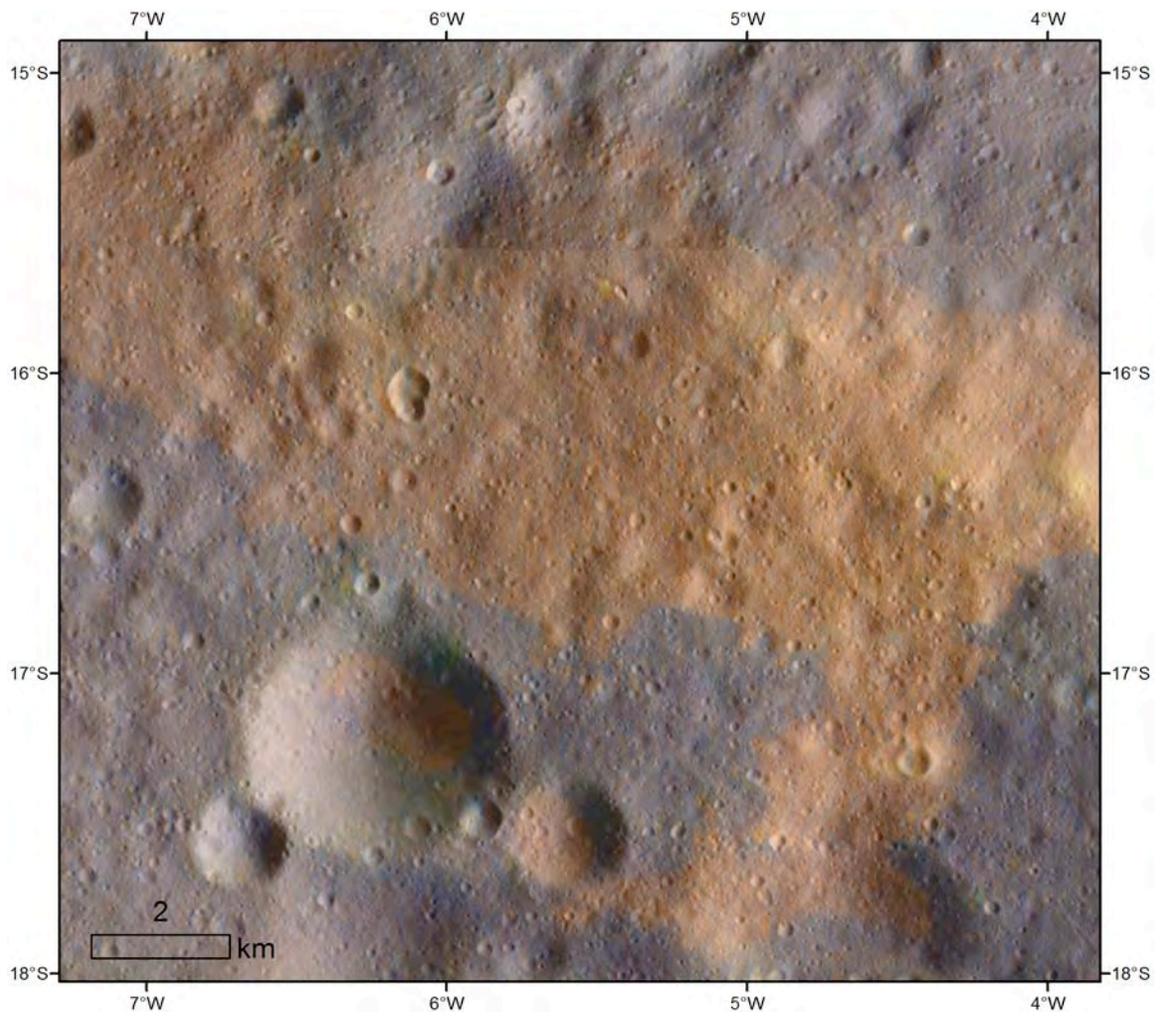

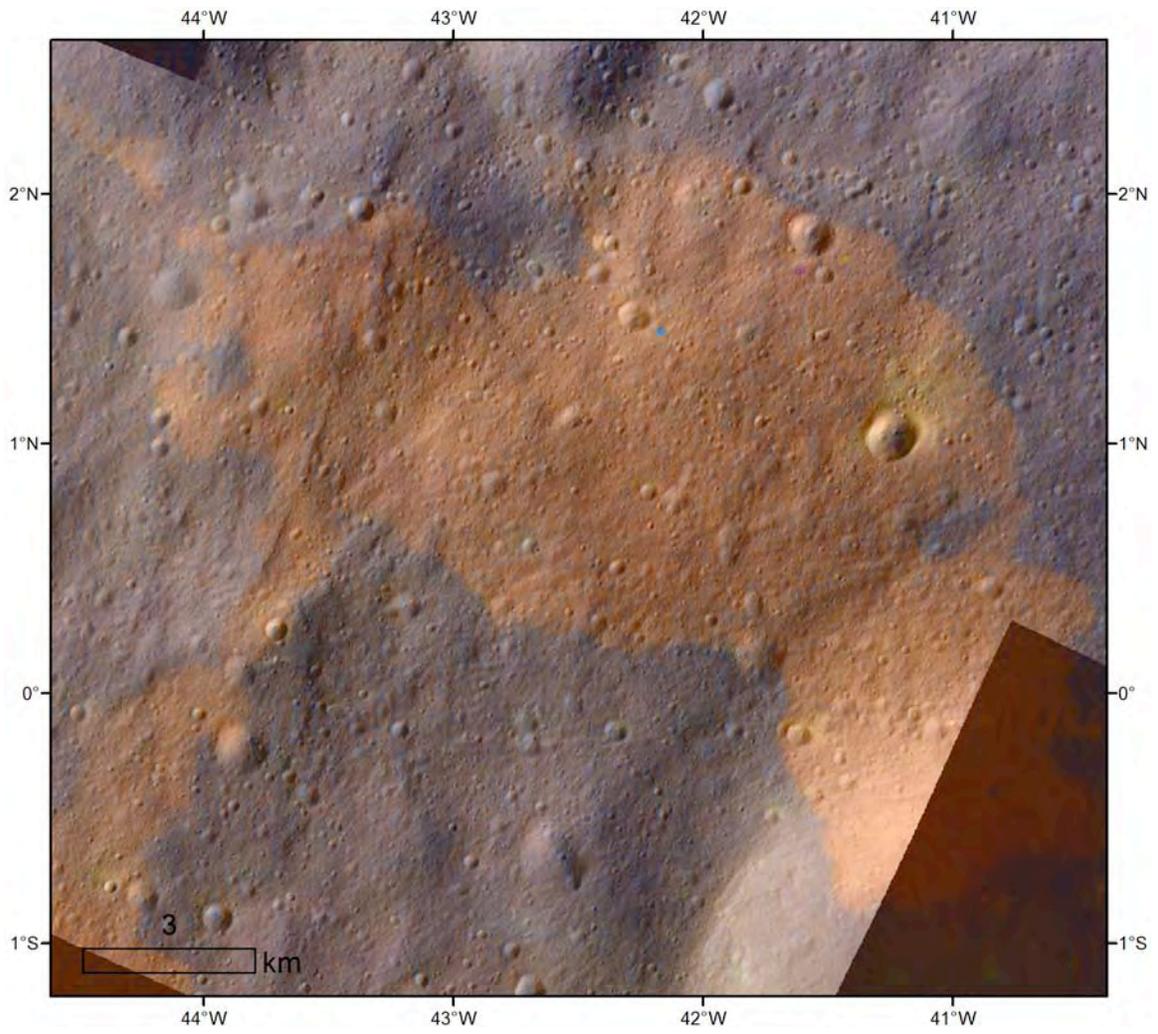

Figure 5.

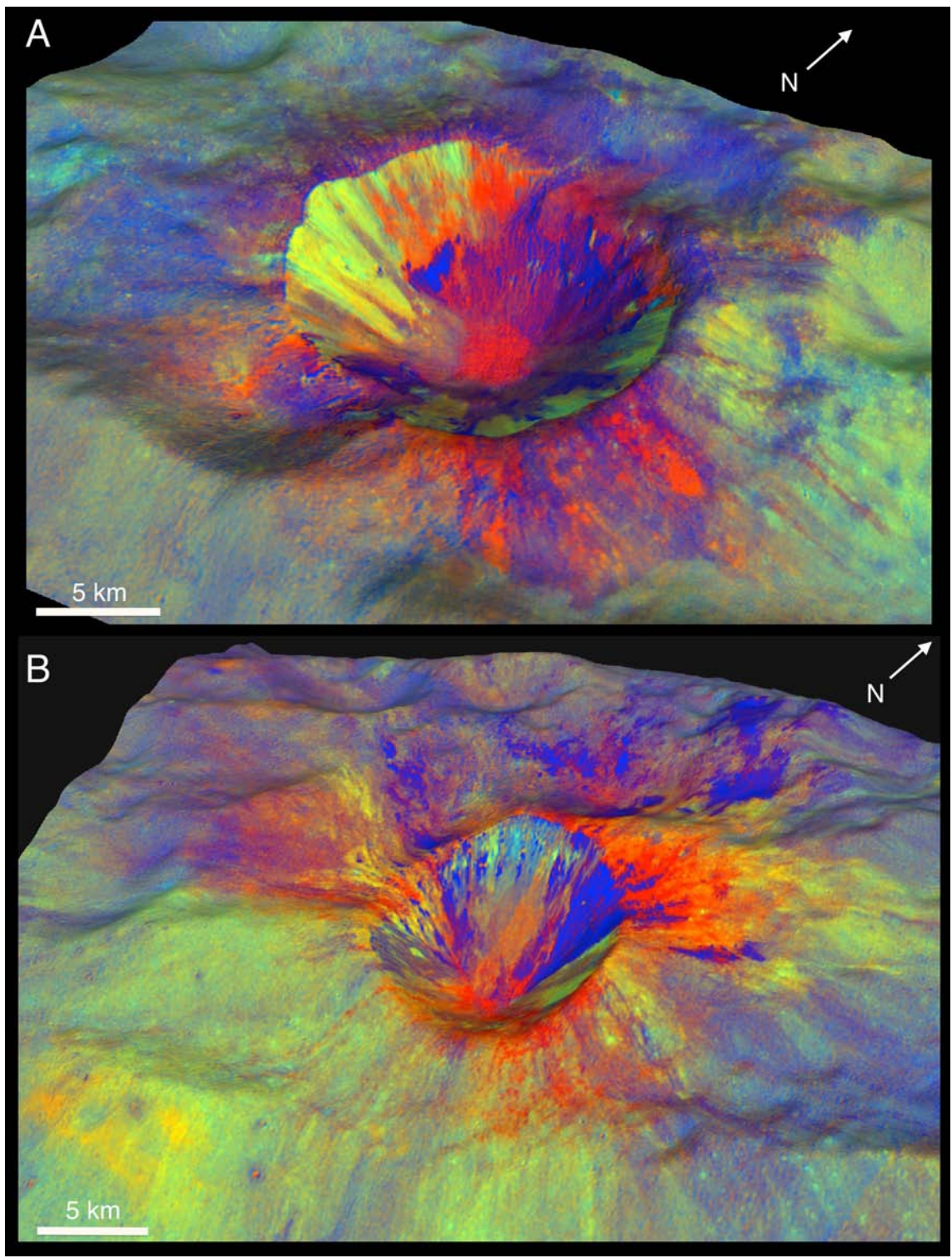

Figure 6 A.

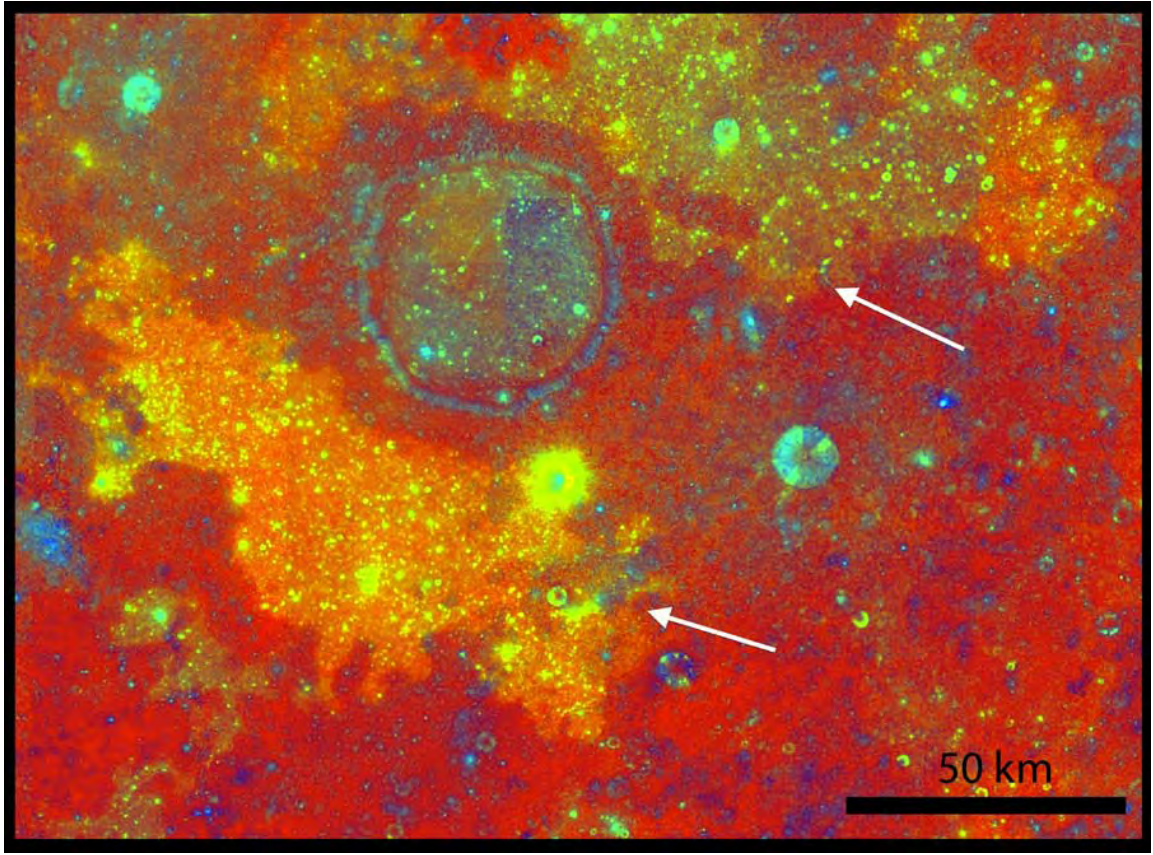

Figure 6 B.

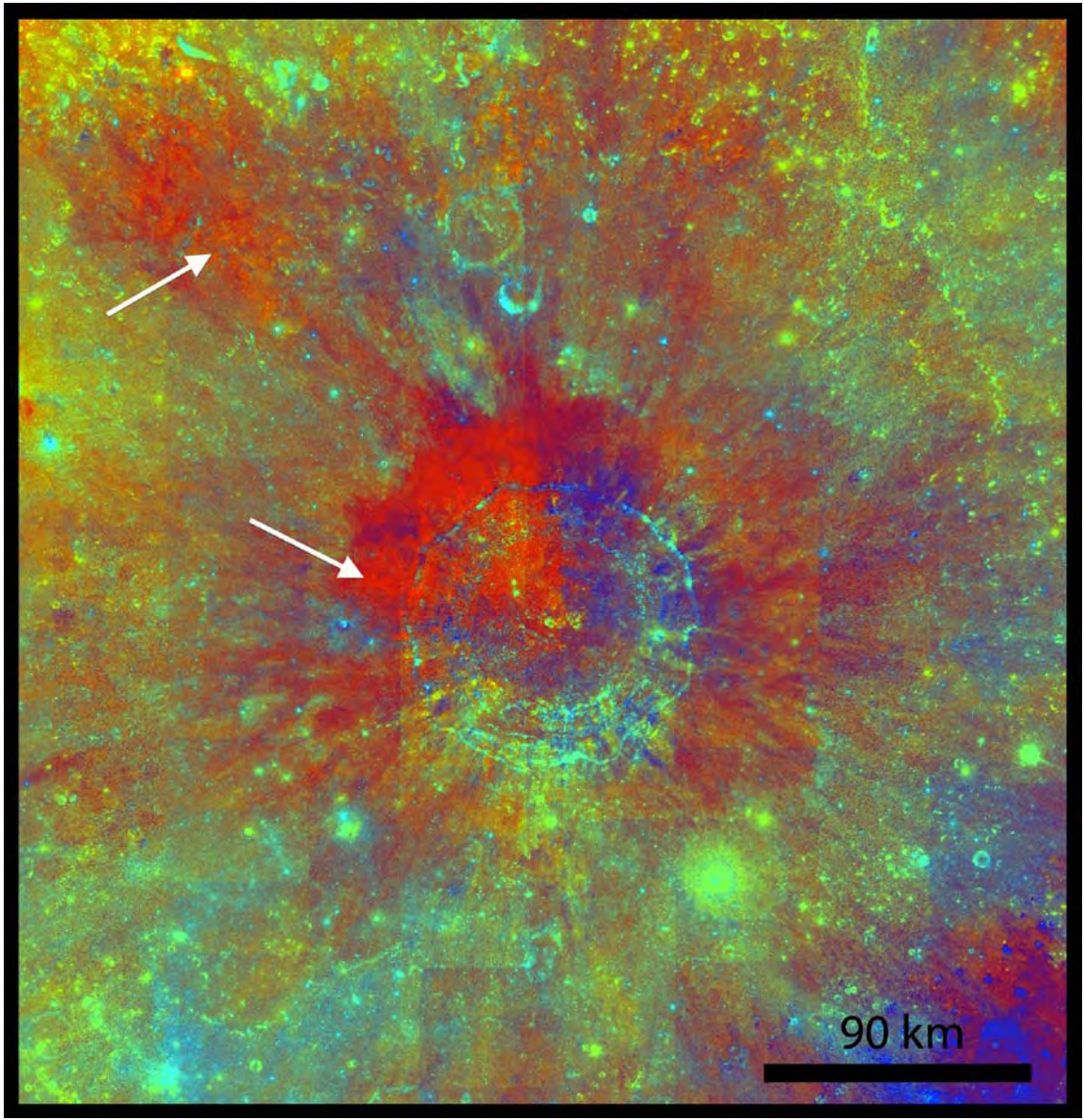

Figure 7 A.

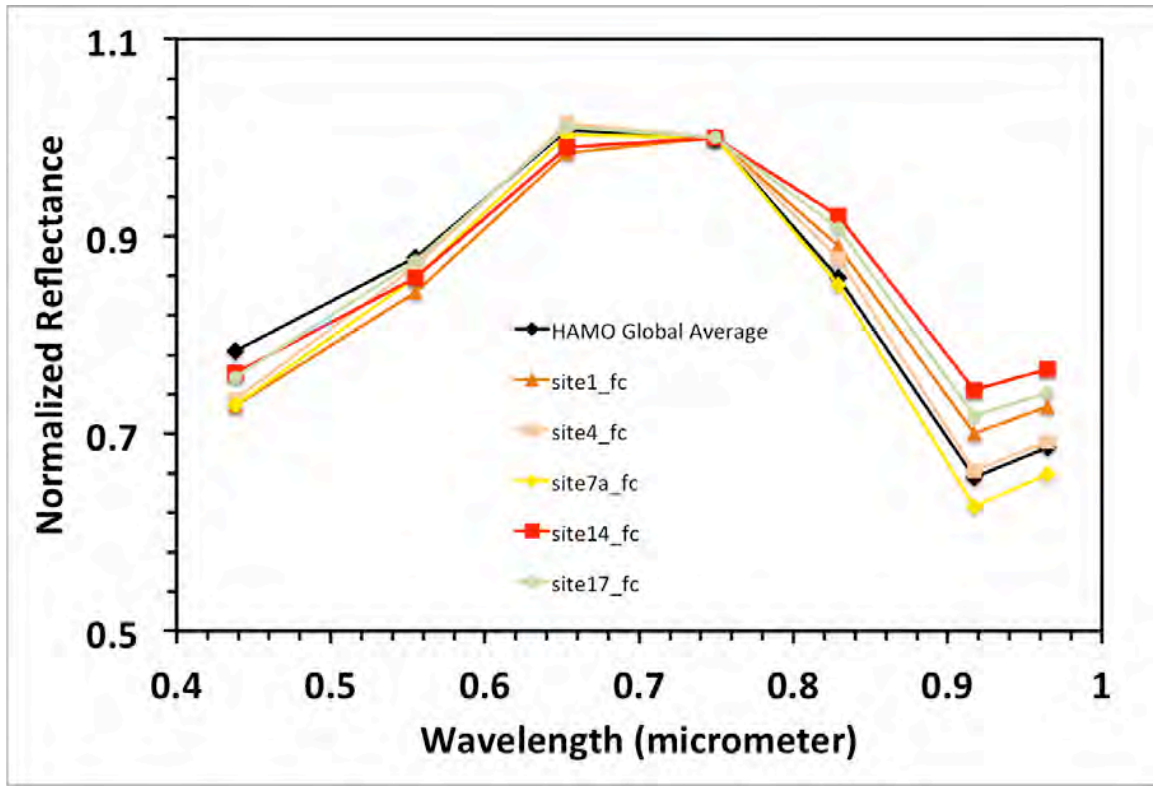

Figure 7 B.

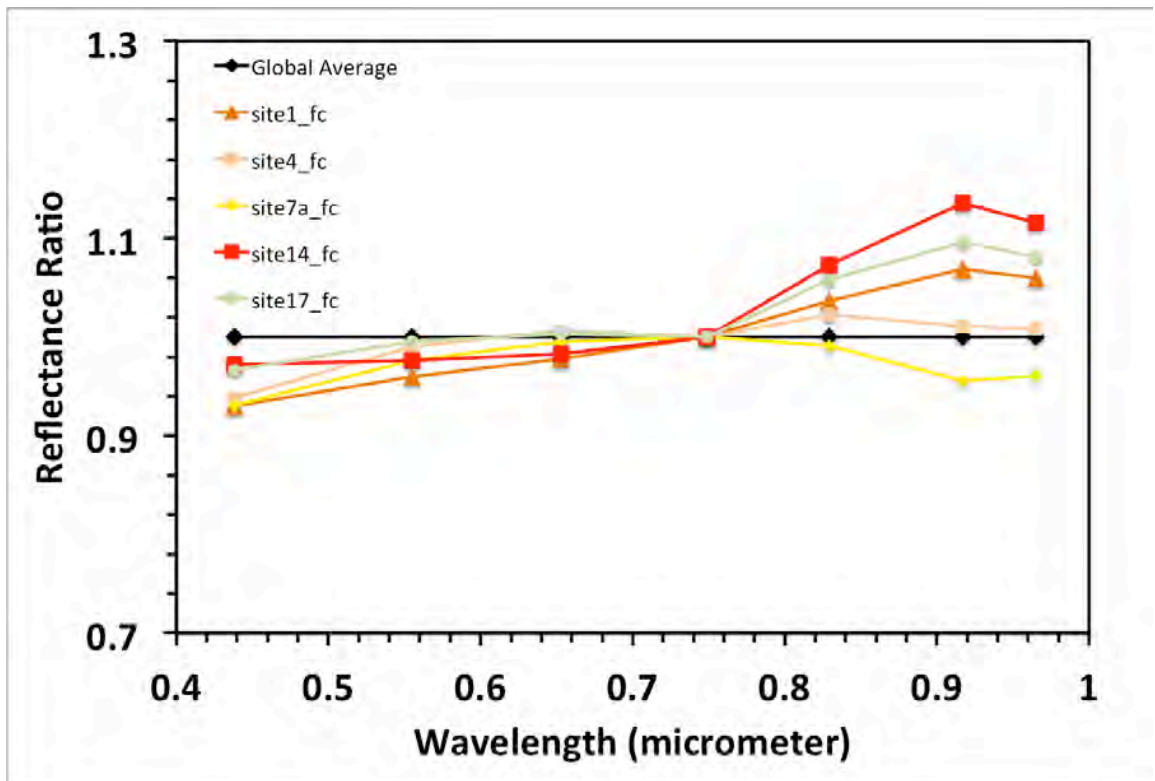

Figure 8.

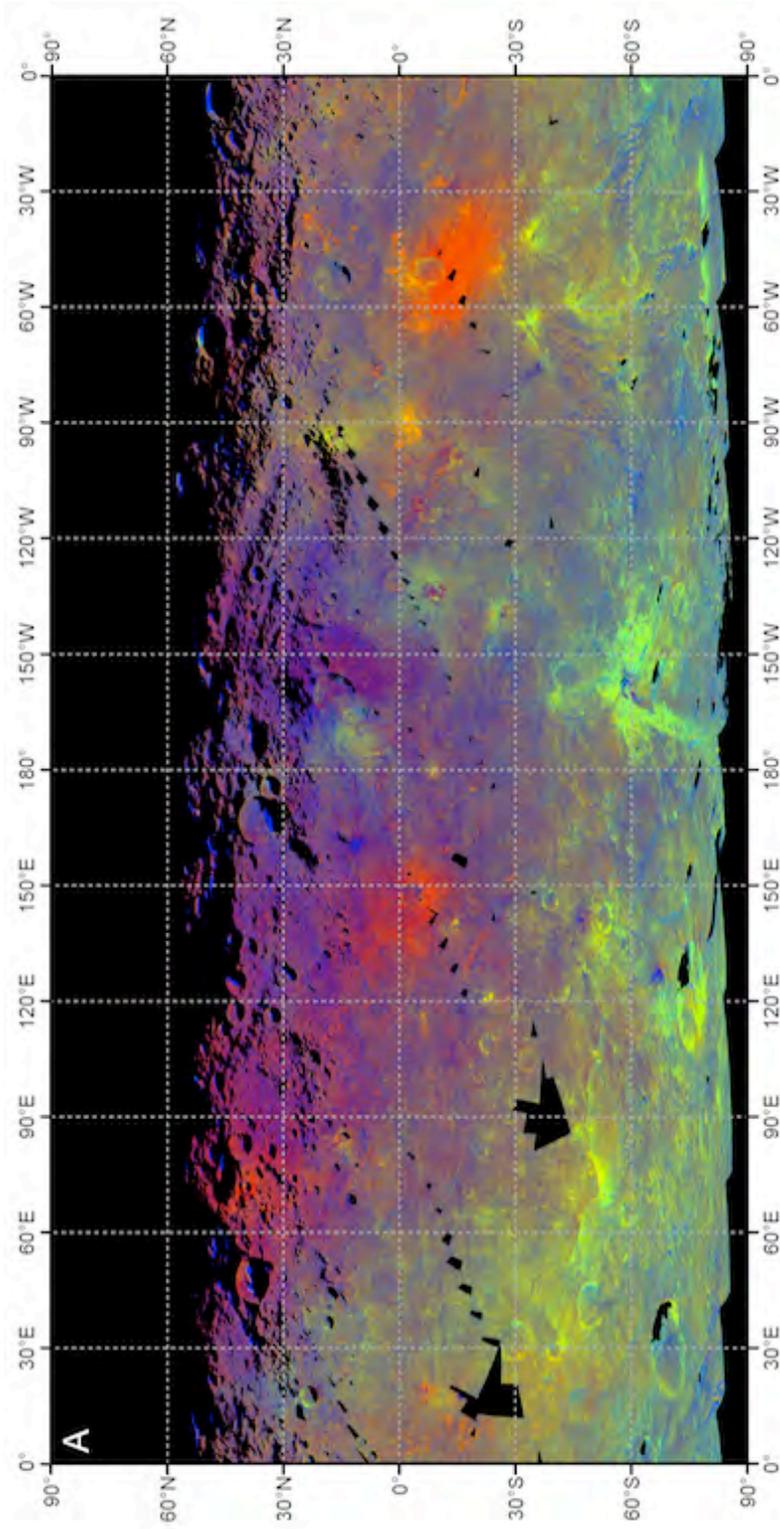

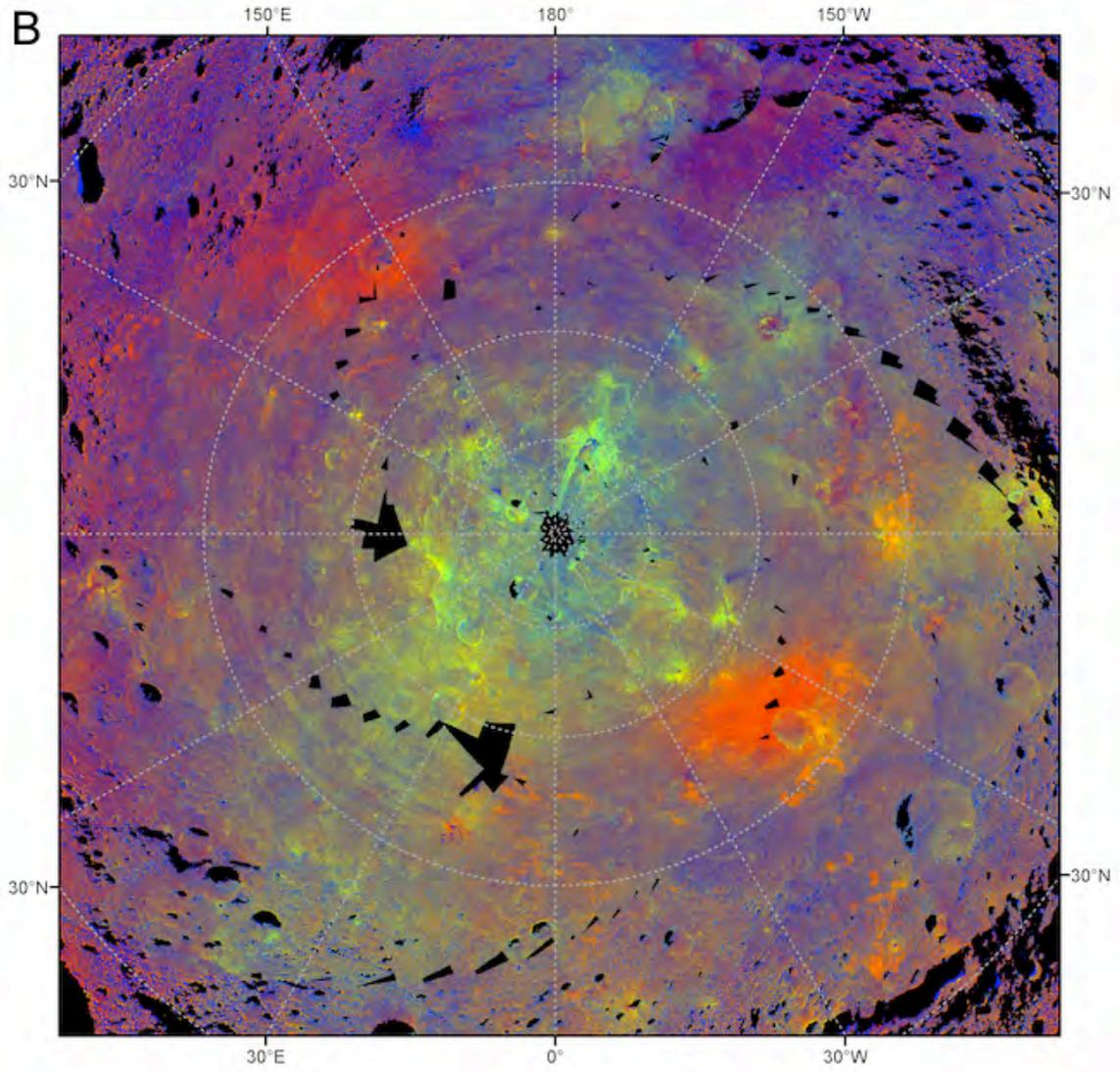

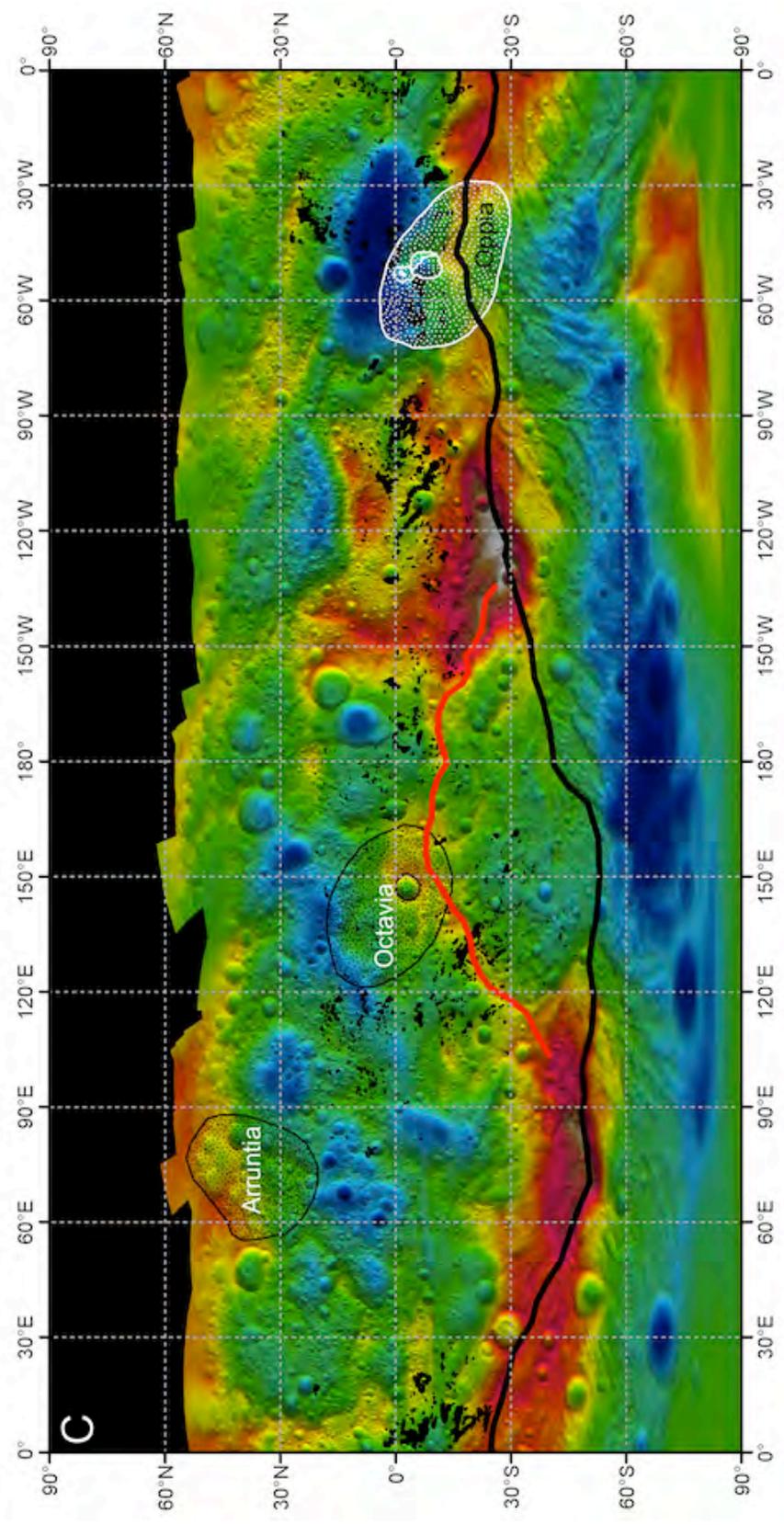

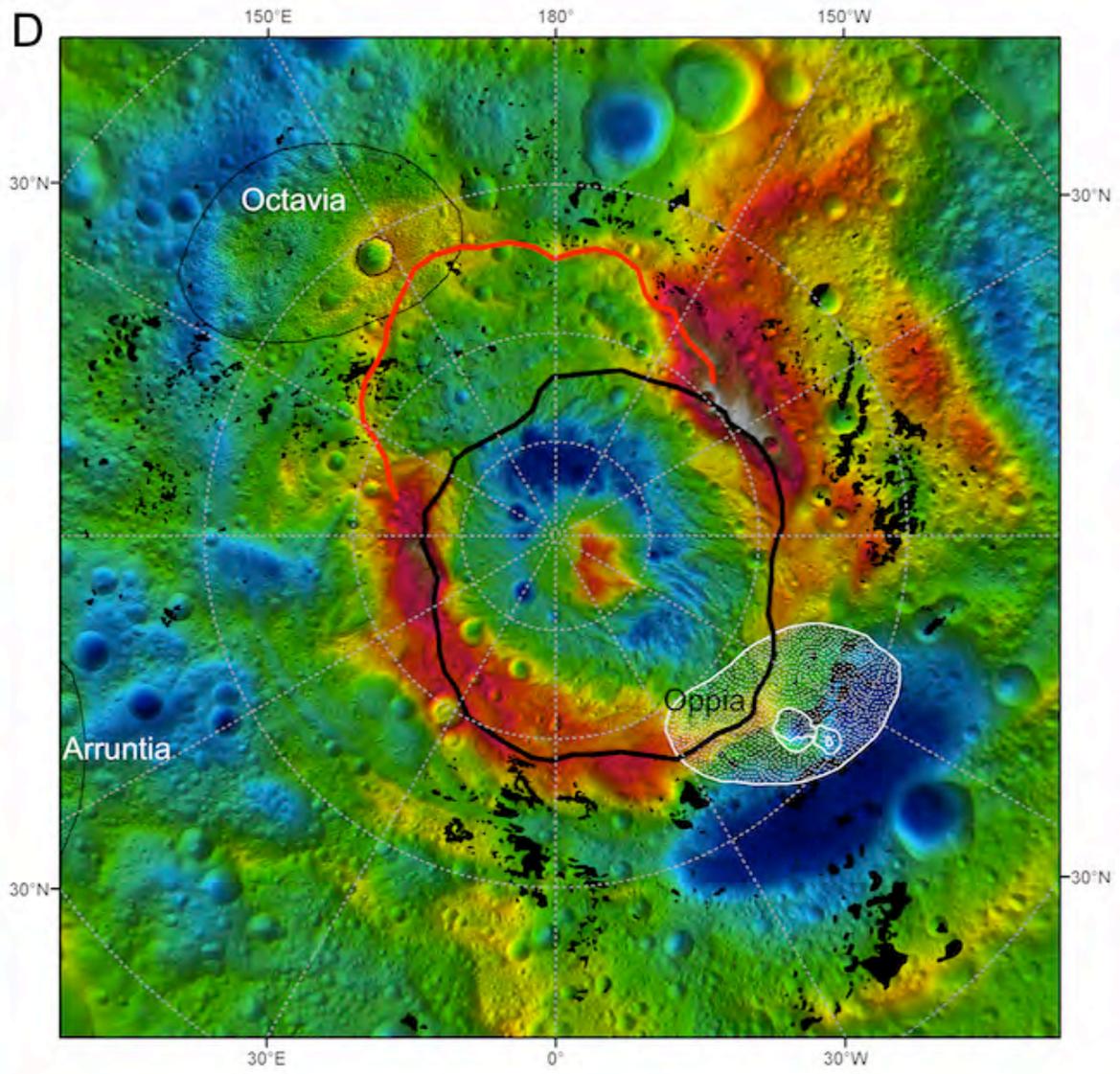

Figure 9.

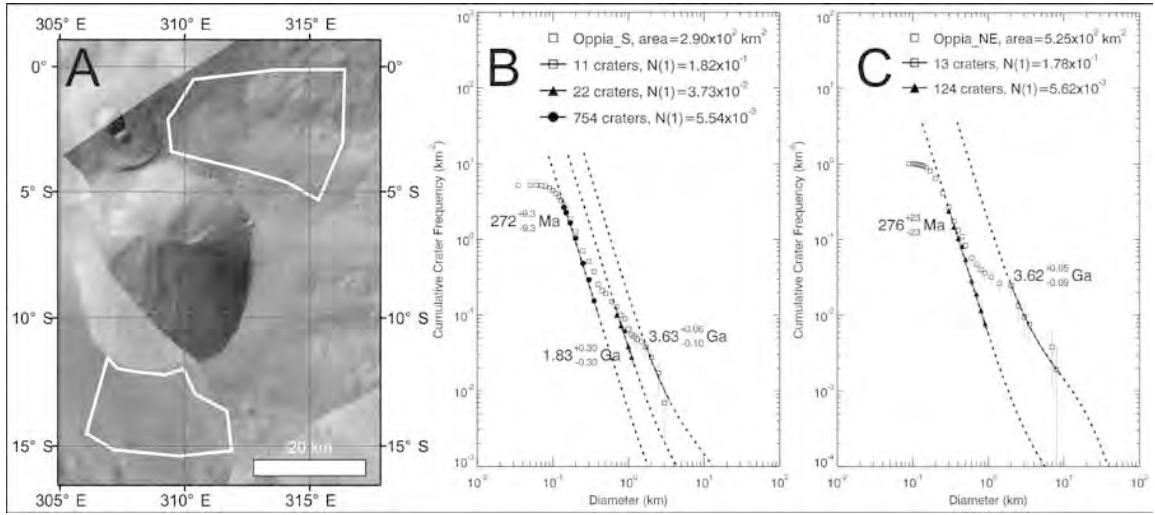

Figure 10.

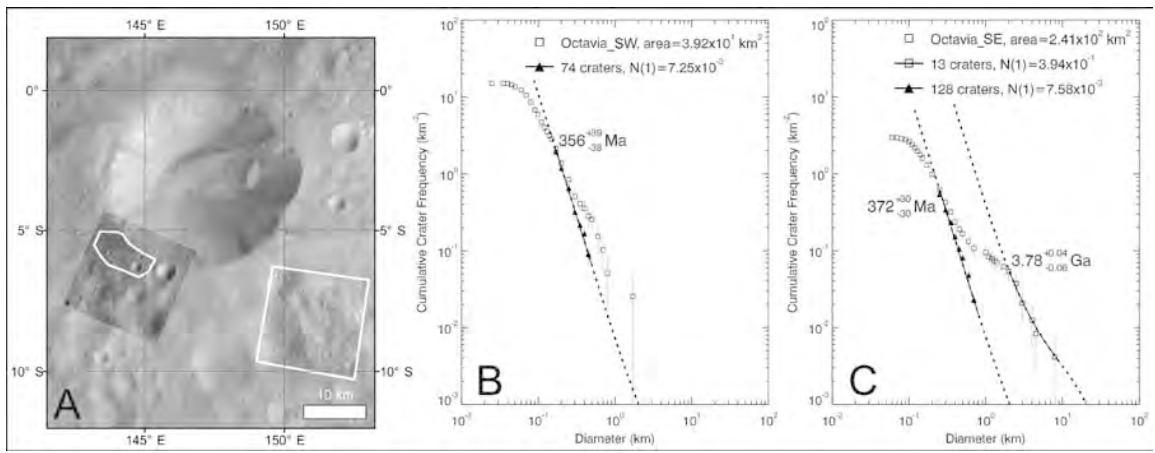

Figure 11.

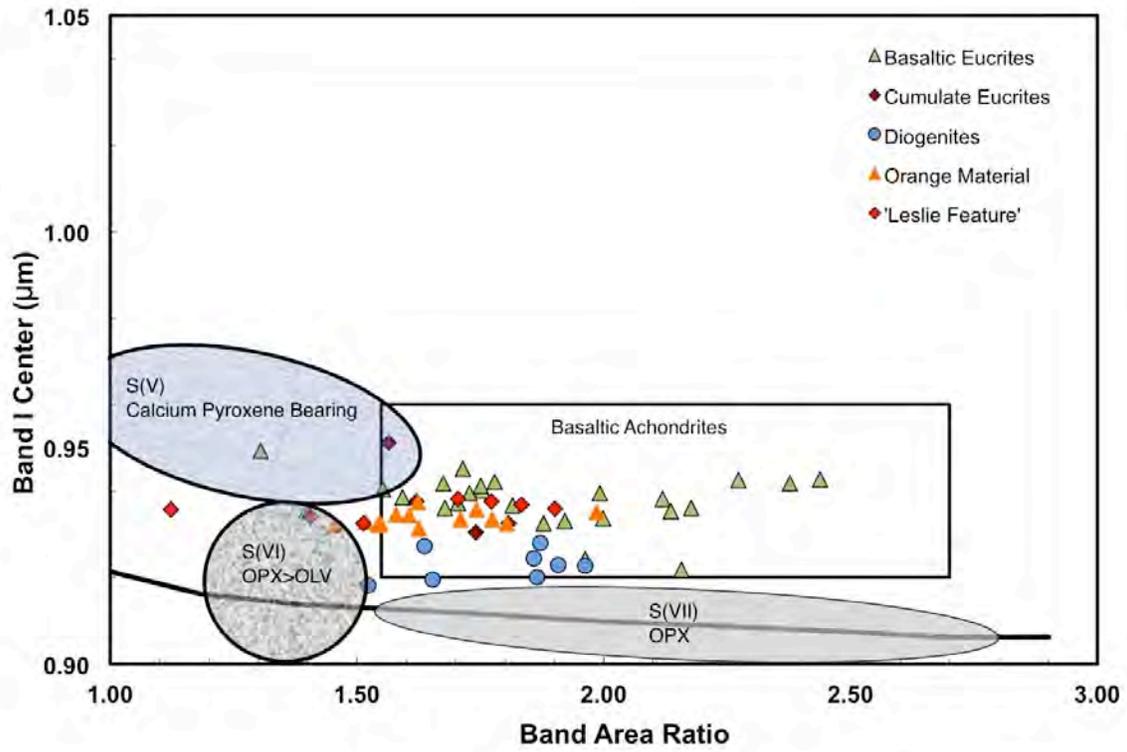

Figure 12 A.

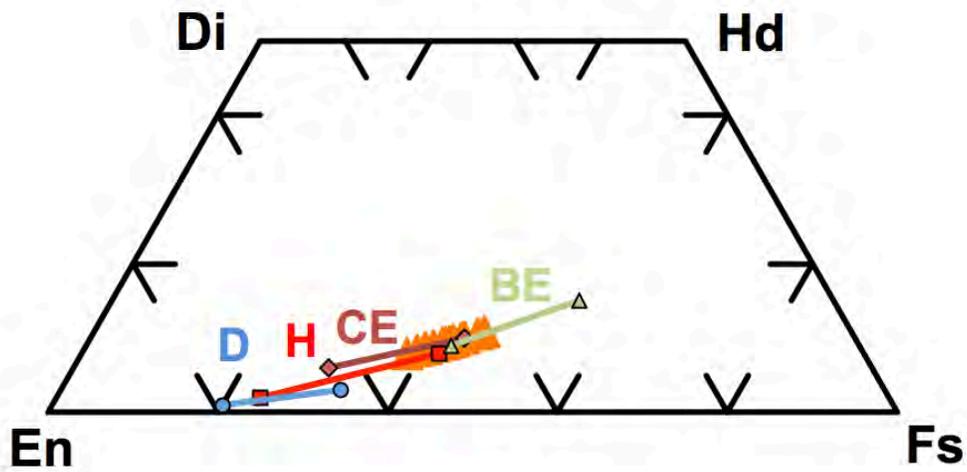

Figure 12 B.

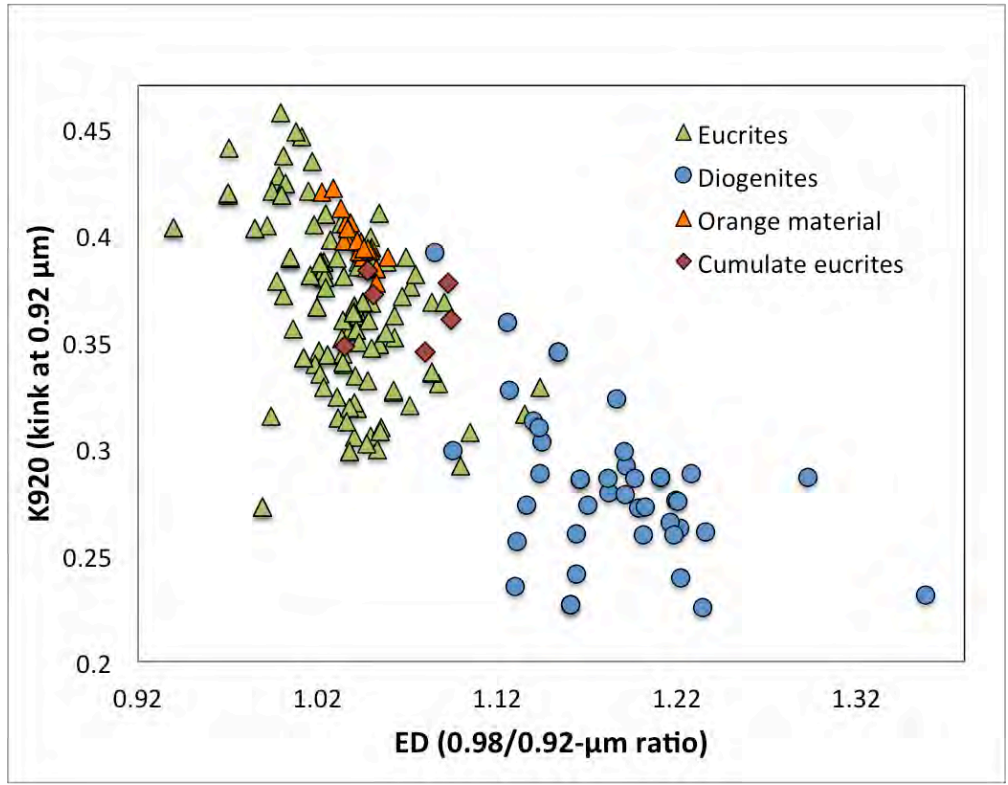

Figure 12 C.

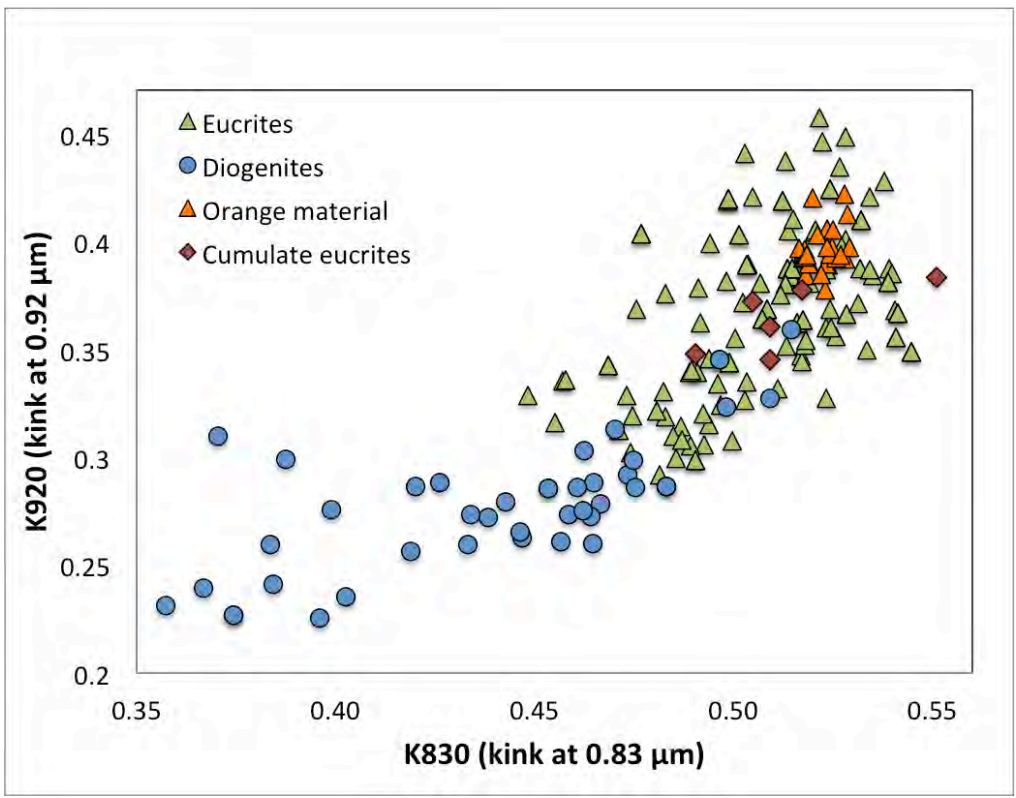

Figure 13 A.

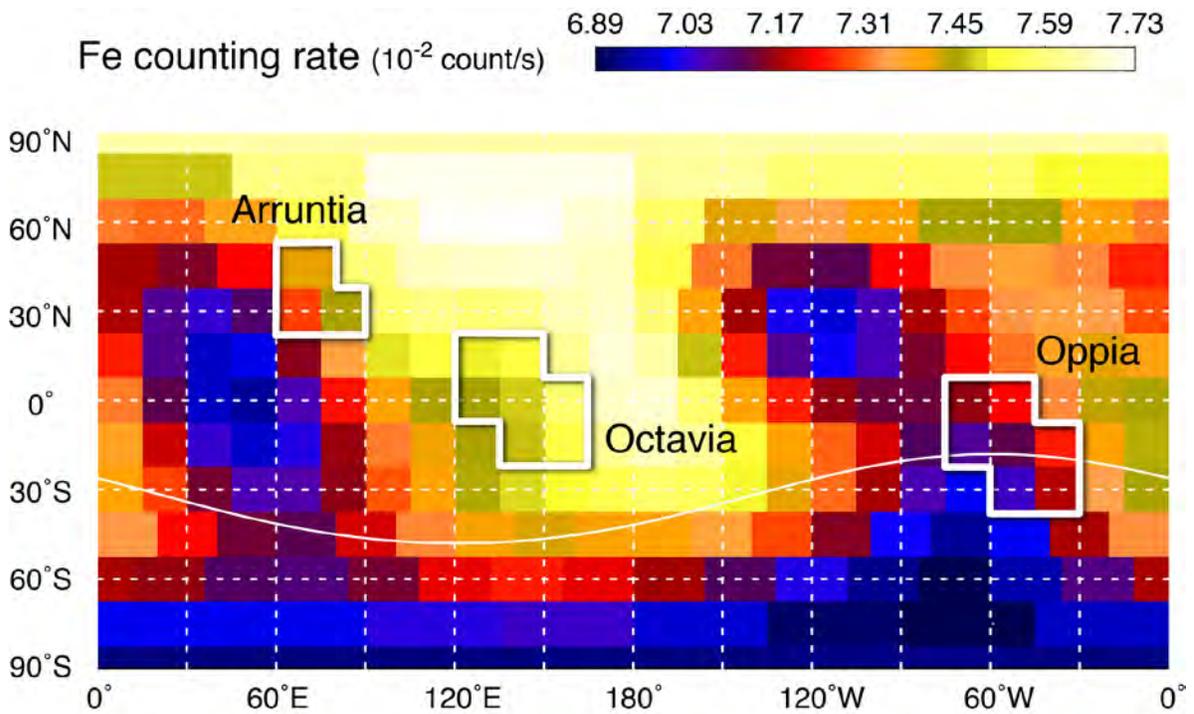

Fe counting rate (10⁻² count/s)

Figure 13 B.

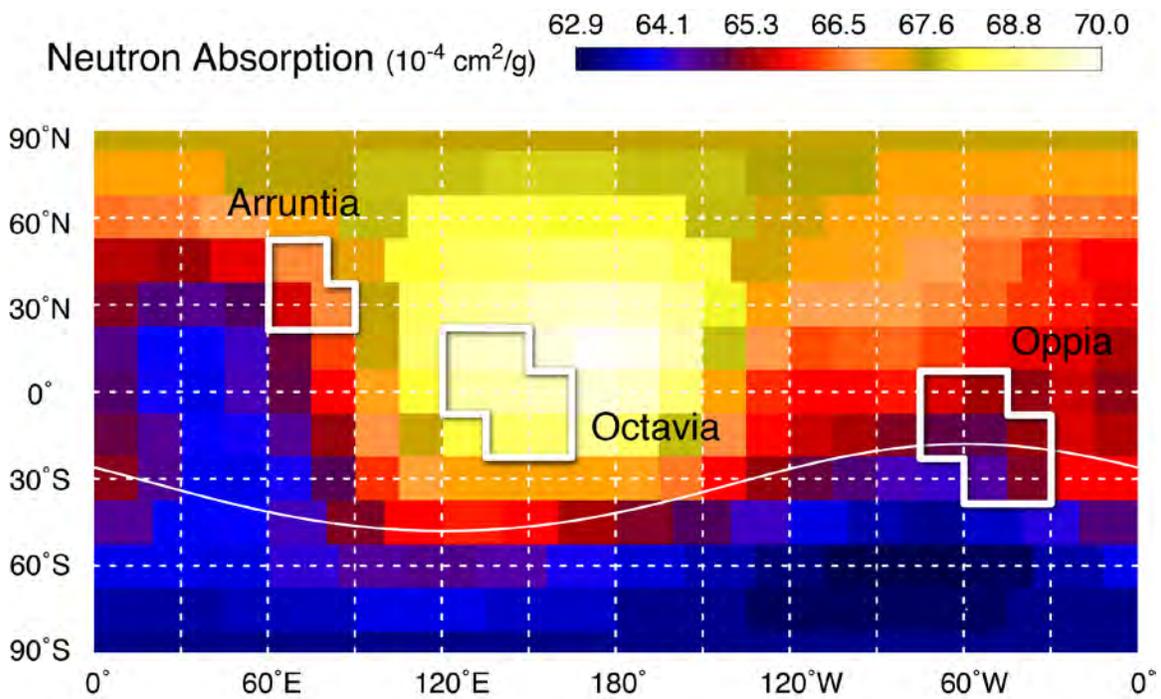

Neutron Absorption (10⁻⁴ cm²/g)

Figure 13 C.

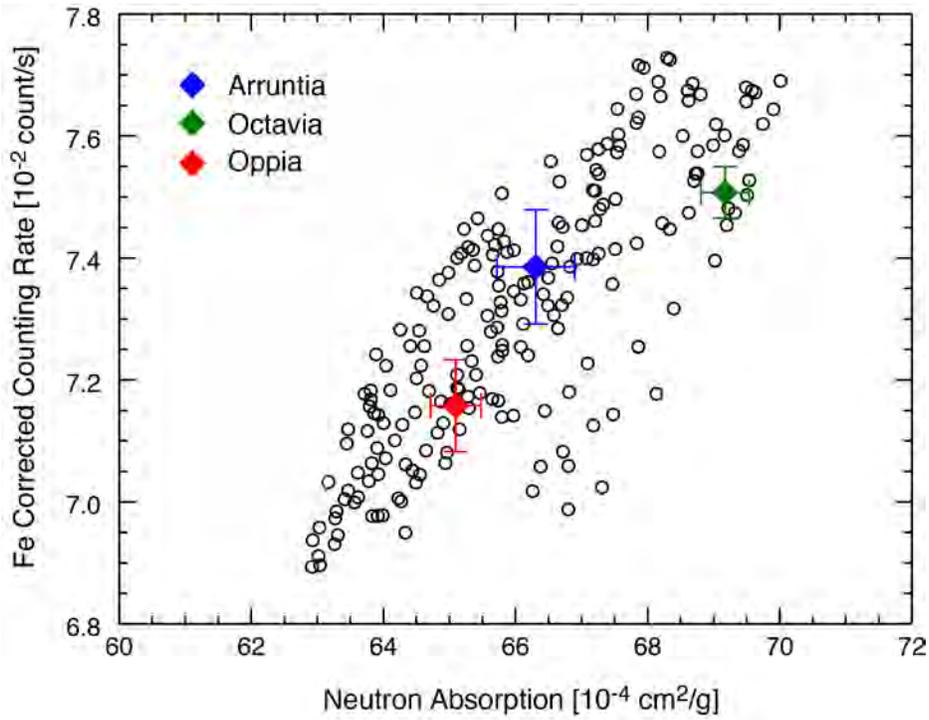

Figure 13 D.

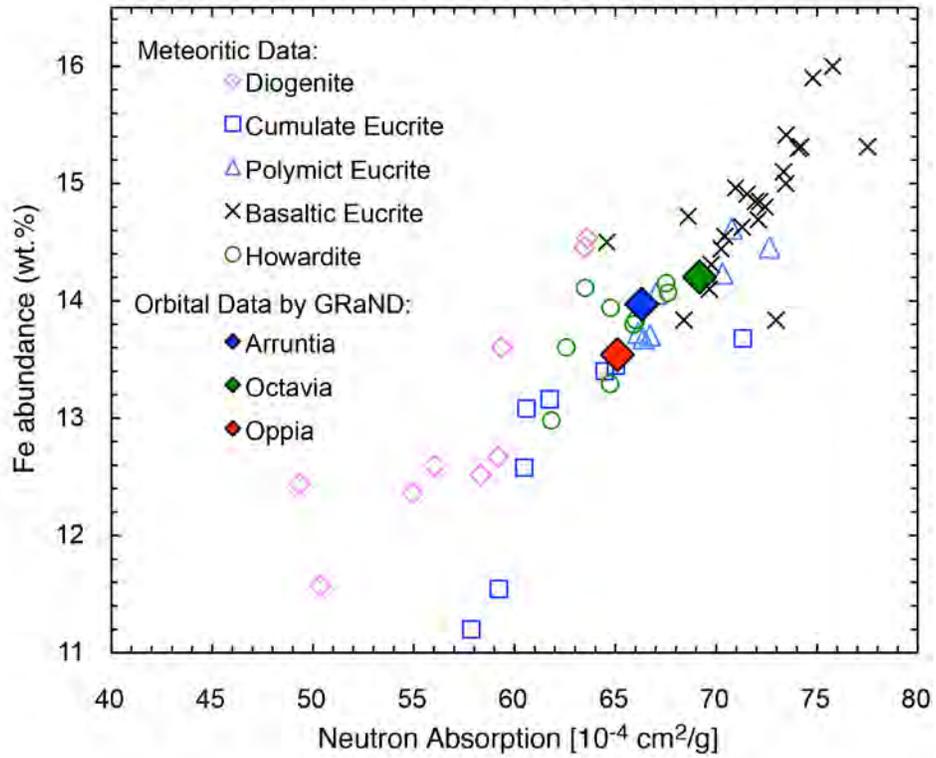

Figure 14.

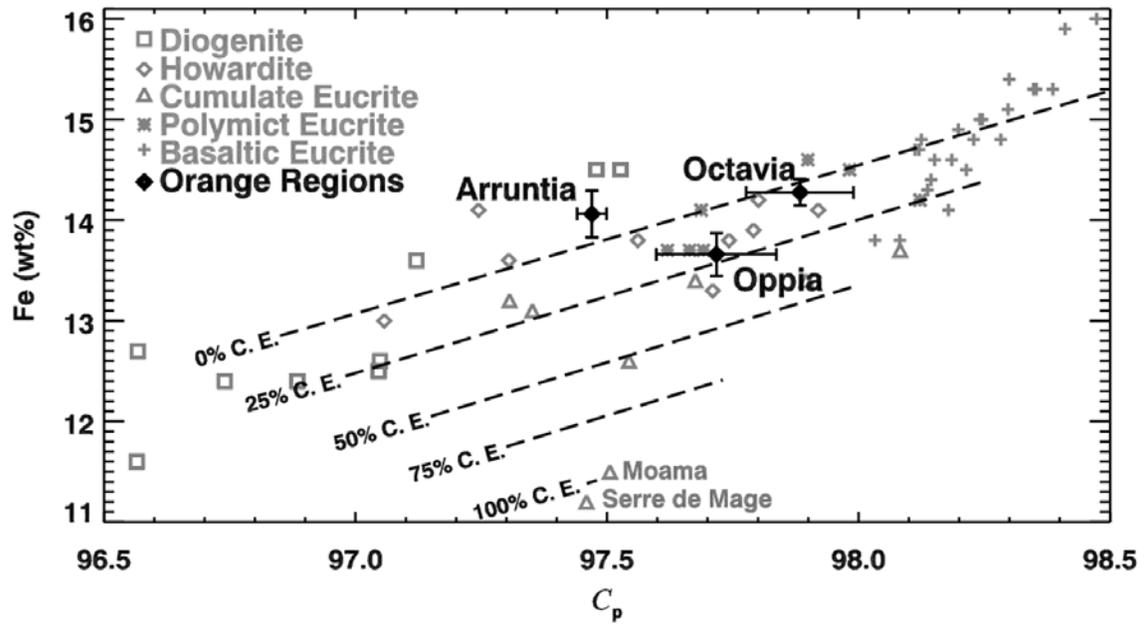

Figure 15 A.

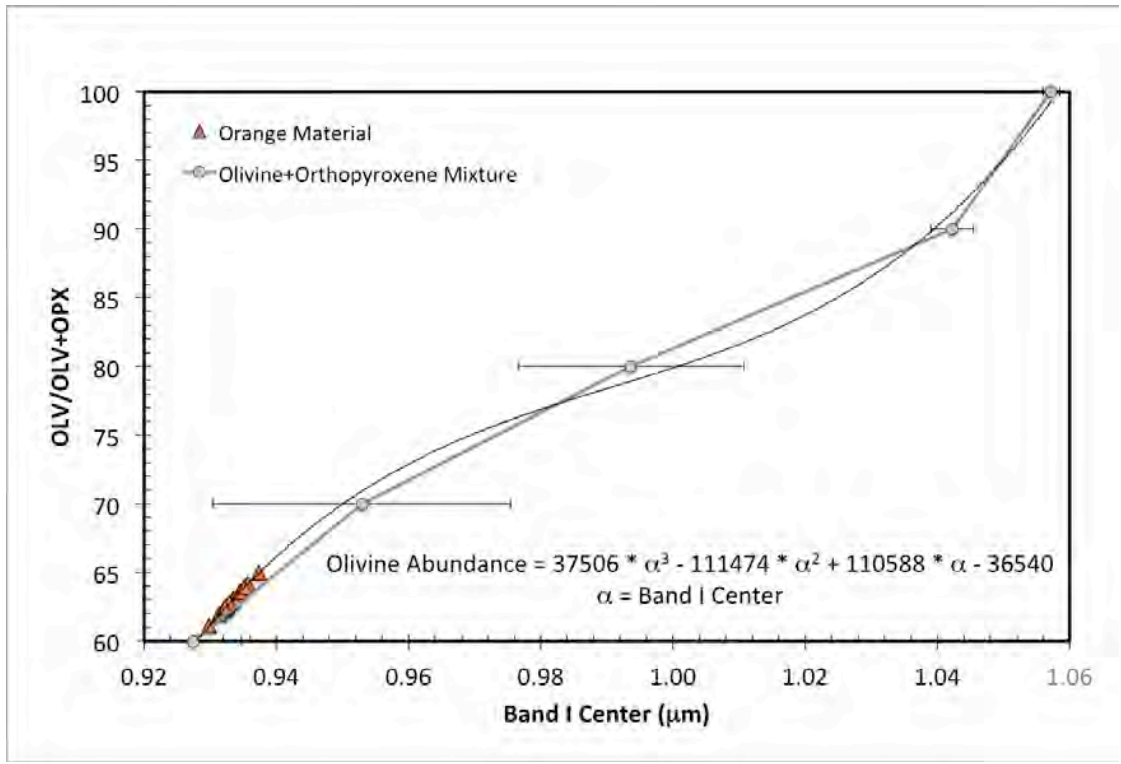

Figure 15 B.

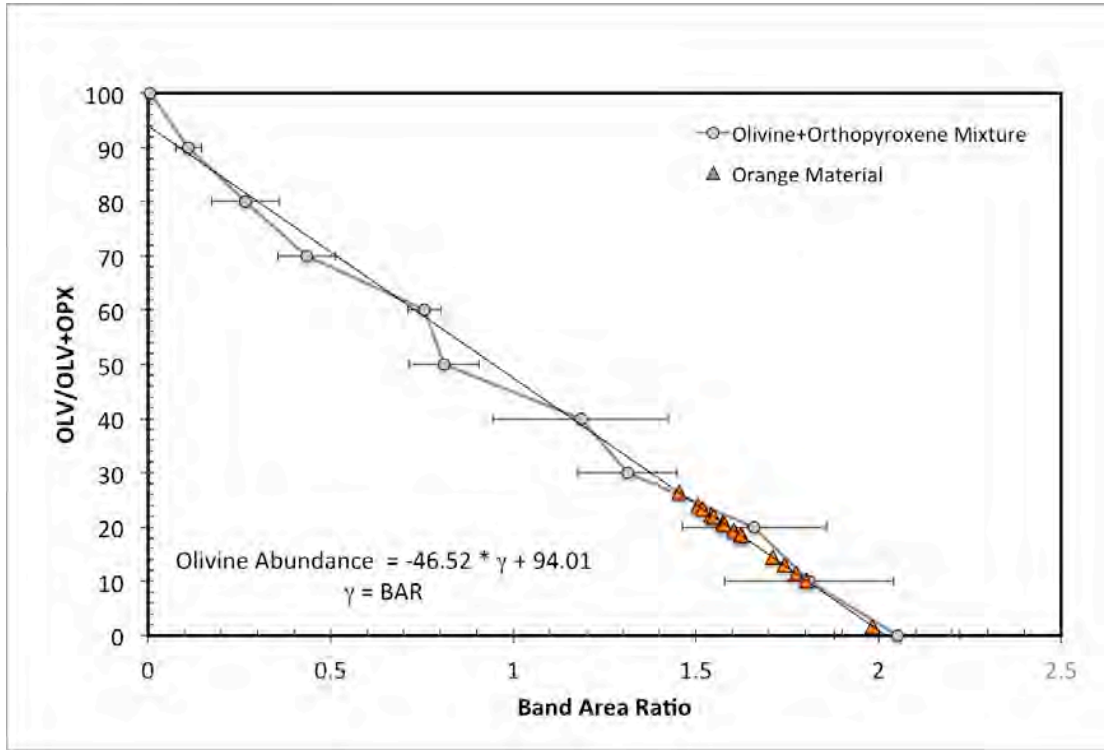

Figure 16 A.

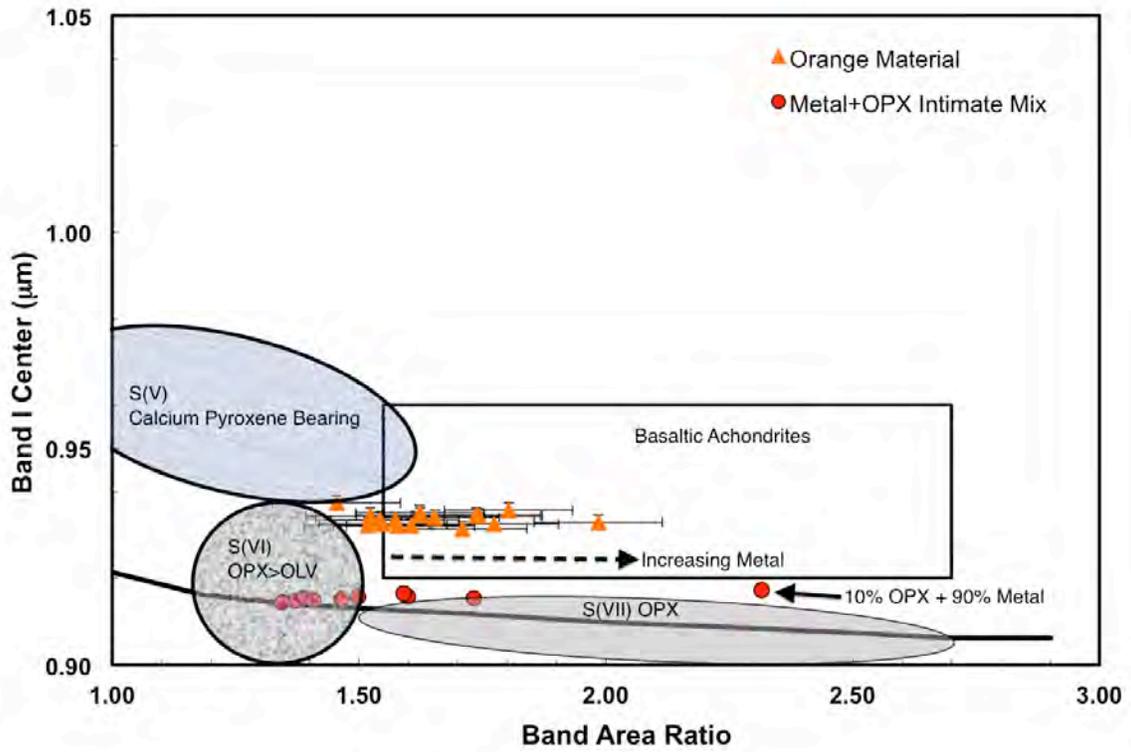

Figure 16 B.

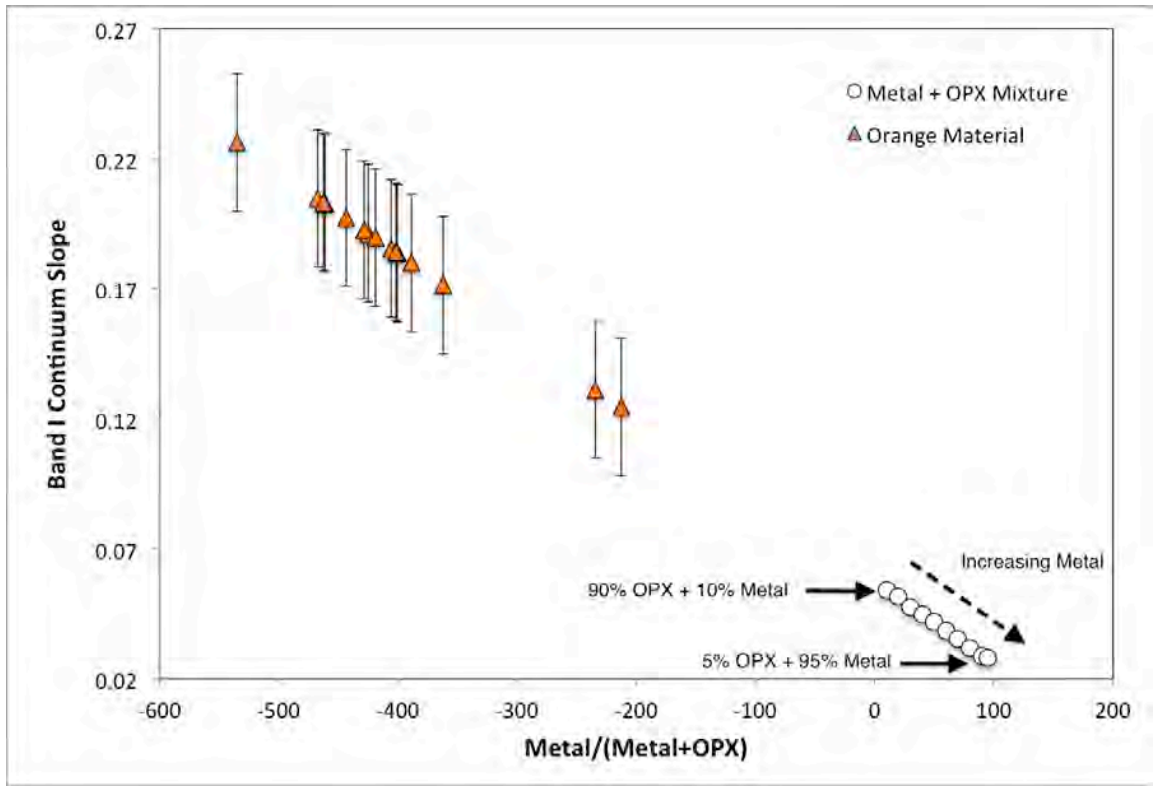

Figure 16 C.

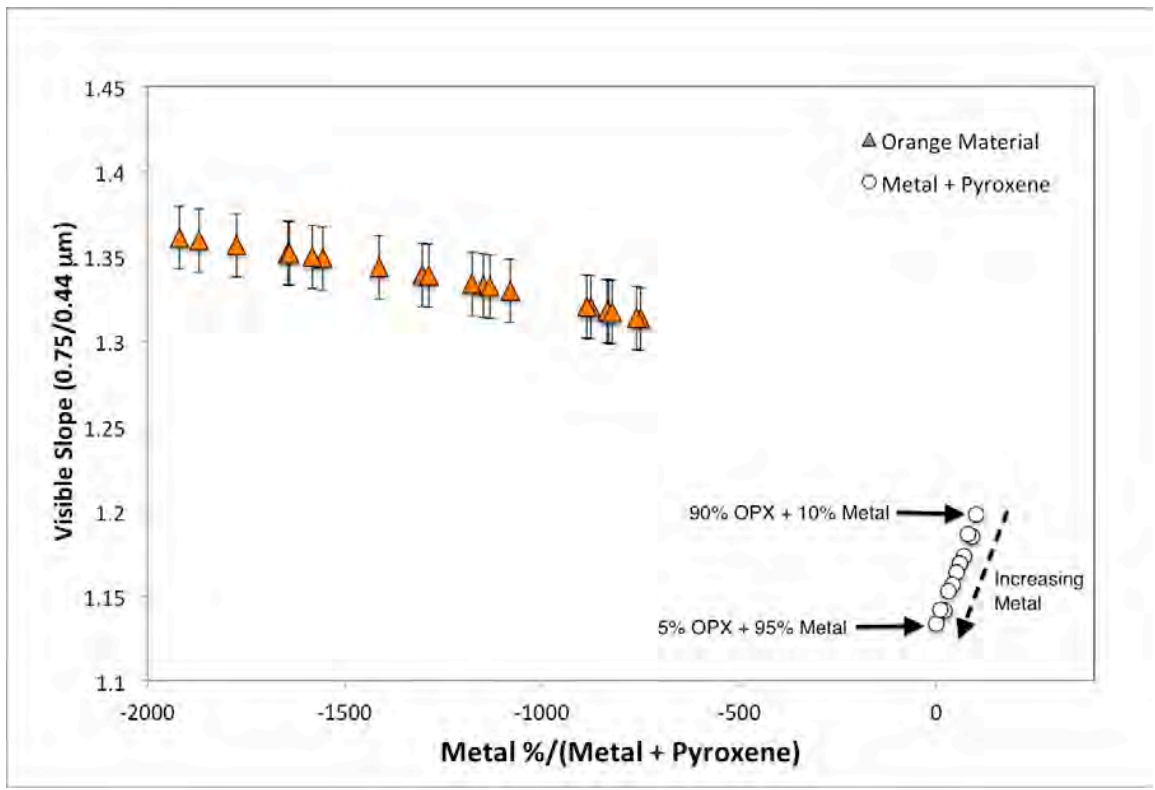

Figure 17 A.

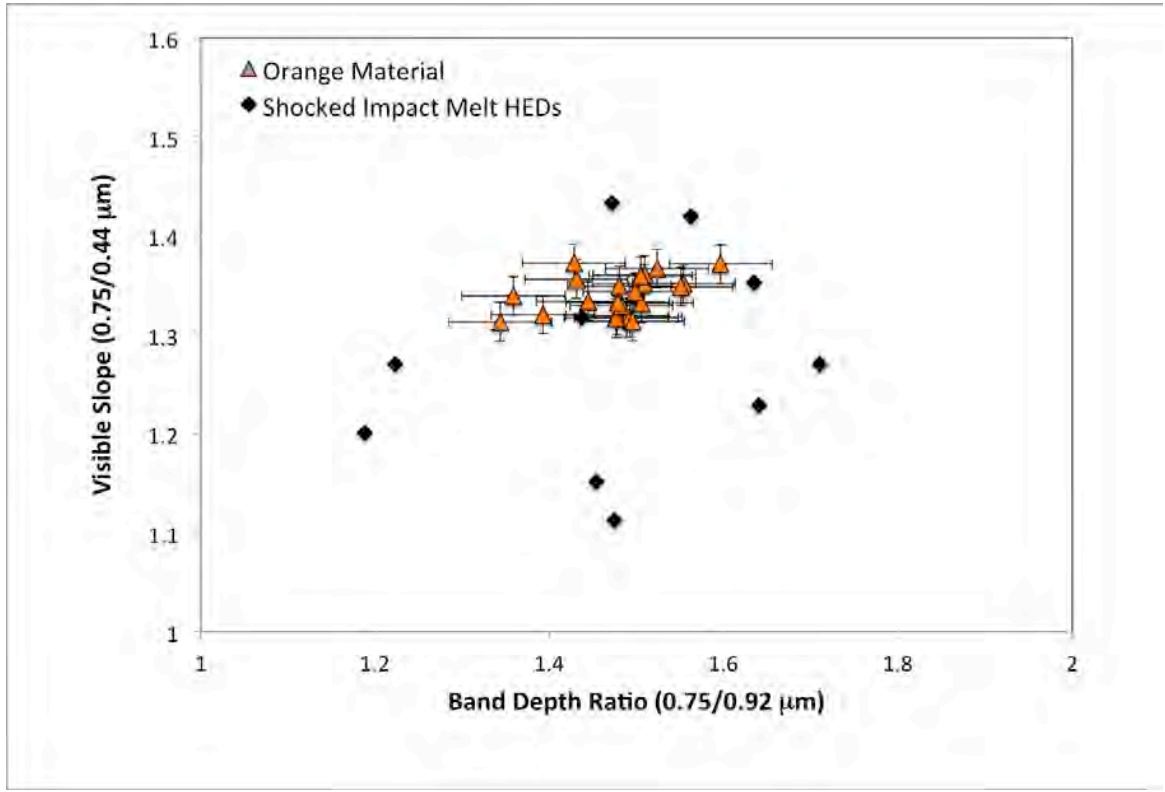

Figure 17 B.

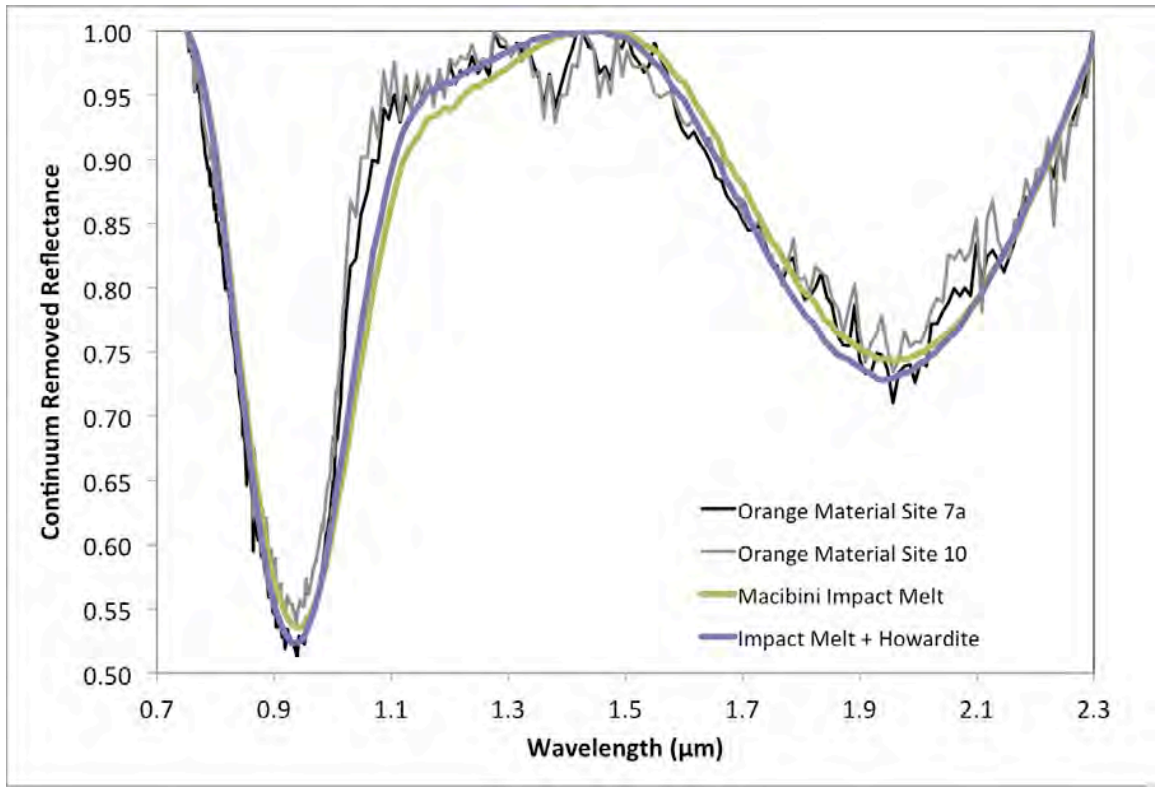

Figure 17 C.

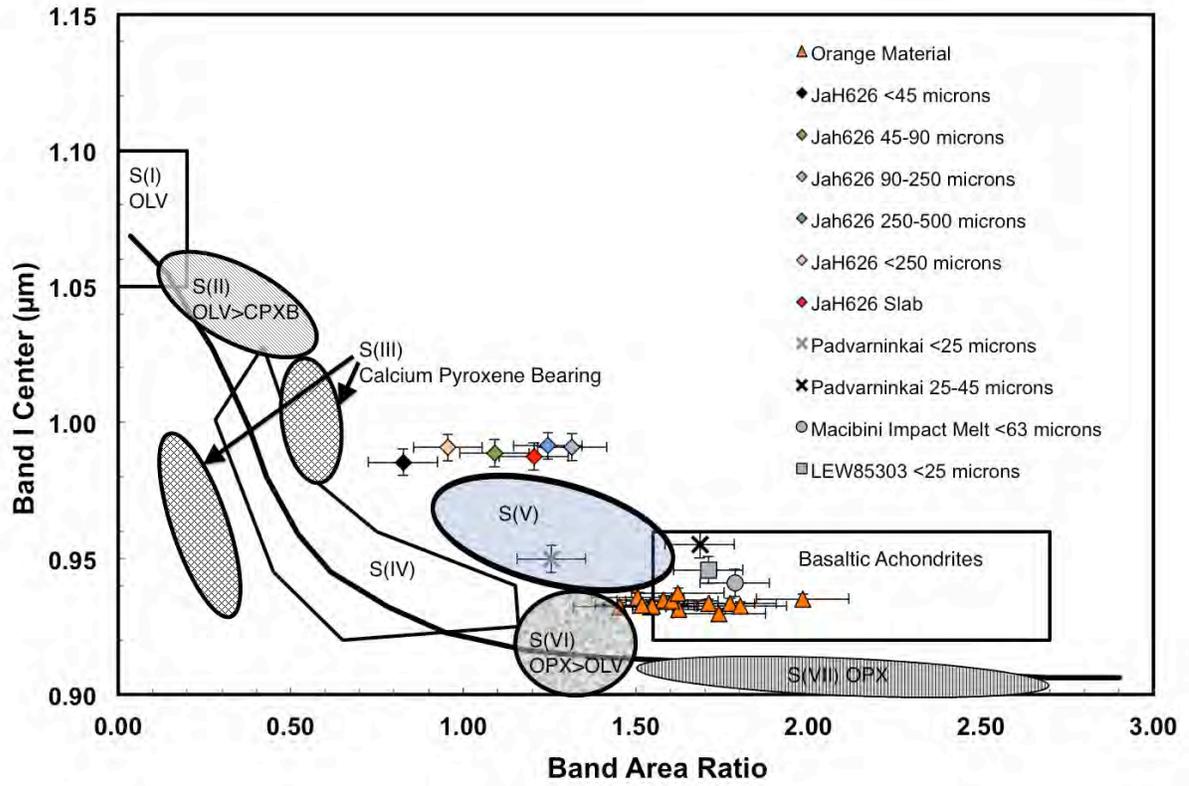

Figure 1: High-resolution version of the Figure 8. Global maps of Vesta in cylindrical projection centered at 180° and in polar projections centered on the south pole. (A) is a global mosaic in cylindrical projection in the Clementine color ratios using data from HAMO 1 observations. (B) is the equivalent but in south polar projection. (C) is a cylindrical map with color coded topography (relative to the Vestan ellipsoid) with the mapping of the orange material: the black polygons represent all the orange patches and orange crater rays (such as seen in Cornelia), black dotted polygons are used for Octavia and Arruntia orange ejecta, and white dotted polygon is used for Oppia orange ejecta. The basin rim of Veneneia is delineated in red and the Rheasilvia basin rim is depicted as a black line. (D) is the same data set and mapping as in (C) but in south polar projection.

Figure 2: (A) Global map of color-coded topography of Vesta in cylindrical projection centered at 180°. Elevation is relative to the Vestan ellipsoid. This map is showing the labels for the orange units used in this study and for which we extracted FC color spectra as well as VIR spectra when available. Mapping of the orange material is in black polygons, black dotted polygons are used for Octavia and Arruntia orange ejecta, and white dotted polygon is used for Oppia orange ejecta. The basin rim of Veneneia is delineated in red and the Rheasilvia basin rim is depicted as a black line. For good visibility of the units' definition, we added image (B) containing the Oppia crater in Clementine color ratio using data from HAMO1 observations.

Figure 1 A.

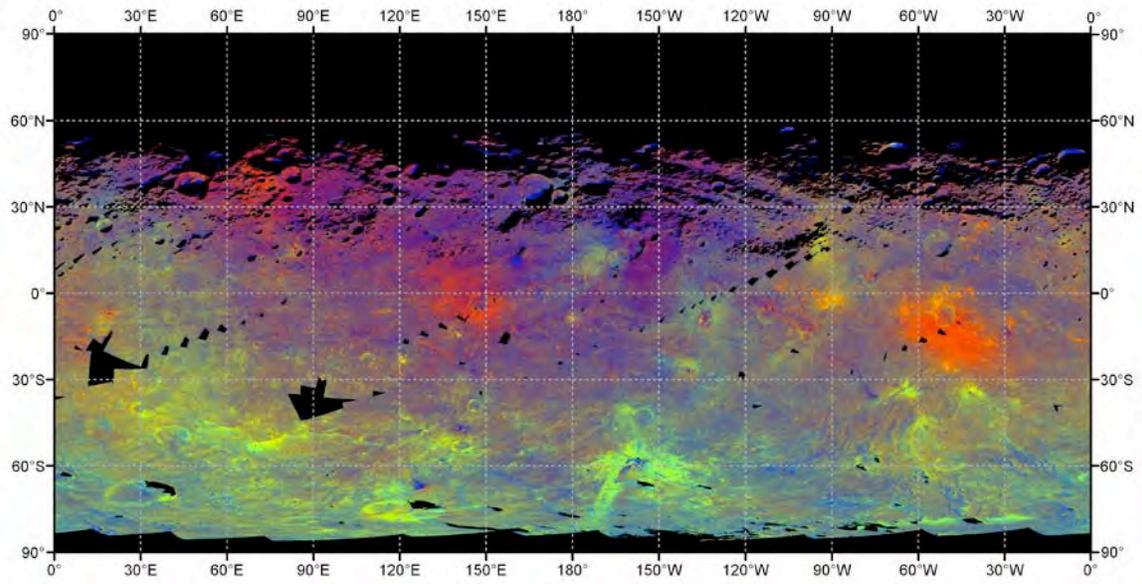

Figure 1B

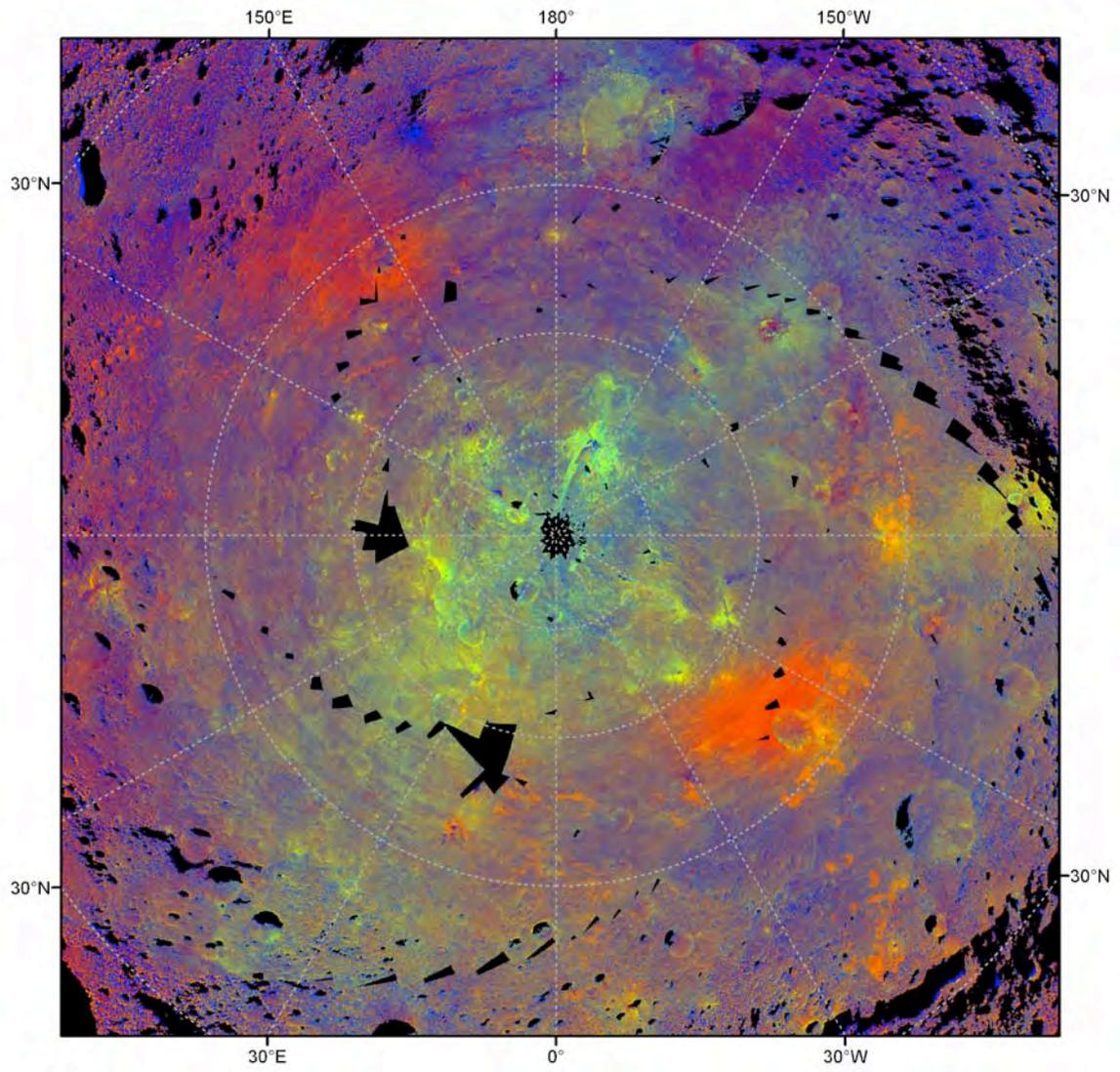

Figure 1C.

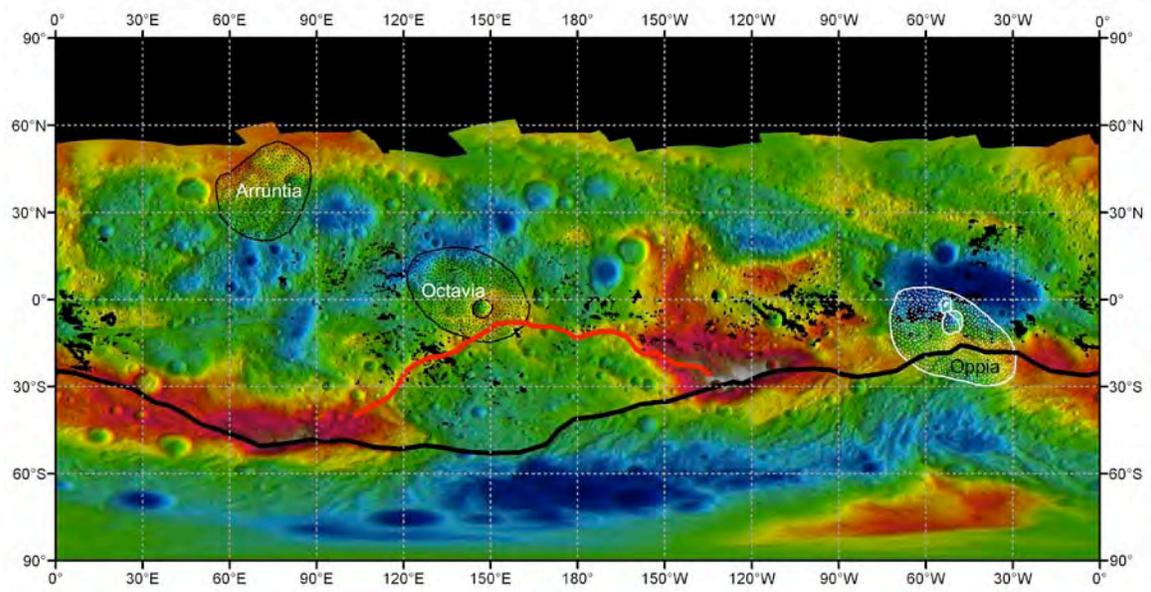

Figure 1D.

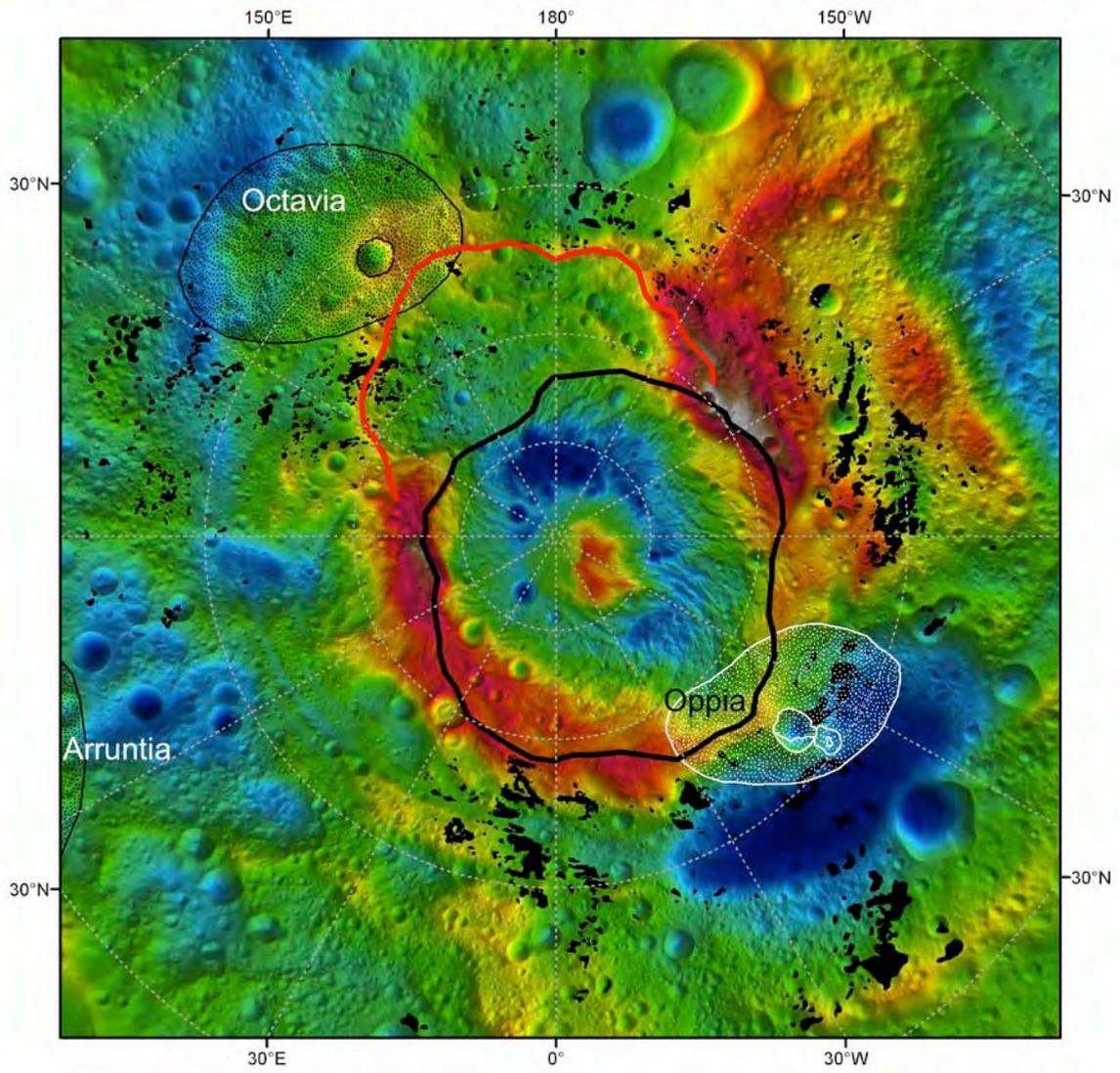

Figure 2A.

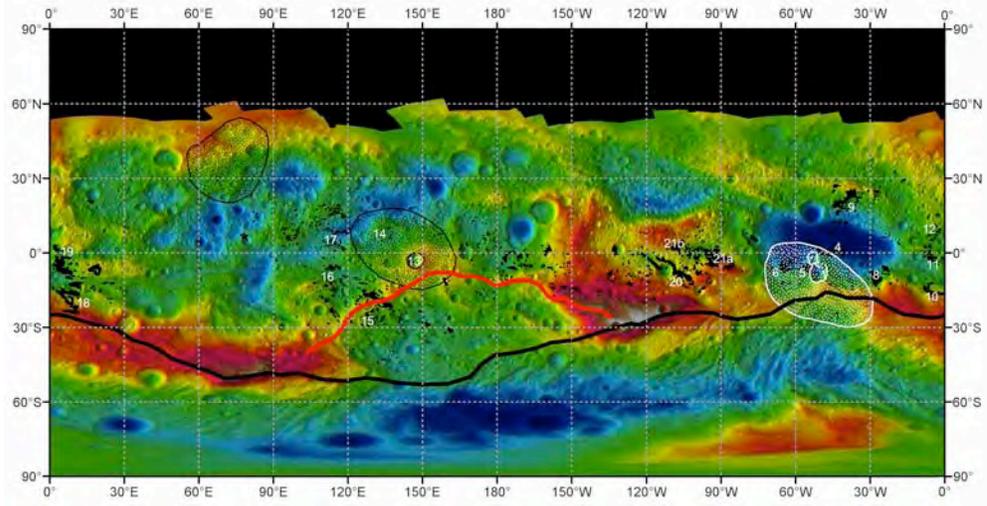

Figure 2B.

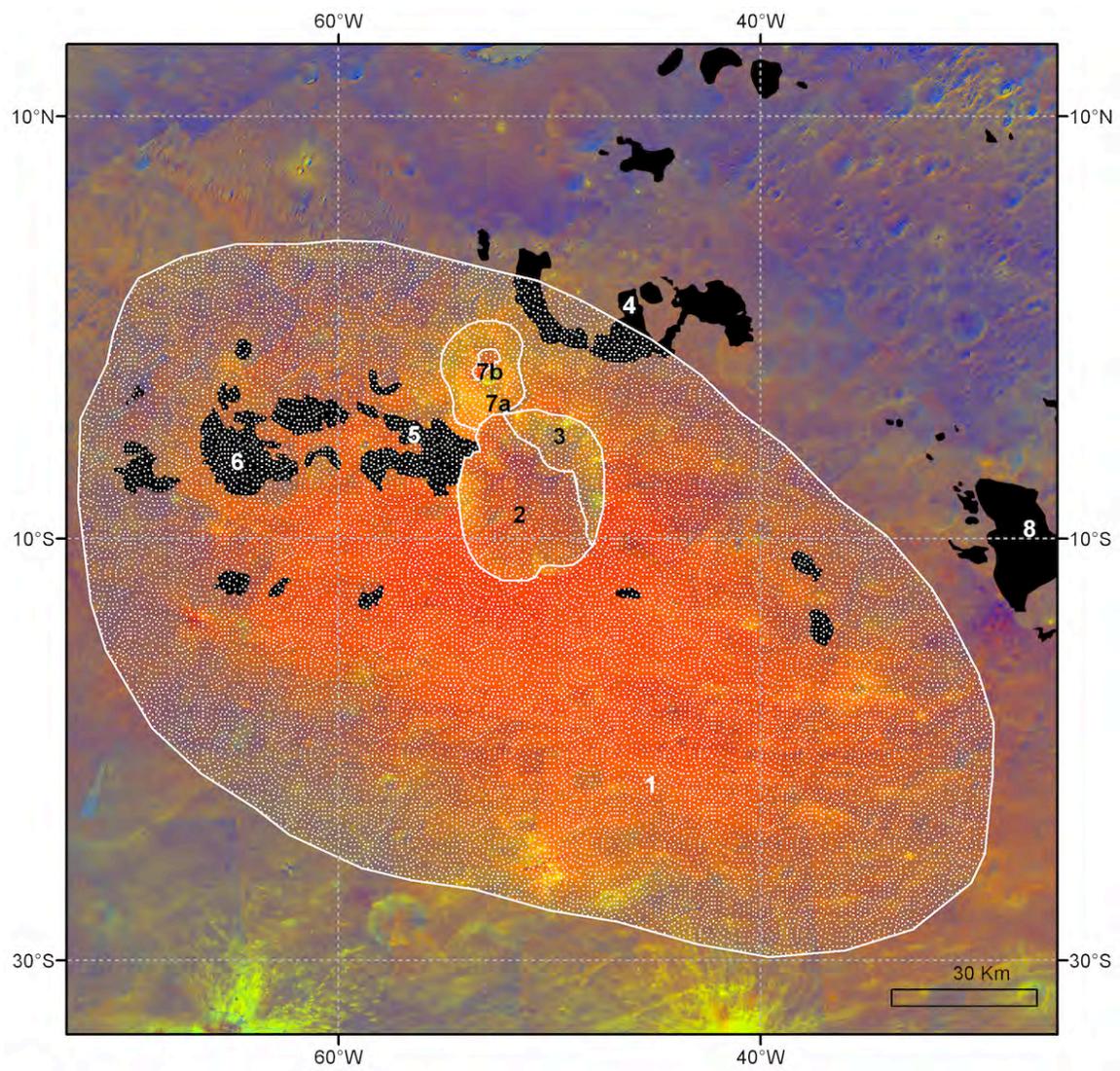